\documentclass[useAMS,usenatbib]{mnras}
\usepackage{amsmath}
\usepackage{amssymb}
\usepackage{graphicx}
\usepackage{hyperref}
\usepackage{longtable}
\usepackage{color}
\usepackage{pdflscape}

\title[Radio Properties of $z\sim5$ LBAs]{Radio observations confirm young stellar populations in local analogues to $z\sim5$ Lyman break galaxies}

\author[Greis et al.]{Stephanie M. L. Greis$^{1}$\thanks{E-mail:
s.m.l.greis@warwick.ac.uk}, Elizabeth R. Stanway$^{1}$, Andrew J. Levan$^{1}$, \and Luke J. M. Davies$^{2}$, J. J. Eldridge$^{3}$
  \newauthor \\
$^{1}$Department of Physics, University of Warwick, Gibbet Hill Road, Coventry, CV4 7AL, UK\\
$^{2}$ICRAR, The University of Western Australia, 35 Stirling Highway, Crawley, WA 6009, Australia\\
$^{3}$The University of Auckland, 38 Princes St, Auckland 1010, New Zealand}
\begin{document}

\date{Accepted 2017 May 18. Received 2017 May 15; in original form 2017 February 23}

\pagerange{\pageref{firstpage}--\pageref{lastpage}} \pubyear{2017}

\maketitle

\label{firstpage}

\begin{abstract}
We present radio observations at 1.5\,GHz of 32 local objects selected to reproduce the physical properties of $z\sim5$ star-forming galaxies. We also report non-detections of five such sources in the sub-millimetre. We find a radio-derived star formation rate which is typically half that derived from H$\alpha$ emission for the same objects. These observations support previous indications that we are observing galaxies with a young dominant stellar population, which has not yet established a strong supernova-driven synchrotron continuum. We stress caution when applying star formation rate calibrations to stellar populations younger than 100 Myr. We calibrate the conversions for younger galaxies, which are dominated by a thermal radio emission component. We improve the size constraints for these sources, compared to previous unresolved ground-based optical observations. Their physical size limits indicate very high star formation rate surface densities, several orders of magnitude higher than the local galaxy population. In typical nearby galaxies, this would imply the presence of galaxy-wide winds. Given the young stellar populations, it is unclear whether a mechanism exists in our sources that can deposit sufficient kinetic energy into the interstellar medium to drive such outflows. 
\end{abstract}

\begin{keywords}
radio continuum:galaxies -- galaxies:high-redshift -- galaxies:starburst
\end{keywords}

\section{Introduction}
Lyman break analogues (LBAs) are local star-forming galaxies selected on their UV fluxes and spectral slopes to mimic those found in Lyman break galaxies (LBGs) in the early Universe. Named after the characteristic break in their spectral flux at 912 and 1216\,\AA\ rest-frame (due to intervening clouds of neutral hydrogen), Lyman break galaxies provide a clear window onto the formation and evolution of galaxies \citep[see e.g.][]{Guhathakurta1990,Steidel1995}. They typically exhibit intense star formation, low metallicities and young dominant stellar populations, making them ideal candidates for the types of galaxies which drove the process of reionization \citep[see e.g.][]{Harford2003, Bunker2010, Rhoads2013}. While such high-redshift objects are difficult to study directly due to their apparent faintness and small angular sizes, detailed insights and interpretation of their physical properties can be gained by studying a local analogue population.

The study of local analogues for $z\sim3$ Lyman break galaxies has advanced rapidly over recent years, building on the work of \citet{Heckman2005} and \citet{Hoopes2007}. \citet{StanwayDavies2014} argue, however, that these samples are unsuitable analogues for the higher redshift ($z\sim5$) LBG population which consists of less luminous, younger and more massive sources, with lower metallicities than their $z\sim3$ counterparts \citep[see e.g.][]{Verma2007}. Building on the pilot sample of \citet{StanwayDavies2014} and further interpreted in \citet{Stanway2014}, \citet{Greis2016} established a sample of 180 galaxies at $0.05<z<0.25$ whose properties, derived from spectral energy distribution (SED) fitting and spectroscopic analyses, reproduce those of $z\sim5$ LBGs in total stellar masses, dust extinction, metallicities, and star formation rates. For a subset of their sources, \citet{StanwayDavies2014} undertook observations at 5.5 and 9\,GHz using the Australia Telescope Compact Array (ATCA). Unlike ultraviolet continuum or optical line emission, radio continuum flux is unaffected by dust attenuation in the source galaxy and thus provides an independent, and potentially more reliable star formation rate indicator. 

Radio continuum emission in galaxies arises from several mechanisms. The brightest radio sources are typically emitting from lobes powered by accretion onto the central supermassive black hole of an active galactic nucleus (AGN) \citep{Minkowski1960, Shields1999, MileyDeBreuck2008}. Similarly, smaller stellar mass black holes may emit in the radio due to accretion from a binary companion \citep{Fender2016}. Finally, star formation provides an additional channel for radio emission through its associated supernovae and their shock waves \citep{Chevalier1982}. Free-free thermal emission, or bremsstrahlung, is emitted when thermal electrons are absorbed in H\,{\sc II} regions or other dense environments, whereas non-thermal or synchrotron emission arises from cosmic-ray electrons accelerated in strong magnetic fields, such as radio jets or supernova shock waves. In the absence of an accreting central engine, both of these processes require massive stars, either to emit the ionizing ultraviolet radiation which heats the HII region, or to explode in supernovae. The radio continuum can thus be used as a tracer for the star formation rate (SFR) \citep[see e.g.][]{Rieke1980, Davies2017}. While thermal radio emission traces star formation almost instantaneously, non-thermal emission is subject to a time delay dependent on the lifespan of massive stars and so takes several tens of Myr to rise to a steady level after the onset of star formation. Hence a lack of synchrotron emission in presently star-forming systems is a good indication of a very young stellar population which has not yet attained a stable supernova-driven radio continuum. In such young galaxies, free-free radio emission dominates the spectrum.

Previous studies have found SFRs in stacked radio and submillimetre observations of $z\sim 3-5$ LBGs between 6$\pm$11 M$_{\odot}$ yr$^{-1}$ \citep[see][]{Ho2010} and $31\pm7$ M$_{\odot}$ yr$^{-1}$ \citep[see][]{Carilli2008}, comparable to their UV and Ly$\alpha$ SFRs, indicating very little, if any, obscured star formation and a very low fraction of AGN hosts. Observations at rest-frame frequencies of 1.5\,GHz (in the L-band) are effectively limited to the low-redshift ($z<0.25$) Universe with current facilities, given the relative insensitivity of meter-wave telescopes. However,  a local LBA sample like ours can be used to explore the radio properties of galaxies with known physical conditions, and thus to interpret high-redshift sources. Using green peas \citep[GPs,][]{Cardamone2009} as local analogues to z$\sim$2-5 LBGs, \citet{Chakraborti2012} suggest that both GPs and LBGs have some non-thermal or synchrotron radio emission, but, as mentioned above, it is far from clear that all LBA samples share the same characteristics.

In this paper we present the results of radio and sub-millimetre observations, taken with the Karl G. Jansky Very Large Array (VLA) and APEX/LABOCA respectively, of a subsample of our Lyman break analogue population established to mirror the properties of galaxies at $z\sim5$.  In section \ref{sec:obs} we describe the observations, before deriving radio star formation rates and star formation densities for our targets in section \ref{sec:results}. In section \ref{sec:Discussion} we consider the interpretation of our derived properties and their implications for high redshift sources, before presenting our conclusions in section \ref{sec:conclusions}. The properties of individual sources are presented in an appendix.

Throughout, we use a standard $\Lambda$CDM cosmology with H$_0$ =70 km s$^{-1}$ Mpc$^{-1}$, $\Omega_M$ = 0.3 and $\Omega_{\Lambda}$= 0.7.

\section{Observations}\label{sec:obs}
\subsection{Sample Selection}

\begin{figure}
\centering
\includegraphics[width=1.1\linewidth]{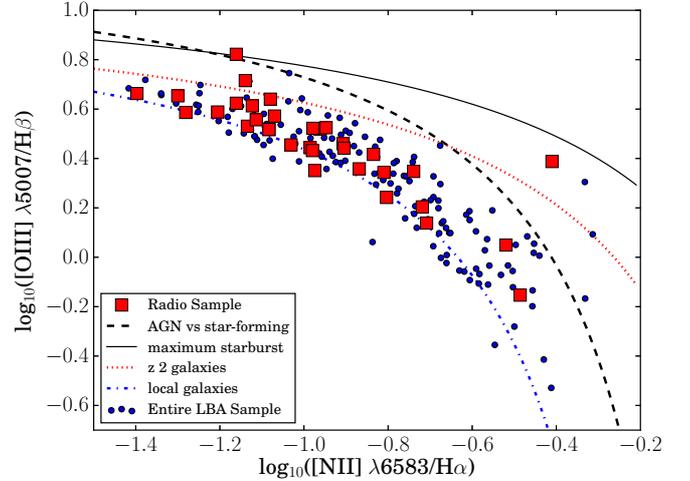}
\caption{BPT diagram showing the loci of our large LBA sample \citep[small circles,][]{Greis2016}, as well as those of the sources discussed in this paper (red squares). The sources for which we have radio observations span the entire range of the parent sample, making it plausible that the radio sources can be regarded as representative of our LBA sample. Note that one of our radio targets may have a substantial AGN contribution to its emission lines. Dashed and solid black lines indicate the AGN vs star forming criterion of \citet{Kauffmann2003} and the maximal starburst line of \citet{Kewley2001} respectively. The locus representing $z\sim2$ galaxies (red dotted line) is from \citet{Steidel2014} while the local galaxy population as detected by SDSS is shown with a dot-dash blue line.}
\label{fig:BPT}
\end{figure}

\begin{figure}
\centering
\includegraphics[width=\linewidth]{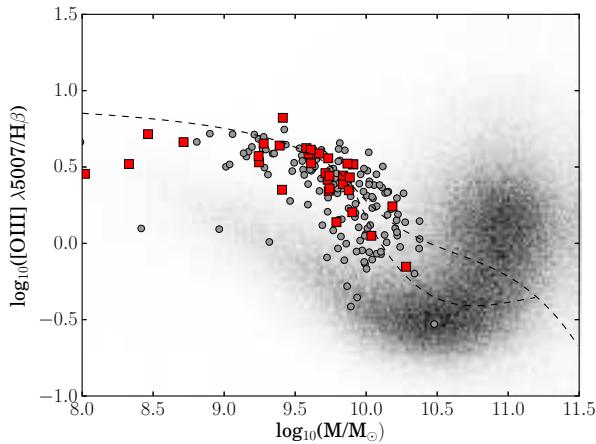}
\caption{Mass-excitation diagram showing our LBA sample \citep[small grey circles,][]{Greis2016}, overlaid on the local SDSS galaxy population (grey distribution). The locations of the galaxies described in this paper are marked by red squares. There is a significant offset between our LBA sample and the bulk of the local galaxy population with a higher ionizing potential observed in the LBAs. The dashed line indicates the proposed separation between star formation (below) and AGN-driven (above) ionizing radiation according to \citet{Juneau2011}. Objects in the bordered region are classified as composites.}
\label{fig:MEx}
\end{figure}

Targets were selected from our catalogue of 180 LBAs as presented in \citet{Greis2016}. Slight preference was given to those objects which had already been observed with ATCA at 5.5 and 9.0\,GHz \citep[see][]{StanwayDavies2014} and those whose SED-derived physical properties made them ideal local analogues to $z\sim5$ LBGs in terms of their masses and ages. Their optically-derived properties are listed in Table \ref{table:PhysProperties}. For brevity, we use the last 5 digits of the SDSS DR7 object identifier as an object designation in the remainder of the paper. Figures \ref{fig:BPT} and \ref{fig:MEx} show the loci of our entire LBA sample as small circles on the BPT \citep{BPT1981} and mass-excitation diagrams \citep{Juneau2011} respectively. The red squares indicate the locations of the radio-observed sources, which can be seen to span the whole range of properties of the LBA sample, making them a representative subsample.

\begin{table*}
\begin{center}
    \hspace*{-1.5cm}
    \small{
    \begin{tabular}{ | l | l | l | l | l | l | l | l | l | l | l | l |}
    \hline
    ObjID & ra & dec & redshift & FUV & mass & dust & age & H$\alpha$ SFR & metal. & observed by\\
     (SDSS DR7) & & & & AB mag & log$_{10}$(M$_{\odot}$) & $E(B-V)$ & log$_{10}$(yrs) & M$_{\odot}$ yr$^{-1}$ & Z$_{\odot}$ & \\
 \hline \hline
587727876380754061 & 355.41867 & -8.71988 & 0.074 & 18.66 & 8.0 & 0.13 & 7.2 & 3.1 & 0.23 & VLA \& ATCA  \\  \hline
587726877273227473 & 336.64663 & -9.68499 & 0.083 & 20.39 & 9.2 & 0.05 & 8.7 & 1.3 & 0.18 & VLA \& ATCA \\  \hline 
587730817902116911 & 344.79825 & -8.77026 & 0.097 & 19.31 & 9.8 & 0.18 & 8.6 & 5.1 & 1.62 & VLA \& ATCA  \\  \hline 
587727226227523734 & 1.16389 & -10.15264 & 0.108 & 19.80 & 8.3 & 0.26 & 7.2 & 10.4 & 0.34 & VLA \& ATCA \\  \hline 
587734841741083073 & 115.62922 & 21.334231 & 0.110 & 20.60 & 9.6 & 0.09 & 8.8 & 2.6 & 0.14 & VLA \\  \hline 

587737827281076428 & 111.6581 & 39.766087 & 0.111 & 19.86 & 9.4 & 0.16 & 8.6 & 8.0 & 0.46 & VLA \\  \hline 
587727178464624784 & 35.15693 & -9.48536 & 0.113 & 19.78 & 8.5 & 0.06 & 6.7 & 6.8 & 0.35 & VLA \& ATCA \& APEX \\  \hline  
587745540508680573 & 128.92412 & 10.299852 & 0.115 & 20.59 & 10.0 & 0.14 & 9.0 & 1.8 & 0.68 & VLA \\  \hline 
587730774950608959 & 332.01197 & 13.226264 & 0.116 & 19.55 & 9.3 & 0.11 & 8.1 & 7.6 & 0.18 & VLA \\  \hline 
587732703947653150 & 178.63296 & 8.5770854 & 0.117 & 18.72 & 9.4 & 0.11 & 8.4 & 10.9 & 0.35 & VLA \\  \hline 

587726877810360392 & 337.13304 & -9.46809 & 0.120 & 20.29 & 10.3 & 0.2 & 9.8 & 4.0 & 0.88 & VLA \& ATCA \\  \hline  
587730818439577821 & 346.21707 & -8.6318 & 0.121 & 19.94 & 9.9 & 0.11 & 8.4 & 3.9 & 0.45 & VLA \\  \hline  
587739381531476079 & 235.02877 & 24.51249 & 0.122 & 19.46 & 9.7 & 0.15 & 8.6 & 8.1 & 0.22 & VLA \\  \hline 
588017979426537518 & 174.90376 & 39.982965 & 0.130 & 19.66 & 9.8 & 0.07 & 8.8 & 2.7 & 0.43 & VLA \\  \hline 
587727179534762100 & 26.84033 & -9.27951 & 0.136 & 19.98 & 8.7 & 0.02 & 7.3 & 5.2 & 0.09 & VLA \& ATCA \\  \hline 
587733431922327825 & 251.40989 & 28.985973 & 0.136 & 20.30 & 9.8 & 0.17 & 8.7 & 3.8 & 0.25 & VLA \\  \hline 

587739648346357993 & 143.6293 & 26.285418 & 0.138 & 20.44 & 9.7 & 0.13 & 8.5 & 3.2 & 0.40 & VLA \\  \hline 
587736542021091412 & 221.73053 & 7.7560733 & 0.142 & 19.27 & 9.6 & 0.05 & 8.3 & 9.5 & 0.24 & VLA \\  \hline 
587735431232356415 & 212.84126 & 47.48035 & 0.145 & 19.73 & 9.7 & 0.08 & 8.7 & 7.6 & 0.16 & VLA \\  \hline 
587737809027334524 & 112.65454 & 39.143992 & 0.145 & 20.73 & 9.9 & 0.14 & 8.9 & 2.8 & 0.44 & VLA \\  \hline
588017949897523218 & 211.19882 & 40.646364 & 0.145 & 20.08 & 9.9 & 0.10 & 8.8 & 6.6 & 0.26 & VLA \\  \hline 

587727230523605083 & 27.9936 & -9.38417 & 0.146 & 20.36 & 9.9 & 0.12 & 8.7 & 8.1 & 0.34 & VLA \& ATCA \& APEX \\  \hline 
587730816826671294 & 340.88031 & -9.44735 & 0.146 & 19.96 & 9.2 & 0.11 & 7.7 & 8.3 & 0.16 & VLA \& ATCA  \\  \hline 
587739827674808586 & 238.20283 & 16.985618 & 0.149 & 20.48 & 9.9 & 0.10 & 8.7 & 5.2 & 0.27 & VLA \\  \hline 
587727178463772856 & 33.14848 & -9.63883 & 0.150 & 20.06 & 9.4 & 0.11 & 8.0 & 4.9 & 0.24 & VLA \& APEX \\  \hline
587735430151929942 & 189.89012 & 49.909903 & 0.150 & 20.43 & 9.7 & 0.12 & 8.7 & 2.0 & 0.27 & VLA \\  \hline 

587727178997432420 & 25.70536 & -9.60746 & 0.161 & 20.1 & 10.2 & 0.34 & 8.7 & 41.8 & 0.58 & VLA \& APEX \\  \hline
587727180071108755 & 25.58712 & -8.76605 & 0.164 & 20.41 & 9.7 & 0.21 & 8.1 & 15.8 & 0.35 & VLA \& ATCA \\  \hline 
587732771035938857 & 137.37213 & 6.5544124 & 0.181 & 19.73 & 9.7 & 0.14 & 8.0 & 19.6 & 0.28 & VLA \\  \hline 
587724240687792239 & 37.28071 & -8.95727 & 0.183 & 20.00 & 9.8 & 0.06 & 8.5 & 10.4 & 0.17 & APEX \\  \hline 
587741601493155849 & 179.62691 & 27.123929 & 0.183 & 19.92 & 9.7 & 0.12 & 8.4 & 7.1 & 0.25 & VLA \\  \hline 
587728668808315004 & 128.45934 & 45.825978 & 0.188 & 20.18 & 9.6 & 0.15 & 8.3 & 21.9 & 0.26 & VLA \\  \hline  
588017704028995952 & 237.44271 & 7.9204591 & 0.198 & 20.18 & 9.6 & 0.10 & 8.2 & 10.4 & 0.43 & VLA \\  \hline 
\hline
    \end{tabular}}
    \caption{Some of the physical properties of our radio targets, derived from SDSS photometry and spectroscopy. Columns give the SDSS identification number, the object's sky position and redshift, and observed FUV magnitude from GALEX. The masses and ages for each object were determined using SED fitting (based on the {\sc BPASS} spectral synthesis model set, see \citet{Greis2016}), while the dust continuum $E(B-V)$ value was calculated from the Balmer decrement. The star formation rates are based on extinction-corrected H$\alpha$ line fluxes, and quoted metallicities use the \citet{Dopita2016} calibration. The last column indicates by which telescope(s) each object was observed. For brevity, we use the last 5 digits of the SDSS object identifier as an object designation in the remainder of the paper.}
    \label{table:PhysProperties}
\end{center}
\end{table*}

\subsection{VLA}
The observations were carried out with the Karl G. Jansky Very Large Array (VLA) in programmes 14A-130, 15A-134, and 16B-104 (PI: Stanway) between March and May 2014, July and September 2015, and September 2016 and January 2017 respectively (see Table \ref{table:VLAObservations}). Each object was observed in a 45 minute long scheduling block with a typical on source time of $\sim$ 20 minutes, split into 3 independent scans of near-equivalent length. All observing programmes used the `A' configuration and identical setups with the exception of targets observed in September 2016 which were observed in a transitional `BnA' configuration and those observed in January 2017, observed in the `AnD' configuration. Observations were taken in the L-band, centred at 1.5\,GHz, and making use of the 1\,GHz bandwidth of the VLA. Radio frequency interference (RFI) reduces the effective frequency bandwidth to $\sim0.5$\,GHz but with substantial variation from source to source. One observation had to be excluded due a bright nearby source which rendered imaging of the faint target impractical. In no other cases was a suffienctly bright target in the field to preclude detection of the science target. As all targets have previously been detected in the ultraviolet and optical, their positions are well-constrained. For each target, therefore, the area to be searched for counterparts consistent with the optical location is very small and the possibility of the observed sources arising from chance alignment is minimal. The 32 observed sources span a redshift range of $0.07<z<0.20$. 

For each source, phase and gain calibration was achieved using a well-established secondary phase calibrator from the VLA calibrator database\footnote{\href{http://www.vla.nrao.edu/astro/calib/manual/csource.html}{VLA calibrator database}}. The calibrator was observed for $\sim2$ minutes between each $\sim8$ minute on-source integration and was typically of order 1Jy in flux and between 5 and 10 degrees from the science target. Absolute flux, delay and bandpass calibration was accomplished through observations of one of the standard VLA calibration sources, 3C48, 3C147, or 3C286 with the appropriate object chosen for the target RA. These sources are regularly monitored by the National Radio Astronomy Observatory (NRAO). Observations were analysed using v4.7.1 of the \textsc{Common Astronomy Software Application} (\textsc{CASA}) package. Each observing block was pipeline-reduced using the VLA data calibration pipeline\footnote{more information on the \href{https://science.nrao.edu/facilities/vla/data-processing/pipeline}{VLA CASA calibration pipeline}}, which is integrated into current versions of \textsc{CASA}. The pipeline is optimised for Stokes I continuum data and performs data-flagging and Hanning smoothing before calculating delay, gain, flux and bandpass calibrations. These are applied to the target data and diagnostic images of the calibrators are constructed. 

If necessary, the target data were then manually flagged for any remaining RFI which is pervasive at the frequencies probed. Following flagging and pipeline calibration, the measurement sets were `cleaned' interactively. We adopted the standard \textsc{CASA} `clean' procedure to invert the visibility data. We used multi-frequency synthesis (`mfs') mode to create one image in the Stokes I (total flux) parameter for each object. This procedure can be problematic for datasets with a large fractional bandwidth, if there is spectral curvature. The standard procedure assumes that the source has a constant spectral index. For our most robust sources, we sub-imaged the data in different frequency bands and saw no evidence for significant flux variation with frequency. We note that the \textsc{CASA} `clean' algorithm is designed to deal with wideband multi-frequency synthesis imaging and we used the recommended parameters for low signal-to-noise data. `Clean' creates a dirty image and primary beam pattern before iteratively reconstructing the sky image. After imaging test cases using different cell sizes, a cell size of 0.5 arcsec was found to produce the optimum signal-to-noise ratio. Our targets were too faint to consider polarimetric data. We imaged using `Briggs' weighting which implements a flexible weighting scheme and a Briggs' robustness parameter of 0.5. `Cleaning' was performed using the Clark cleaning algorithm and imaged with the phase centre at the pointing centre. 

After successful `cleaning' of both the calibrator and science target, we determined that each frame contained a number of mJy-level sources, although these were typically sparesely distributed through the field-of-view and did not affect flux measurement of the science target. In a few cases, a brighter target was present in the field and its sidelobe pattern apparent. In these cases, the image was carefully inspected and the local noise level used as an estimate of the rms. In no case was the science target detection consistent with a sidelobe detection of a brighter source. After verifying each detection through visual inspection, we split the targets' measurement sets into individual scans and imaged each scan independently to ensure that the object was detected in all three scans. Given the effects of RFI, the sensitivity varied substantially from scan to scan. Objects for which two out of the three scans yielded non-detections were considered to have a detection only if the source was detected in the scan with the lowest rms noise level. Measurements of the radio fluxes and object sizes were obtained from the images combining all 3 scans using \textsc{CASA}'s `imfit' routine with 3 to 5 arcsec radii around the source location (where imfit did not converge, we used `imstat' to derive upper flux limits). 

\subsection{APEX} 
Five LBA targets were observed in the 870 $\mu$m (345 GHz) atmospheric window using the Large Apex BOlometer CAmera (LABOCA) at the Atacama Pathfinder EXperiment (APEX) as part of ESO programme ID 093.B-0414(A) (PI: Stanway). These observations used a compact spiral raster mode designed to optimise coverage in the central arcminute. While the allocated time was 11 hours, our observed programme totalled 6.8 hours of which approximately a third of the time was on source; further details are shown in Table \ref{table:APEXObservations}. All observations used the same setup. Initial pointing and focus observations were obtained on a Solar System planet. The pointing was then refined and the flux scale determined using one of the standard APEX/LABOCA secondary flux calibrators: HLTAU, N2071IR, CRL618 and V883-ORI. Regular skydip observations were performed to calibrate for atmospheric opacity. Observations were reduced using v1.05-3 of the MiniCRUSH software suite\footnote{\href{http://www.astro.caltech.edu/~sharc/crush}{MiniCRUSH software}} \citep{Kovacs2008}, a minimalist version of the data reduction software CRUSH optimised for the APEX bolometers. This removes correlated noise in the time stream, which is then used to construct a source model using a maximum likelihood estimator. MiniCRUSH is a non-interactive software. We selected the `deep' pipeline option, which aggressively filters correlated noise and is optimally filtered (smoothed) to optimise signal-to-noise on point sources. This will underestimate extended structure, but we expect none on this scale. The pixel scale used was 4 arcsec. The APEX beam has a FWHM of 19.5 arcsec and the pipeline smoothes the data to a FWHM of 27.58 arcsec. 

At $z\sim 0.1 - 0.2$, the 870 $\mu$m band of LABOCA is close to the transition between the dust-dominated thermal black-body, which dominates the far-infrared, and the radio continuum emission that dominates at $\sim$1 GHz. Hence the flux density from both SED components is relatively small in the observed band.

\begin{table*}
\begin{center}
    \begin{tabular}{ | c | c | c | c | c |}
    \hline
	ObjID 	& Observing	& Beam size	& Exposure	& rms\\
	 	& date		& arcsec$^2$	& (min)		& $\mu$Jy\\
 \hline \hline
16911 &	2014 Mar 31	& 1.7 $\times$ 1.1	& 27.7	& 30 	 \\ \hline
54061 &	2014 May 16	& 1.8 $\times$ 1.1	& 26.2	& 43 	 \\ \hline
71294 &	2014 May 22	& 1.9 $\times$ 1.1	& 25.5	& 30 	 \\ \hline
77821 &	2014 May 28	& 1.6 $\times$ 1.1	& 25.0	& 37	 \\ \hline 
60392 &	2014 May 31	& 1.9 $\times$ 1.1	& 25.7	& 30     \\ \hline

76428 &	2015 Jul 14	& 3.2 $\times$ 1.0	& 25.6	& 31	 \\ \hline
08959 &	2015 Jul 21 \& Sep 07	&1.4 $\times$ 1.2	& 2 $\times$ 24.6 & 38  \\ \hline
80573 &	2015 Aug 01	& 2.3 $\times$ 1.1	& 24.0	& 40     \\ \hline
23734 &	2015 Aug 11	& 1.6 $\times$ 1.1	& 25.6	& 72	 \\ \hline
24784 &	2015 Aug 13	& 2.0 $\times$ 1.2	& 25.6	& 34	 \\ \hline
27825 &	2015 Sep 01	& 1.5 $\times$ 1.1	& 25.6	& 25	 \\ \hline
15004 &	2015 Sep 16	& 1.4 $\times$ 1.1	& 25.6	& 31	 \\ \hline
27473 &	2015 Sep 20	& 1.8 $\times$ 1.1	& 25.6	& 28	 \\ \hline

34524 &	2016 Sep 18	& 2.2 $\times$ 1.2	& 24.5	& 19 \\ \hline
57993 &	2016 Sep 19	& 2.2 $\times$ 1.1	& 24.6	& 27 \\ \hline
83073 &	2016 Sep 22	& 2.4 $\times$ 1.1	& 24.5	& 32 \\ \hline
37518 &	2016 Sep 23	& 2.9 $\times$ 1.2	& 24.5	& 30 \\ \hline
08586 &	2016 Sep 23	& 1.7 $\times$ 1.2	& 24.6	& 23 \\ \hline
53150 &	2016 Sep 24	& 3.9 $\times$ 1.2	& 24.5 	& 54 \\ \hline
76079 &	2016 Sep 24	& 1.7 $\times$ 1.1	& 24.6 	& 31 \\ \hline
95952 &	2016 Sep 24	& 2.0 $\times$ 1.3	& 24.6	& 17 \\ \hline
91412 &	2016 Sep 28	& 1.6 $\times$ 1.2	& 24.6	& 60 \\ \hline
23218 &	2016 Sep 29	& 2.2 $\times$ 1.0	& 24.6	& 44 \\ \hline
56415 &	2016 Sep 29	& 1.9 $\times$ 1.7	& 24.6	& 45 \\ \hline

55849 & 2017 Jan 18	& 1.2 $\times$ 1.2	& 24.6	& 15 \\ \hline
29942 &	2017 Jan 23	& 1.9 $\times$ 1.0 	& 24.6	& 15\\ \hline
38857 &	2017 Jan 24	& 1.4 $\times$ 1.11	& 24.6	& 28 \\ \hline

05083 &	2017 Jan 24	& 1.7 $\times$ 1.1 	& 24.6	& 29 \\ \hline
08755 &	2017 Jan 24	& 1.6 $\times$ 1.2 	& 24.6	&  27 \\ \hline
32420 &	2017 Jan 24	& 2.5 $\times$ 1.0	& 24.6	& 39 \\ \hline
62100 & 2016 Dec 15 \& 2017 Jan 24	& 2.0 $\times$ 1.1 	& 2 $\times$ 24.6	&  37\\ \hline
72856 &	2017 Jan 24	& 2.9 $\times$ 1.1 	& 24.6	& 18 \\ \hline
\hline
    \end{tabular}
    \caption{VLA observations. Exposure time given is on-source.}
    \label{table:VLAObservations}
\end{center}
\end{table*}

\begin{table*}
\begin{center}
    \begin{tabular}{ | c | c | c | c | c |}
    \hline
	ObjID 	& Observing	& Total Observing 	& rms\\
	 	& date		& Time  (hrs)		&  $\mu$Jy \\
 \hline \hline
92239 &	2014 Mar 24	& 1.6	& 5.3	 \\ \hline
72856 &	2014 Mar 25	& 0.5	& 14.2	 \\ \hline
24784 &	2014 Mar 26	& 2.3	& 4.6	 \\ \hline
05083 &	2014 Jul 29	& 1.2	& 3.9	 \\ \hline
32420 &	2014 Jul 30	& 1.2	& 5.7	 \\ \hline \hline
    \end{tabular}
    \caption{APEX observations.}
    \label{table:APEXObservations}
\end{center}
\end{table*}

\section{Results}\label{sec:results}
When imaging all three VLA scans combined, 27 objects out of the 32 observed with the VLA were detected at signal-to-noise ratios above 3. The remaining 5 were undetected above 3$\sigma$. Excluding the non-detections, the VLA observations were detected at a mean SNR of 6.3. 
Of the sources observed with the VLA, 10 had previously been observed with ATCA, making it possible to determine their radio spectral slope (see Fig. \ref{fig:VLA_ATCA5500_alpha} in section \ref{sec:SpecSlope}). \\
All APEX observations yielded non-detections and hence upper flux limits (see Fig. \ref{fig:APEX_VLA}).  These were consistent with expectations from their H$\alpha$-derived star formation rates and suggests none of these five sources contains a heavily obscured, powerful AGN -- as would be expected given their modest Balmer decrements.

\begin{table*}
\begin{center}
    \begin{tabular}{ | l | l | l | l | l | l |}
    \hline
    ObjID & 1.5 GHz flux & SNR & Angular size & 1.5 GHz SFR & $\Sigma_{SFR}$ \\
      & $\mu$Jy &  & arcsec$^2$ & M$_{\odot}$ yr$^{-1}$ & M$_{\odot}$ yr$^{-1}$ kpc$^{-2}$ \\
 \hline \hline
83073$^*$ & 54 $\pm$ 17 & 3.2 & (6.1 $\pm$ 2.8)$\times$(1.9 $\pm$ 1.5)& 1.18 $\pm$ 0.37 & $0.03^{+0.3}_{-0.02}$ \\ \hline
76428 & 90 $\pm$ 20 & 4.5 & (3.08 $\pm$ 1.14)$\times$(1.10 $\pm$ 0.15) &  2.0 $\pm$ 0.4 & $0.18^{+0.23}_{-0.09}$  \\  \hline 
80573$^*$ & 113 $\pm$ 20 & 5.6 & (2.77 $\pm$ 1.37)$\times$(0.74 $\pm$ 0.48) & 2.7 $\pm$ 0.5 &  $0.39^{+2.2}_{-0.26}$ \\  \hline 
08959$^*$ & 131 $\pm$ 18 & 7.3 & (2.28 $\pm$ 0.59)$\times$(0.27 $\pm$ 0.42) &  3.2 $\pm$ 0.4 & $1.5^{+18.9}_{-1.1}$  \\ \hline 
53150$^*$ & 132 $\pm$ 15 & 8.8 & (2.78 $\pm$ 1.66)$\times$(1.22 $\pm$ 0.67)& 3.29 $\pm$ 0.37 & $0.28^{+1.4}_{-0.18}$ \\ \hline
77821$^*$ & 246 $\pm$ 41 & 6.0 & (1.9 $\pm$ 0.23)$\times$(1.59 $\pm$ 0.17) & 6.7 $\pm$ 1.1 &  $0.6^{+0.3}_{-0.2}$  \\  \hline
76079$^*$ & 113 $\pm$ 10 & 11.3 & (2.11 $\pm$ 0.45)$\times$(0.48 $\pm$ 0.45) & 3.09 $\pm$ 0.27 & $0.8^{+17.1}_{-0.5}$ \\  \hline
37518$^*$ & 76 $\pm$ 13 & 5.8 & (3.41 $\pm$ 1.6)$\times$(0.99 $\pm$ 0.87) & 2.40 $\pm$ 0.41 & $0.17^{+2.9}_{-0.12}$ \\  \hline
27825 & 75 $\pm$ 17 & 4.4 & (1.90 $\pm$ 0.54)$\times$(1.20 $\pm$ 0.23) & 2.7 $\pm$ 0.6 & $0.25^{+0.29}_{-0.13}$ \\  \hline 
57993$^*$ & 45 $\pm$ 9 & 4.7 & (4.4 $\pm$ 1.3)$\times$(3.3 $\pm$ 1.2) & 1.62 $\pm$ 0.34 & $0.02^{+0.04}_{-0.01}$ \\ \hline
91412$^*$ & 127 $\pm$ 21 & 6.0 & (2.97 $\pm$ 0.6)$\times$(2.75 $\pm$ 0.65) & 4.96 $\pm$ 0.82 & $0.12^{+0.11}_{-0.05}$\\  \hline
56415 & $<$135 & $<3$ & - & $<5.5$ & - \\ \hline
34524 & $<57$ & $<3$ & - & $<2.5$ & - \\ \hline
23218$^*$ & 79 $\pm$ 16 & 4.9 & (3.6 $\pm$ 1.1)$\times$(3.2 $\pm$ 1.3)& 3.24 $\pm$ 0.66 & $0.06^{+0.1}_{-0.03}$ \\ \hline
08586$^*$ & 62 $\pm$ 11 & 5.6 & (2.17 $\pm$ 0.72)$\times$(1.74 $\pm$ 0.9) & 2.69 $\pm$ 0.48 & $0.13^{+0.36}_{-0.08}$ \\ \hline
72856 & $<$54 & $<3$ & - & $<2.4$ & - \\ \hline
29942 & $<45$ & $<3$ & - & $<2$ & - \\ \hline
32420 & 211 $\pm$ 42 & 5.0 & (2.39 $\pm$ 0.36)$\times$(1.36 $\pm$ 0.13) & 11.0 $\pm$ 2.2 & $0.6^{+0.3}_{-0.2}$ \\ \hline
38857$^*$ & 175 $\pm$ 26 & 6.7 & (1.17 $\pm$ 0.29)$\times$(0.12 $\pm$ 0.52) & 12.0 $\pm$ 1.8 & $11.7^{+0.5}_{-10.2}$ \\ \hline
55849 & $<45$ & $<3$ & - & $< 5.8$ & -\\ \hline
15004 & 147 $\pm$ 13 & 11.0 & (1.4 $\pm$ 0.15)$\times$(1.09$ \pm$ 0.09) & 11.1 $\pm$ 1.0 & $0.9^{+0.3}_{-0.2}$  \\  \hline  
95952 & 39 $\pm$ 12 & 3.3 & (4.55 $\pm$ 1.91)$\times$(1.47 $\pm$ 0.32) & 3.30 $\pm$ 1.01 & $0.06^{+0.11}_{-0.04}$\\ \hline
\end{tabular}
    \caption{Results for sources which were observed with the VLA only, giving their measured fluxes and signal-to-noise ratios. The angular sizes of the sources were estimated using \textsc{CASA} imfit. In most cases, \textsc{CASA} suggested that the objects were not resolved, and their estimated size measurements are hence upper limits. Objects for which `imfit' provided a size measurement deconvolved from the beam are marked with an asterisk. Radio star formation rates use the \citet{KennicuttEvans2012} conversion. Star formation rate densities $\Sigma_{SFR}$ for our sample are derived from radio data alone. It should be noted again that since observed angular sizes are convolved with the beam, the values found for $\Sigma_{SFR}$ are lower limits.}
    \label{table:Results}
\end{center}
\end{table*}

\begin{table*}
\begin{center}
    \begin{tabular}{ | l | l | l | l | l | l | l | l | l |}
    \hline
    ObjID & 1.5 GHz flux & VLA & 5.5 GHz flux & ATCA & $\alpha$ & Angular size & 1.5 GHz SFR & $\Sigma_{SFR}$ \\
      & $\mu$Jy & SNR & $\mu$Jy & SNR & & arcsec$^2$ & M$_{\odot}$ yr$^{-1}$ & M$_{\odot}$ yr$^{-1}$ kpc$^{-2}$ \\
 \hline \hline
54061$^*$ & 471 $\pm$ 70 & 6.7 & 215 $\pm$ 38 & 5.6  & -0.60 $\pm$ 0.25 & (4.12 $\pm$ 0.68)$\times$(1.67 $\pm$ 0.46) & 4.3 $\pm$ 0.6 &  $0.4^{+0.4}_{-0.2}$ \\
 \hline
27473 & 70 $\pm$ 21 & 3.3 & 57 $\pm$ 23 & 2.5 & -0.16 $\pm$ 0.57$^{\dagger}$ & (3.97 $\pm$ 1.96)$\times$(0.97 $\pm$ 0.18) & 0.8 $\pm$ 0.2 & $0.1^{+0.2}_{-0.07}$ \\
 \hline 
16911$^*$ & 171 $\pm$ 26 & 6.6 & 91 $\pm$ 20 & 4.6 & -0.49 $\pm$ 0.29 & (1.31 $\pm$ 0.51)$\times$(0.54 $\pm$ 0.44)  & 2.8 $\pm$ 0.4 &  $1.57^{+14.5}_{-1.0}$  \\
  \hline 
23734$^*$ & 283 $\pm$ 32 & 8.8 & 220 $\pm$ 34 & 6.4 & -0.19 $\pm$ 0.21 & (1.60 $\pm$ 0.43)$\times$(0.35 $\pm$ 0.45) & 6.0 $\pm$ 0.7 & $3.5^{+5.6}_{-2.4}$  \\
 \hline 
24784 & 132 $\pm$ 31 & 4.3 & 151 $\pm$ 27 & 5.6 & 0.10 $\pm$ 0.32 & (2.00 $\pm$ 0.56)$\times$(1.33 $\pm$ 0.26) & 3.1 $\pm$ 0.7 & $0.3^{+0.4}_{-0.2}$  \\
  \hline 
60392$^*$ & 92 $\pm$ 13  & 7.1 &  54 $\pm$ 26 & 2.1 & -0.41 $\pm$ 0.51$^{\dagger}$ & (1.81 $\pm$ 0.34)$\times$(1.09 $\pm$ 0.12) &  2.5 $\pm$ 0.4 &  $0.33^{+0.2}_{-0.12}$ \\ \hline 
62100 & 19 $\pm$ 4.8 &  4.0 & 61 $\pm$ 25 & 2.4 & 0.90 $\pm$ 0.53$^{\dagger}$ & (1.39 $\pm$ 0.07)$\times$(0.04 $\pm$ 0.00) & 0.7 $\pm$0.2 & $2.6^{+0.8}_{-0.8}$ \\
  \hline 
05083 & 149 $\pm$ 36 & 4.1 & 76 $\pm$ 22 & 3.5 & -0.52 $\pm$ 0.42 & (1.82 $\pm$ 0.31)$\times$(1.4 $\pm$ 0.19) & 6.1 $\pm$ 1.5 & $0.5^{+0.2}_{-0.3}$ \\
  \hline 
71294$^*$& 157 $\pm$ 38 & 4.1 & 72 $\pm$ 21 & 3.4 & -0.60 $\pm$ 0.42 &(1.36 $\pm$ 0.72)$\times$(0.93$\pm$0.5) & 6.5 $\pm$ 1.6 & $1.0^{+4.7}_{-0.7}$ \\  
\hline 
08755$^*$ & 254 $\pm$ 33 & 7.7 & 159 $\pm$ 22 & 7.1 &  -0.36 $\pm$ 0.20 & (2.10 $\pm$ 0.38)$\times$(0.61 $\pm$ 0.37) & 13.7 $\pm$ 1.8 & $1.7^{+4.3}_{-0.9}$ \\
  \hline 
   \hline
    \end{tabular}
    \caption{As in table \ref{table:Results} for those objects with both VLA and ATCA data. Objects marked with $\dagger$ have ATCA observations with SNR$<3$ and their derived radio spectral slopes hence have large associated uncertainties. Since the ATCA observations effectively provide upper limits, the true spectral slopes of these sources are likely steeper than the values indicated here, see section \ref{sec:SpecSlope} for discussion.}
    \label{table:Results_withSpecSlope}
\end{center}
\end{table*}

\subsection{Star Formation Rate} 
\label{sec:SFR}
Using the 1.5 GHz radio luminosity to SFR conversion of \citet{KennicuttEvans2012}, the mean star formation rate derived for the galaxies in our sample is $4.8\pm0.7$\,M$_\odot$\,yr$^{-1}$. This compares to a mean derived from H$\alpha$ line emission in the same targets of $8.2\pm1.3$\,M$_\odot$\,yr$^{-1}$. 
Excluding any non-detections, we find a 1.5 GHz SFR ranging from 0.7 M$_{\odot}$ yr$^{-1}$ to 13.7 M$_{\odot}$ yr$^{-1}$. These relatively low values are consistent with those inferred from the optical/ultraviolet and with previous work in the radio and submillimetre, both on high redshift targets directly, and on similar local analogue samples.

The H$\alpha$-derived SFR presented in \citet{Greis2016} is very sensitive to the ionization conditions and metallicity of the nebular gas as well as potential dust obscuration. The most massive stars - and hence the most likely progenitors of supernovae and radio continuum flux - form in dusty molecular clouds and might therefore be missed when only considering the UV and optical emission of the galaxy. We dust-correct the H$\alpha$ flux using the Balmer decrement and assuming case B recombination and a standard ratio of H$\alpha$ to H$\beta$ of 2.86. Having found LBAs to be generally dust-poor both from SED-fitting as well as via Balmer decrement measurements, we do not expect obscured star formation to play a large part in our sample. However, if such star formation were present, it should lead to an {\it excess} in the radio emission relative to H$\alpha$ predictions.

\begin{figure}
\centering
\includegraphics[width=\linewidth]{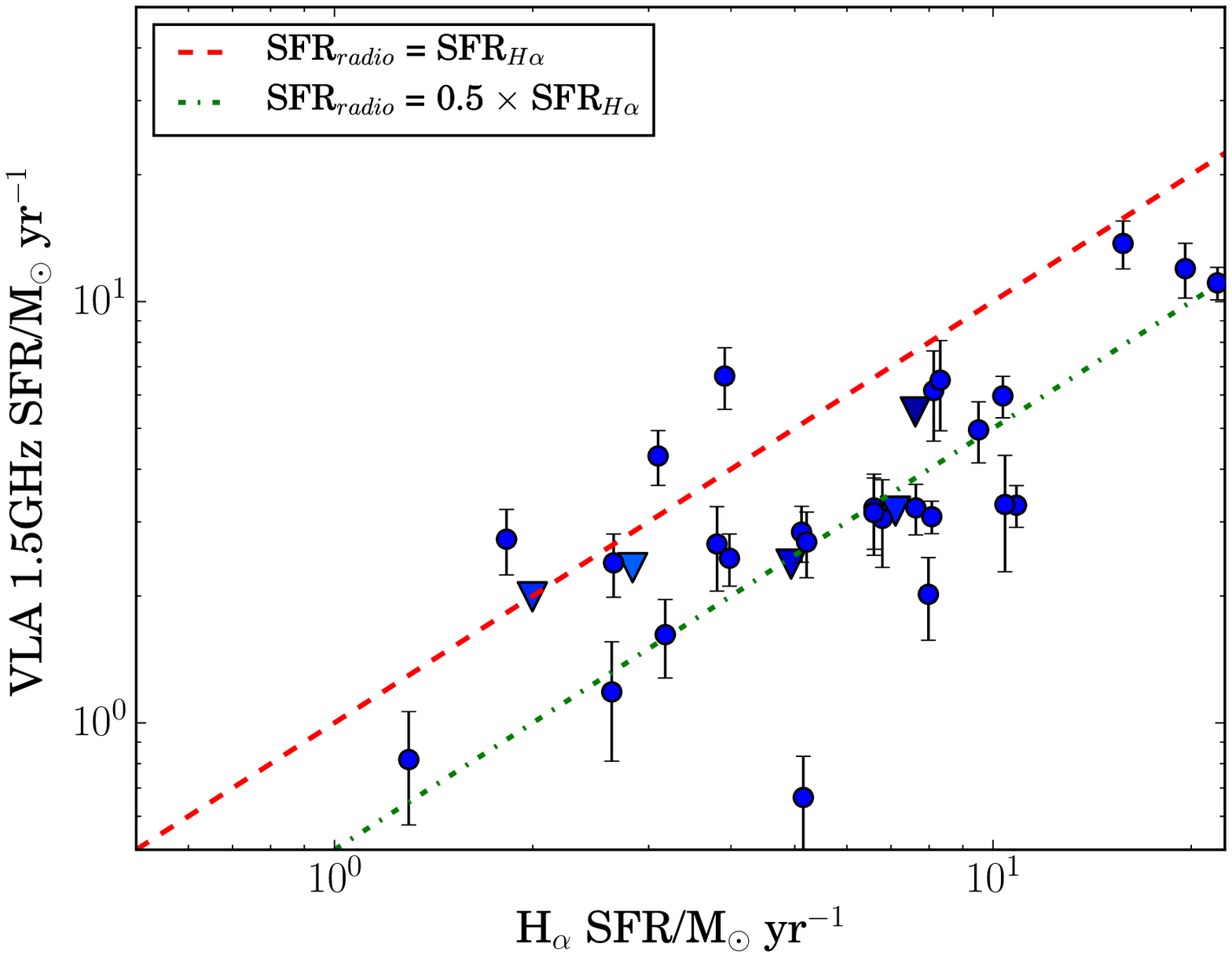}
\caption{Comparison of the star formation rates calculated from radio (using \citet{KennicuttEvans2012}) and (Balmer-decrement) dust-corrected H$\alpha$ (using \citet{Kewley2004}, see also \citet{Kennicutt1998}) fluxes. The red dashed line indicates the line of equality, while the green dot-dashed line shows radio SFR as half of the H$\alpha$. Triangles indicate 3$\sigma$ upper limits on non-detections. It can clearly be seen that most objects in our sample have lower radio than H$\alpha$-inferred SFRs. The error bars on H$\alpha$ are too small to be seen. H$\alpha$ is not corrected for Milky Way dust.}
\label{fig:HalphaRadioSFR}
\end{figure}

Fig. \ref{fig:HalphaRadioSFR} shows the inferred 1.5 GHz radio star formation rates of our sample as well as those found from H$\alpha$. The red dashed line indicates a one-to-one relation, while the green dot-dashed one shows a radio SFR that is half as high as the one found in H$\alpha$. The majority of our sources lie well below the one-to-one line, indicating that the galaxies have significantly higher H$\alpha$-inferred SFRs than radio SFRs. We find that these results remain unchanged when using the more recent \citet{Tabatabaei2017} radio-continuum to H$\alpha$ SFR calibration. 
The only objects which display stronger radio than H$\alpha$ SFRs are Objects 80573, 54061, 77821, and 29942. Apart from Obj 54061, these have all been found to be among the most massive sources in our sample and older than the typical source (see appendix \ref{App:IndividualObjs} for more details). Thus it would be unsurprising if those sources were more likely to host dust-obscured star formation.

The {\it deficit} in emission relative to H$\alpha$ in the bulk of the sample is more surprising and is discussed in section \ref{sec:Ages_SFR}. 

\subsection{Star Formation Rate Density}
\label{sec:SFRD}
We determine measurements or upper limits on the angular sizes of the objects using \textsc{CASA}'s `imfit' procedure. This enables us to improve the size, and hence density constraints, of the sources compared to previous (unresolved) optical observations. In order to measure the objects' sizes, `imfit' assumes an elliptical fit to the source, and provides major and minor axis measurements, as well as uncertainties on both. In about half of the sources (14 out of 32), the procedure was unable to deconvolve the source from the beam. In these cases, we use the measured value, which is likely a substantial overestimate of the true source size.  In the remainder (18 out of 32), the radio source is resolved, and here we use the deconvolved size as reported by the software; these sources are marked with an asterisk in tables \ref{table:Results} and \ref{table:Results_withSpecSlope}. The deconvolved measurements are approximately 10-30\% smaller (in both minor and major axis) than the unresolved sizes quoted by `imfit' for a given source. We use these size measurements to calculate an upper limit on the area of the source, and define an effective lower value of the star formation rate surface density as
$$\Sigma_{SFR}\equiv \frac{\textrm{SFR}_{1.5\,GHz}}{\pi a b}\quad \textrm{[}\textrm{M}_{\odot}\,\textrm{yr}^{-1}\textrm{kpc}^{-2}\textrm{]}$$
where $\textit{a}$ and $\textit{b}$ give the semi-major and semi-minor axis of the galaxy in kpc, and $\textit{SFR}_{1.5\,GHz}$ is the 1.5 GHz radio-derived star formation rate discussed above. 
We also calculate stellar mass densities based on the SED-fitting procedure presented in \citet{Greis2016}. 

\begin{figure}
\centering
\includegraphics[width=\linewidth]{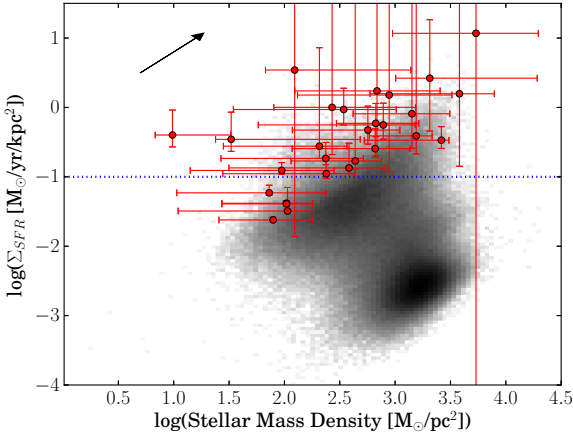}
\caption{Lower limits on stellar mass density and the star formation rate density for our sample using the radio-derived SFRs. Since the sources are unresolved in both the optical and radio observations, the true sizes are likely smaller than the ones used here, and hence the derived densities are  lower limits (the arrow in the top left corner of the figure indicates the direction in which the sample would move with decreasing size). The underlying grey-scale represents the local galaxy population as sampled by the SDSS in the same redshift range as our sample, and clearly seperates into passive (low $\Sigma_{SFR}$) and star forming galaxies. The blue dotted line indicates the criterion of 0.1 M$_{\odot}$ yr$^{-1}$ kpc$^{-2}$, suggested by  \citet{Heckman2002} as indicating the onset of starburst-driven galactic winds, or superwinds (see section \ref{sec:leakybox_winds}).}
\label{fig:SFRD_BPASSmass}
\end{figure}

In Fig \ref{fig:SFRD_BPASSmass} we compare the stellar mass density (in M$_{\odot}$/pc$^2$) and $\Sigma_{SFR}$ (using size limits derived from 1.5 GHz observations) of our sample to  those of local SDSS galaxies. The SDSS sample was selected to be in the same redshift range as our sample, and sizes, masses and SFRs are derived from the optical. We exclude AGN.

Two loci representing passive and actively star forming galaxies can be seen, with all of our sources lying well above the star-forming locus and displaying significantly higher $\Sigma_{SFR}$ and potentially lower mass densities than other local galaxies. The sample clearly represents an extreme subsample of the larger population.

\subsection{Spectral Slope}
\label{sec:SpecSlope}
By measuring the radio spectral slope, $F_{\nu}\propto \nu^{\alpha}$, in the sources it is possible to consider the origin of their radio emission. A steeper slope is associated with non-thermal, or synchrotron, emission, while a flat spectrum is indicative of thermal radio emission arising from thermal electrons in HII regions. Emission associated with AGN and their jets can span a wide range of spectral indices. 

Ten of our targets were observed at 5.5 GHz with ATCA \citep{StanwayDavies2014}. Of these, three reported fluxes have a signal-to-noise ratio $<3$, and are thus likely to represent upper limits, rather than robust detections. An additional three sources were detected at $3<$SNR$<5$. Given the extended beam of ATCA at this frequency for objects at these declinations, these detections might also be suspect in a blind field survey in which the false detection rate rises rapidly as signal-to-noise ratios fall below 5. However, \citet{StanwayDavies2014} inspected each source and found it to be coincident with the targetted optical galaxy, decreasing the likelihood of a false positive detection. The 5.5 GHz fluxes for these targets were also consistent with, or deficient relative to, the expected star formation rate in each case. If the low signal-to-noise detection, in fact, represents a misclassification of background noise, then the deficiency in radio flux seen in both in the \citet{StanwayDavies2014} sample and the observations discussed in this paper, is more dramatic still. In the following, we treat sources with a SNR $<3$ in the ATCA data as upper limits, while quoting the measured value.

\begin{figure}
\centering
\includegraphics[width=\linewidth]{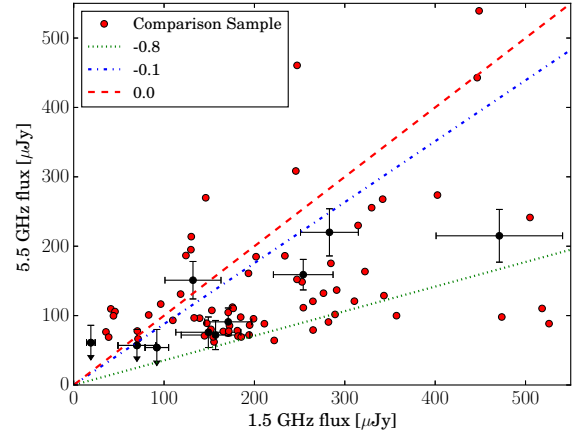}
\caption{The fluxes and spectral slopes for objects observed with both VLA and ATCA. All spectral slopes are consistent with star formation driven synchrotoron or non-thermal emission. Sources with 5.5GHz measurements at less than $3\sigma$ are shown at the measured values with an upper errorbar indicating a $5\sigma$ limit, and the lower uncertainty unbounded. A comparison population is taken from observations of sub-mJy sources by \citet{Miller2013} and \citet{Huynh2012} at 1.4 GHz and 5.5 GHz respectively.}
\label{fig:VLA_ATCA5500_alpha}
\end{figure}

Comparing the fluxes in the ATCA and VLA radio bands, we find that about half of the sources show spectral slopes, or upper limits, consistent with dominant thermal emission. The spectral slopes of the remainder suggest a mixture of thermal and star-formation driven synchrotron radiation (see Table \ref{table:Results_withSpecSlope} and Fig. \ref{fig:VLA_ATCA5500_alpha}). For comparison, we show the fluxes of galaxies taken from observations in the Extended Chandra Deep Field South by \citet{Miller2013} and \citet{Huynh2012} at 1.4 GHz and 5.5 GHz respectively. These targetted sub-mJy, and hence predominantly star-forming, galaxies likely similar to our sample. We find that our sample and this previous work are consistent in both their distribution and the range of their spectral slopes. 

In normal star-forming galaxies with constant star formation rates, one would expect a radio spectral slope of $\sim - 0.7$ to $-0.8$. However, we find significantly shallower spectral slopes in our sources, suggesting that they have a high fraction of thermal radio flux. Interestingly, this is what would be expected of a recent short-duration starburst, which produced thermal radio as well as H$\alpha$ emission, but very little non-thermal synchrotron emission. Both the radio-derived results as well as the previously found stellar population ages are thus consistent with recent starbursts in the majority of our sources. 

\begin{figure}
\centering
\includegraphics[width=\linewidth]{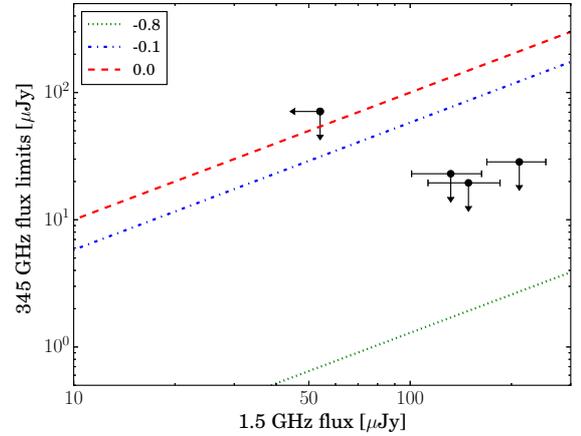}
\caption{Like Fig. \ref{fig:VLA_ATCA5500_alpha}, showing 5$\sigma$ upper flux limits for APEX 345 GHz non-detections and VLA 1.5 GHz fluxes (or 3 $\sigma$ limit on one source). The spectral slopes inferred for these sources are consistent with shallow slopes indicative of a large thermal fraction.}
\label{fig:APEX_VLA}
\end{figure}

In addition to the spectral slopes derived from VLA and ATCA observations, we also plot upper limits for the 4 objects for which we obtained both VLA observations and APEX non-detections, assuming they would have been detected at 5$\sigma$ above the rms noise (see Fig. \ref{fig:APEX_VLA}). In agreement with the spectral slopes found above, the limits on their 1.5 - 345 GHz spectral fluxes are consistent with a combination of high thermal fraction and synchrotron emission due to recent star-formation.

\begin{figure}
\centering
\includegraphics[width=\linewidth]{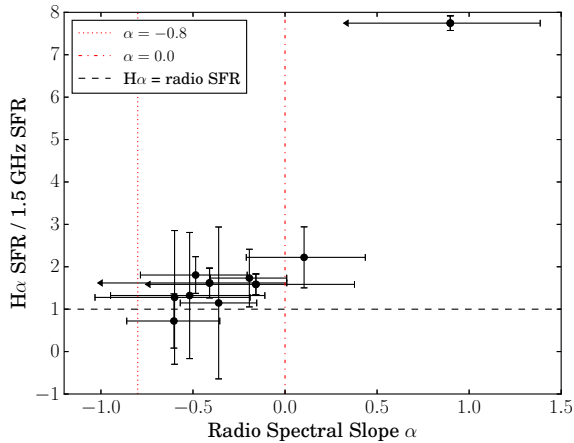}
\caption{The spectral slopes and ratios of H$\alpha$ to 1.5GHz radio SFRs for objects observed with both VLA and ATCA. This provides further confirmation that these sources do not contain a strong non-thermal component. Sources with 5.5GHz measurements at less than $3\sigma$ are shown with a spectral slope derived from the measured values with an upper errorbar indicating a $5\sigma$ 5.5 GHz limit, and the lower uncertainty unbounded. The systems with the highest H$\alpha$ to 1.5GHz ratio, indicative of a recent starburst, show a clear trend towards the flatter radio spectral slopes found in galaxies dominated by non-thermal radio emission. The most extreme system (Object 62100) has some of the least well constrained spectral slope measurements.}
\label{fig:alpha_SFRratio}
\end{figure}

Fig. \ref{fig:alpha_SFRratio} shows the ratio of H$\alpha$ to 1.5GHz radio SFRs against the spectral slopes found within them. A clear correlation between shallower (more positive) slopes and higher H$\alpha$ to 1.5 GHz SFR ratios can be seen, indicating that systems with a recent starburst (those with high H$\alpha$ to 1.5 GHz SFR ratios) are deficient in supernova-driven synchrotron emission which would give them steeper spectral slopes. 

\section{Discussion}
\label{sec:Discussion}

\subsection{Do our Sources Contain AGN?}
\begin{figure}
\centering
\includegraphics[width=\linewidth]{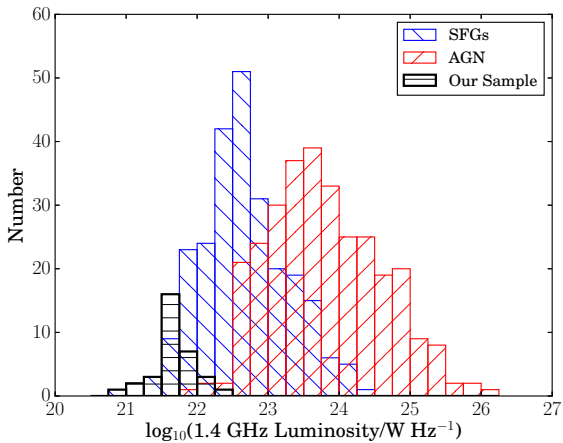}
\caption{Comparing the observed luminosities of AGN (red) and star-forming galaxies (blue) in the \citet{BestHeckman2012} sample, cross-matched with the MPA-JHU catalogue, to the galaxies presented here (black). By calculating the fractional probability of AGN in each luminosity bin, and comparing it to the luminosity distribution of our sample, we predict that $\sim 0.57$ AGN may be present within our observations.}
\label{fig:BestHeckman_compar}
\end{figure}

At the typical luminosity of our sample, $L_{1.4GHz}\sim5\times10^{21}$\,W\,Hz$^{-1}$, star forming galaxies are an order magnitude more abundant than AGN powered sources \citep{Condon2002, Sadler2016}, so we would expect very few such contaminants, even without our careful pre-selection based on optical and ultraviolet data. In Fig. \ref{fig:BestHeckman_compar}, we illustrate the luminosity distributions of star-forming galaxies and those identified as AGN within the \citet{BestHeckman2012} sample, as well as that of our sources. By calculating the probability of AGN in each luminosity bin, and multiplying it by the number of our observed targets in each bin, we expect the total number of AGN in our sample to be $\sim 0.57$. This prediction is, of course, subject to small number statistics and the associated Poisson uncertainty. \citet{Sadler2016} suggest that compact accretion-powered radio sources may still be significant at low luminosities, but uses a limit several orders of magnitude brighter than our target sample. The AGN population is poorly constrained at the very low luminosities relevant here, with the best constraints coming from optically-selected samples \citep[e.g.][]{BestHeckman2012}, but the decline in the AGN luminosity function appears to be steep in current deep surveys rendering it unlikely that there are significant numbers of concealed AGN in our sample. However, even adopting a conservative estimate of 1$\pm$1 AGN contaminants entering our sample, our primary conclusions remain unchanged, with the bulk of our targets both powered by star formation and deficient in radio flux. In selecting our large LBA sample, we excluded any objects which were classified as AGN in SDSS (though a few lie in the overlap region, see Fig. \ref{fig:BPT}). Thus, while it is conceivable that a small number of our targets might host a radio-quiet AGN, this seems unlikely given the optical properties of the sources, which indicate star-formation driven optical emission lines and low dust-obscuration. 

Uncertainty arises for two objects; their strong lines place Objects 76428 and 16911 in or close to the AGN region of the BPT diagram.
However, the line ratio diagnostic is sensitive to the object's redshift, star formation rate densities, and metallicity (among other factors), all of which are more extreme in our sample than in the normal galaxies for which the BPT diagnostics are calibrated. The validity of the BPT diagram for high-redshift sources - and hence potentially their local analogue populations - has been questioned \citep[see e.g.][]{Bian2016, Steidel2014}, showing that the locus of high-redshift sources is shifted upwards compared to local galaxies.  Hence the location of Objects 76428 and 16911 on the BPT diagram may be an indication of their extreme physical properties, rather than AGN activity. The spectral slopes of AGN range from -1 to flat, showing a broad variety of behaviours with frequency. Therefore, the spectral slope has little diagnostic ability in this case. We note that, admittedly at brighter radio luminosities, AGN-dominated galaxies in the SDSS tend to show a far higher radio-to-H$\alpha$ flux ratio than our targets. An AGN sufficiently obscured to not modify the optical emission line ratios in the BPT diagram typically adds to the radio continuum, providing an additional component on top of the host galaxy's star formation, while an unobscured AGN produces relatively little H$\alpha$ for a given radio flux (and indeed relative to other optical emission lines, hence the location of such sources in the BPT diagram). Thus the presence of an AGN in one or more of our targets does not provide an explanation for the observed radio deficit.  If the UV-optical emission represents star formation (as suggested by SED fitting and the emission line spectra), an AGN origin for any component of the radio emission would only increase the deficit discussed in section \ref{sec:SFR}.

\subsection{Determining Ages}
\label{sec:Ages_SFR}
Most standard calibrations for SFR, including those of \citet{KennicuttEvans2012} which we use here, assume star formation to have been continuous over the last 100 Myr. By contrast, only the stars formed in the last 10 Myr contribute to H$\alpha$ flux, while stars over a much larger range of ages contribute to the supernova rate and hence power any 1.5 GHz radio synchrotron emission. Thus, a precise conversion from flux to star formation rate will be dependent on the stellar population age and star formation history. By using a stellar population synthesis model, it is possible to derive such conversions. It is important to note that for a given flux, the standard conversion rates underestimate the true rate of star formation in populations younger than a few hundred Myr.

We calculate the stellar population using the Binary Population and Spectral Synthesis \citep[{\sc BPASS v2.0},][Eldridge et al in prep]{Stanway2016} population synthesis code for a system forming stars at a constant rate of 1\,M$_\odot$\,yr$^{-1}$, but which has yet to stabilise at young ages. From this, we process the output stellar spectrum through v13.04 of the radiative transfer code {\sc CLOUDY} \citep{Ferland2013,Ferland1998} to calculate the [O\,{\sc II}] and H$\alpha$ fluxes. We also use the predictions of BPASS to calculate a core-collapse supernova rate, and correct this for type Ia supernovae which contribute $\sim \frac{1}{3}$ of the SN rate at late times, and which track the overall star formation rate and thermal contribution with a delay of 100\,Myr. The radio-SFR calibration assumes $\sim 10 -20\%$ thermal emission in a mature stellar population, with the remainder due to synchrotron emission which is proportional to the rate of core-collapse supernovae.

\begin{figure}
\centering
\includegraphics[width=\linewidth]{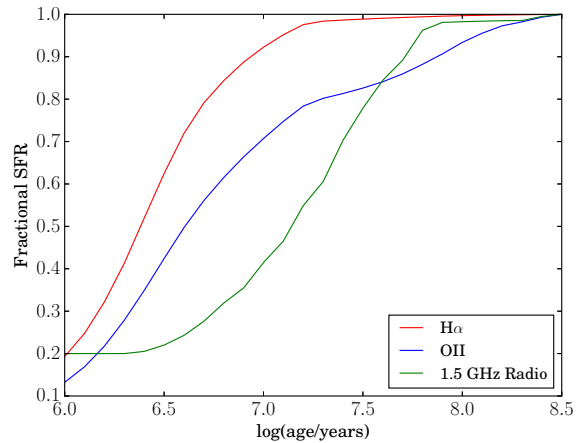}
\caption{An estimate of the fraction of `true' star formation rate that would be inferred given the standard calibrations, for young stellar populations. Fractions are calculated based on a population forming stars at a constant rate in the BPASS stellar population synthesis code, as described in section \ref{sec:Ages_SFR}.  Here, we show the fractional SFR as a function of stellar population age for the H$\alpha$ (red line), OII (blue line) and radio (green line) indicators. Radio continuum takes longer to establish than the nebular line emission leading to underestimates in the radio-derived star formation rate.}
\label{fig:SFR_age_general}
\end{figure}

Given that the standard calibrations are calculated for a population age of 100\,Myr, we scale our outputs by the values determined for populations at that age, to calculate a fraction of the steady-state line flux and supernova rate that will be measured at each age. By applying the standard calibrations to these outputs, we calculate the fraction of the `true' SFR that would be inferred at a given age, were the standard calibration to be applied. In Fig. \ref{fig:SFR_age_general}, we illustrate this, and hence show the age dependence of each star formation rate indicator. As expected, the supernova rate (and hence inferred synchrotron emission) only becomes significant at ages above $\sim 10$ Myr, increasing steeply thereafter, while H$\alpha$ and [O\,{\sc II}] line emission rise more rapidly after the onset of star formation. While the (statistically unlikely) presence of an AGN would complicate this analysis, its action would be to indicate still younger stellar populations: the nebular and radio flux from an AGN are both rapidly established and would act to increase the radio deficit relative to the optical emission lines.

\begin{figure}
\centering
\includegraphics[width=\linewidth]{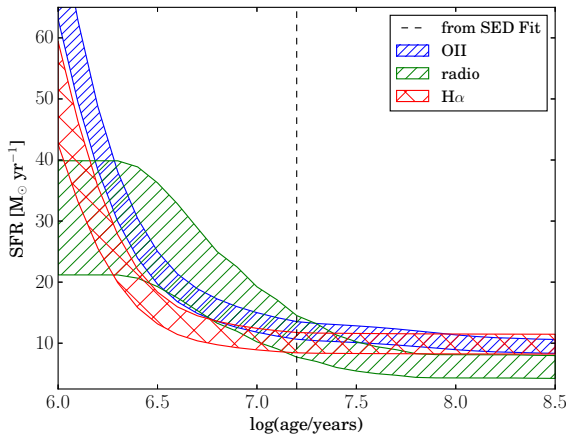}
\caption{Using the conversion rates in Fig. \ref{fig:SFR_age_general}, we show the inferred SFR for a given stellar population age for Obj~23734,  assuming continous star formation. The green dashed line indicates the radio SFR for a given age, while the solid blue and dotted red lines show the OII and H$\alpha$ inferred SFRs respectively. As the standard calibrations assume a population of $\sim$ 100 Myr, a younger population with the same observed flux has a higher inferred SFR. The best-fitting age for this object, where the three SFR indicators intersect, is in good agreement with the SED-derived age of log(age)$\sim 7.2$.}
\label{fig:SFR_age}
\end{figure}

Thus, in principle, finding the overlap region in which the calculated SFRs coincide makes it possible to constrain the age of the stellar population. In Fig.   \ref{fig:SFR_age} we provide an illustration of this technique for one of our targets, demonstrating good agreement with the young SED-derived age for this object. We caution however that uncertainties in dust correction of observed fluxes, star formation history and in handling of the thermal component of the radio continuum (discussed below) suggest that overinterpretation of these results at this stage is unwise.

As discussed in section \ref{sec:SpecSlope}, the spectral indices of our sample are largely consistent with the flat slopes indicative of thermal emission and star-formation driven synchrotron radiation. Thermal radio emission arises from H\,{\sc II} regions and is proportional to the photoionization rate in the source. As such we might expect it to trace similar timescales to H$\alpha$ and OII. This component typically contributes only $\sim10 - 20\%$ to the radio continuum at 1.4\,GHz and thus is often neglected in calculating calibrations.

Interestingly,  \citet{Tabatabaei2017} present not only a calibration from total radio continuum to star formation rate, but also its decomposition into thermal and non-thermal components, calibrated against the local KINGFISH galaxy sample \citep{Kennicutt2011}.  If we apply their derived calibration based only on thermal emission to our data, we find that it is consistent with the H$\alpha$ expectation. In other words, for some cases we infer a thermal fraction in our observed sources of 100\%, well above the typical fraction in local sources of 10\% at 1.4\,GHz, and higher than the maximum fraction in the KINGFISH sample (35\%). We note that \citet{Tabatabaei2017} found that the thermal fraction was highest in dwarf irregular galaxies and at high star formation rates.  If this trend holds, certain sources of our sample of very compact, irregular, starburst sources represents an extreme case. A thermal fraction of $\sim$100\% in these objects would be consistent with the picture described above in which the non-thermal emission expected to be present simply has not yet had time to develop in these young galaxies. It is also in good agreement with the low gas-phase metallicities found in these sources (see Table \ref{table:PhysProperties}), which indicate that relatively little metal-enrichment has occured.

\subsection{Leaky Box Model \& Winds}
\label{sec:leakybox_winds}
An alternative explanation for the relative deficit of radio emission in our sample might be the `leaky box' model in which the near-relativistic electrons providing the magnetic field of a galaxy are ejected from the galactic gravitational well, causing a decrement in observed radio emission compared to the true flux \citep{Kaiser2005}. From the star formation rate densities found in Section \ref{sec:SFRD}, we might expect that galaxy-wide superwinds are present in most of our sources. Such winds arise from the energy and momentum injected into the ISM by stellar winds and supernovae \citep[][ see also discussion in Chen et al 2010]{ChevalierClegg1985,Heckman1990}.
While there are likely to have been supernovae in our sample, their stellar populations are not old enough to have produced a substantial number of Wolf-Rayet or AGB stars (which drive strong stellar winds).

There are conflicting studies correlating the velocity of outflows with related parameters such as star formation rate and its density. \citet{Heckman2002} shows that outflow speeds with wind velocities of up to $\sim 3000$ km s$^{-1}$ are ubiquitous in galaxies with SFRDs above 0.1 M$_{\odot}$ yr$^{-1}$ kpc$^{-2}$  and correlate with  SFRD, while \citet{Rubin2014} find no such minimum threshold. \citet{Tanner2016} attempt to reconcile these findings with simulations that show  correlations between SFR (and SFRD) and outflow velocity break down at a point dependent on the entrained cold gas component, which varies from galaxy to galaxy. These factors are also likely to be metallicity dependent and so, again, local correlations may not be entirely applicable to our sample.

\citet{Davies2017} find a sub-linear relation between the SFR and radio luminosity of local galaxies. They interpret this as indicative of a leaky box model but note that this may conflict with their observed FIR-radio relation. A large thermal component may help reconcile this tension.

Spatially resolved morphological studies may provide insights into outflow structures and geometry. Such imaging, or resolved spectroscopy, of our sources would shed further light on the question of outflows, particularly by probing for an excess of ionized gas along the galaxy's minor axis (along which outflows may be more likely to escape), or broader emission line profiles (higher wind velocities) along the minor axis than the major axis. Such imaging would also provide a clear discriminator between distributed star formation occuring throughout the galaxy and a more localised event such as a nuclear starburst or any AGN contribution.

\section{Summary and Conclusions}\label{sec:conclusions}
We have presented the results of our analysis of radio and sub-millimetre observations of a sample of local analogues to $z\sim5$ Lyman break galaxies, taken with the VLA at 1.5 GHz and LABOCA on APEX at 870 $\mu$m. The observed sources span the range of physical properties of our larger Lyman break analogue sample \citep{Greis2016} and can therefore be considered as representative.
Out of 32 objects observed with the VLA, 27 were detected with a signal-to-noise ratio of at least 3 (when taking into account all three scans combined), while none of the 5 APEX observations yielded detections. We improve the size measurements, and hence star formation rate and stellar mass density, constraints of the sources. Our main results are:

\begin{enumerate}
\item We find that there does not appear to be any strongly dust obscured star formation in these systems, agreeing with the Balmer decrement measurements and SED-derived dust values found previously in \citet{Greis2016}.

\item Given the radio luminosities of our targets we predict $\sim$0.6 contaminating AGN sources in our sample of 32. This is consistent with their optical line ratios and the possibility that $\sim1-2$ of our targets may show AGN-like characteristics. The possible presence of a single AGN would strengthen our conclusion that the radio luminosity is deficient in these sources.

\item The mean star formation rate derived from radio observations of galaxies in our sample is $4.8\pm0.7$\,M$_\odot$\,yr$^{-1}$. This compares to a mean derived from H$\alpha$ line emission in the same targets of $8.2\pm1.3$\,M$_\odot$\,yr$^{-1}$. 

\item The observed low radio fluxes and spectral slopes derived (where possible) are consistent with young stellar populations that have not yet established strong supernova-driven synchrotron emission. This is in good agreement with previously derived results from SED fitting \citep[see][]{Greis2016}.

\item We present a method for constraining the age of a young stellar population using star formation rate indicators which traces stars at different timescales, such as OII, H$\alpha$, and radio continuum.
\end{enumerate}

Using the galaxies' SED-derived masses and radio-inferred star formation rate, together with (loose) constraints on their radio-derived physical sizes, it is possible to determine lower limits on the mass- and star formation rate densities within these sources. We find a moderate to significant offset between our radio LBA sample and more typical local SDSS star-forming galaxies selected at the same redshifts (Fig. \ref{fig:SFRD_BPASSmass}). The star formation rate densities of the LBAs are potentially indicative of galactic superwinds typically found in local starbursts or high-z LBGs. If high redshift galaxies mirror their local analogues, this would have interesting consequences for the chemical enrichment of the surrounding intergalactic medium. However, their relatively young ages might preclude the existence of a strong AGB population which could drive such winds. Hence it is unclear whether such superwinds actually occur in our sources. In order to better understand the impact the galaxies in our sample are having on their surroundings, spatially resolved spectroscopy is needed.

\section{Acknowledgements}
SMLG is funded by a research studentship from the UK Science and Technology Facilities Council (STFC). ERS also acknowledges support from STFC consolidated grant ST/L000733/1 and from the University of Warwick's Research Development Fund. 

This paper makes use of data obtained from the Karl G. Jansky Very Large Array, which is operatred by the NRAO. The National Radio Astronomy Observatory (NRAO) is a facility of the National Science Foundation operated under cooperative agreement by Associated Universities, Inc.

We make use of Ned Wright's very useful online cosmology calculator \citep{Wright2006} and the {\sc TOPCAT} table operations software \citep{Taylor2005}.

\medskip 

\bibliographystyle{mnras}  
\bibliography{Radio_bib}

\newpage
\appendix
\section{Descriptions of Individual Objects}
\label{App:IndividualObjs}

\begin{figure*}
\begin{center}
    \begin{tabular}{ c c c c }
\includegraphics[width=0.45\textwidth]{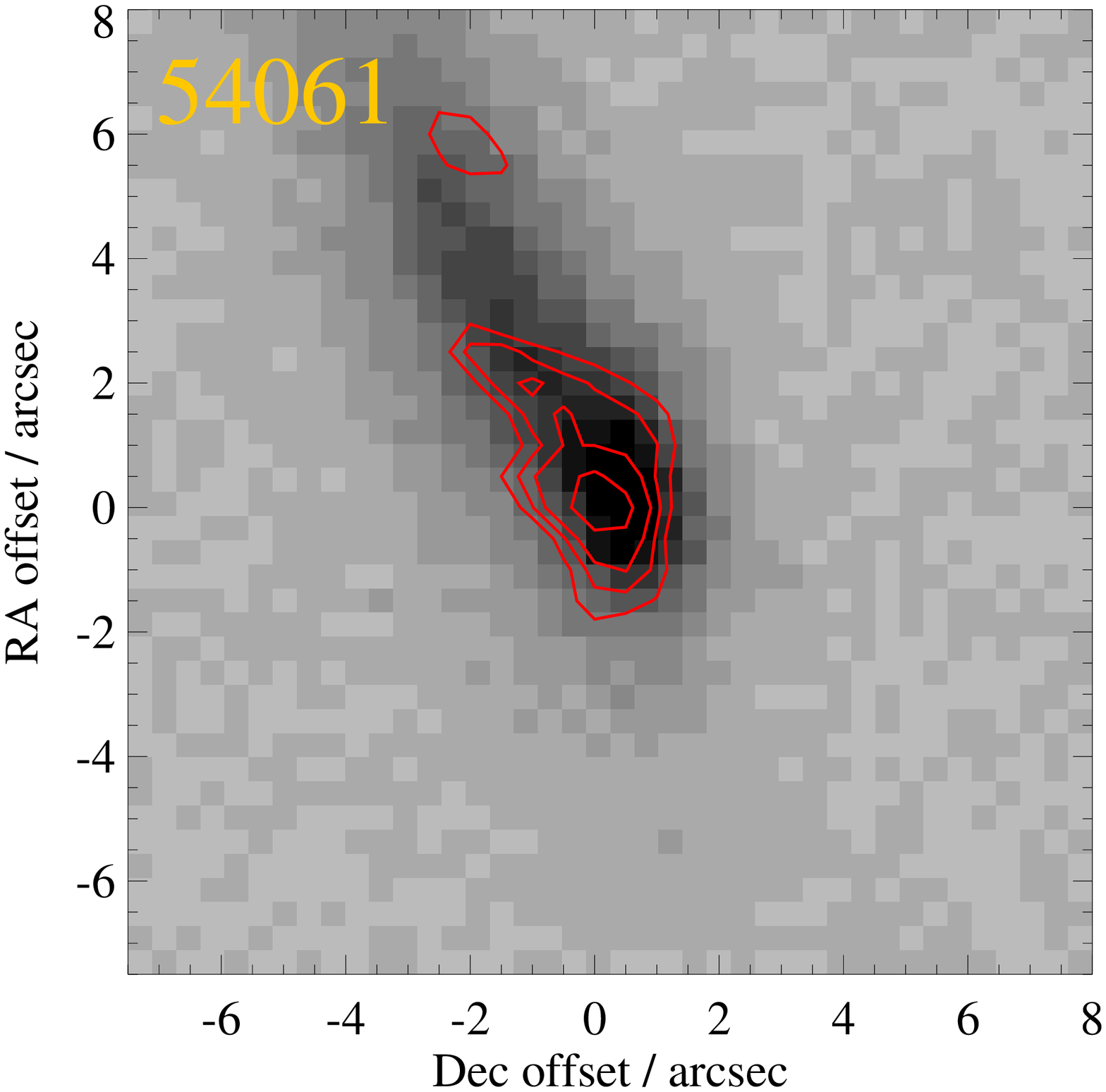} &
\includegraphics[width=0.45\textwidth]{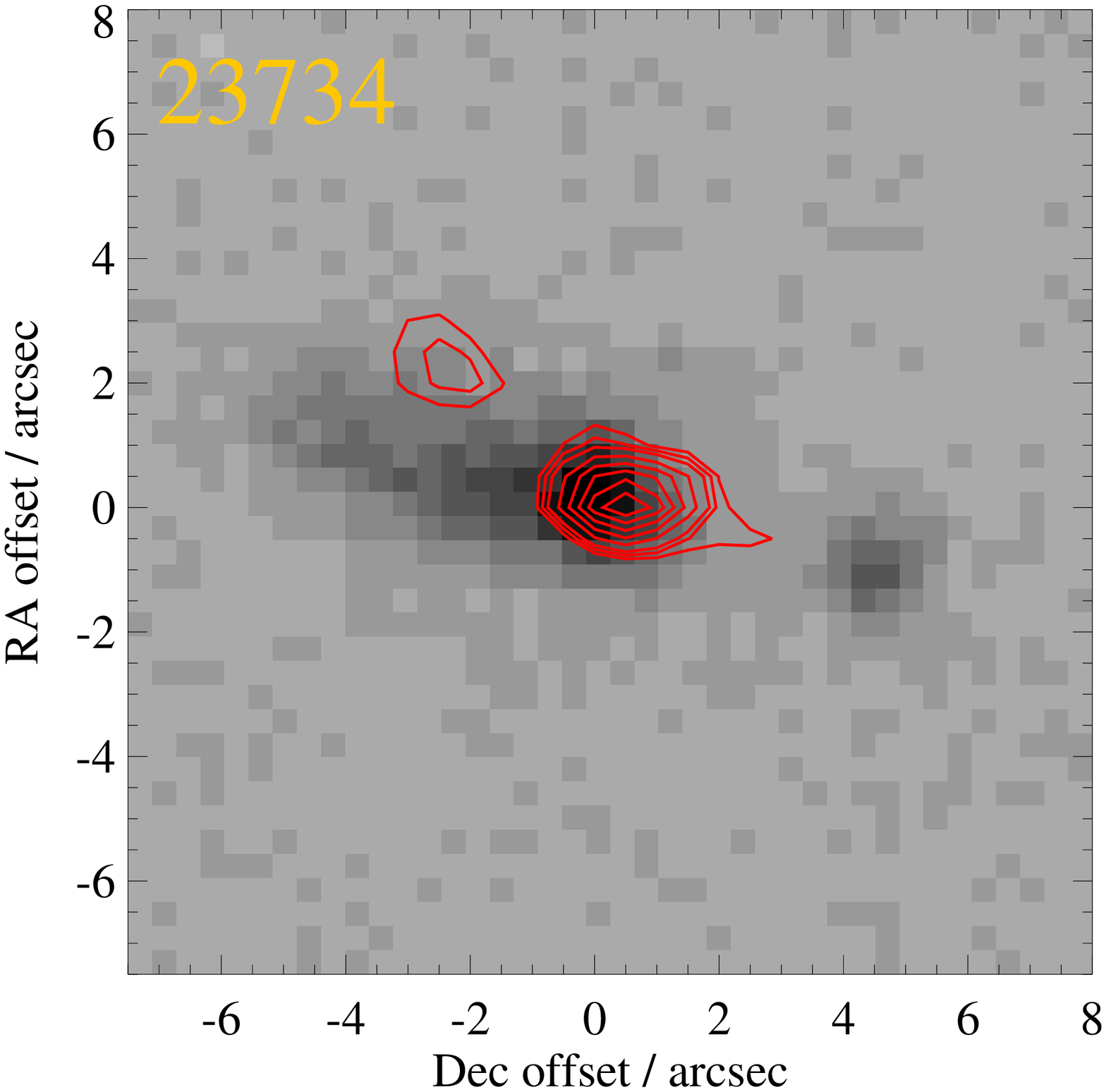} & \\
    \end{tabular}
    \caption{Radio contour plots overplotted on SDSS detections for the most extended optical sources in our sample. The contour levels are indicated at 2, 2.5, 3, 4, and 5$\sigma$.}
    \label{tab:extendedContours}
\end{center}
\end{figure*}

\begin{figure*}
\begin{center}
    \begin{tabular}{ c c c c }
    \includegraphics[width=0.32\textwidth]{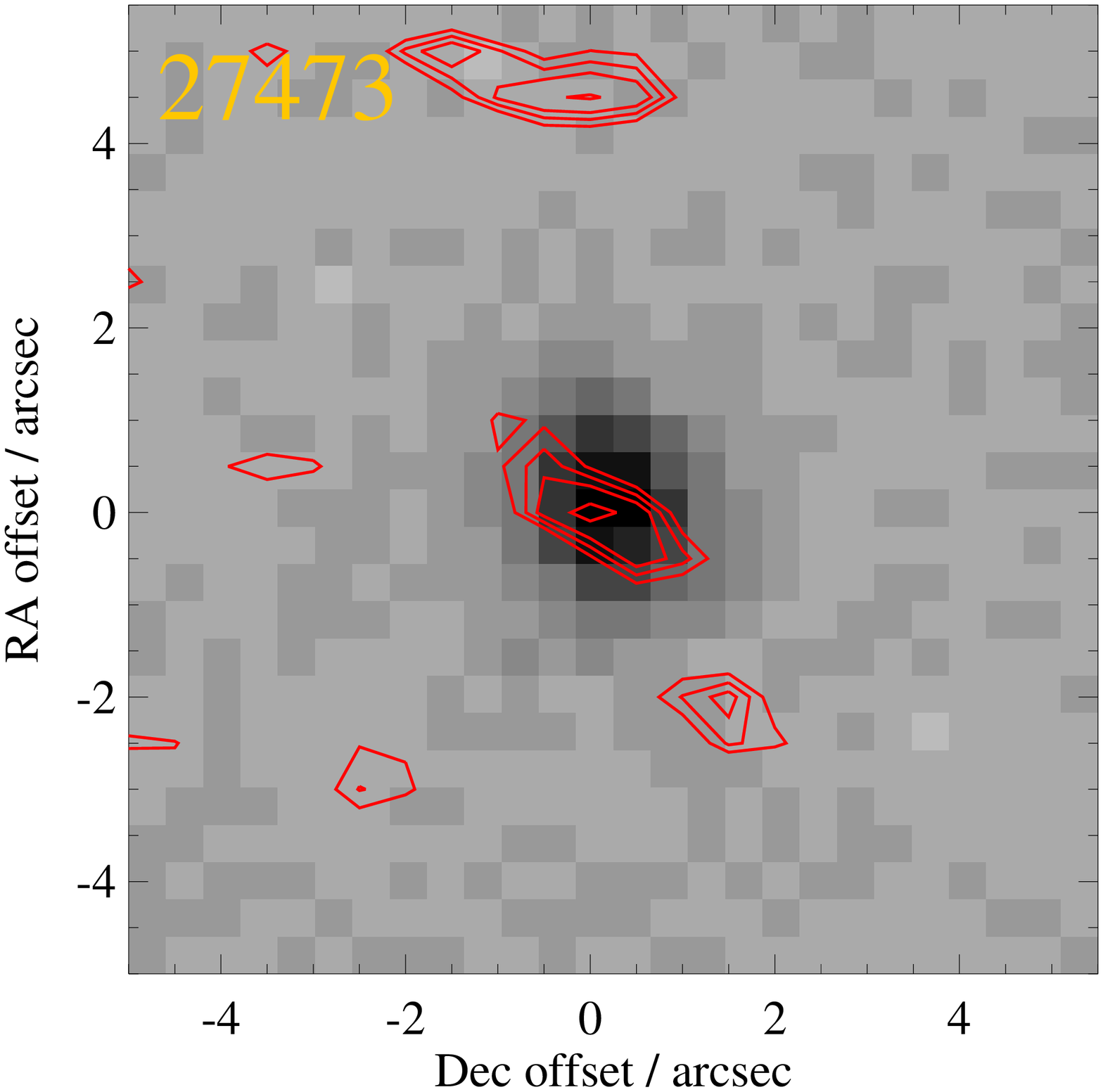} & 
    \includegraphics[width=0.32\textwidth]{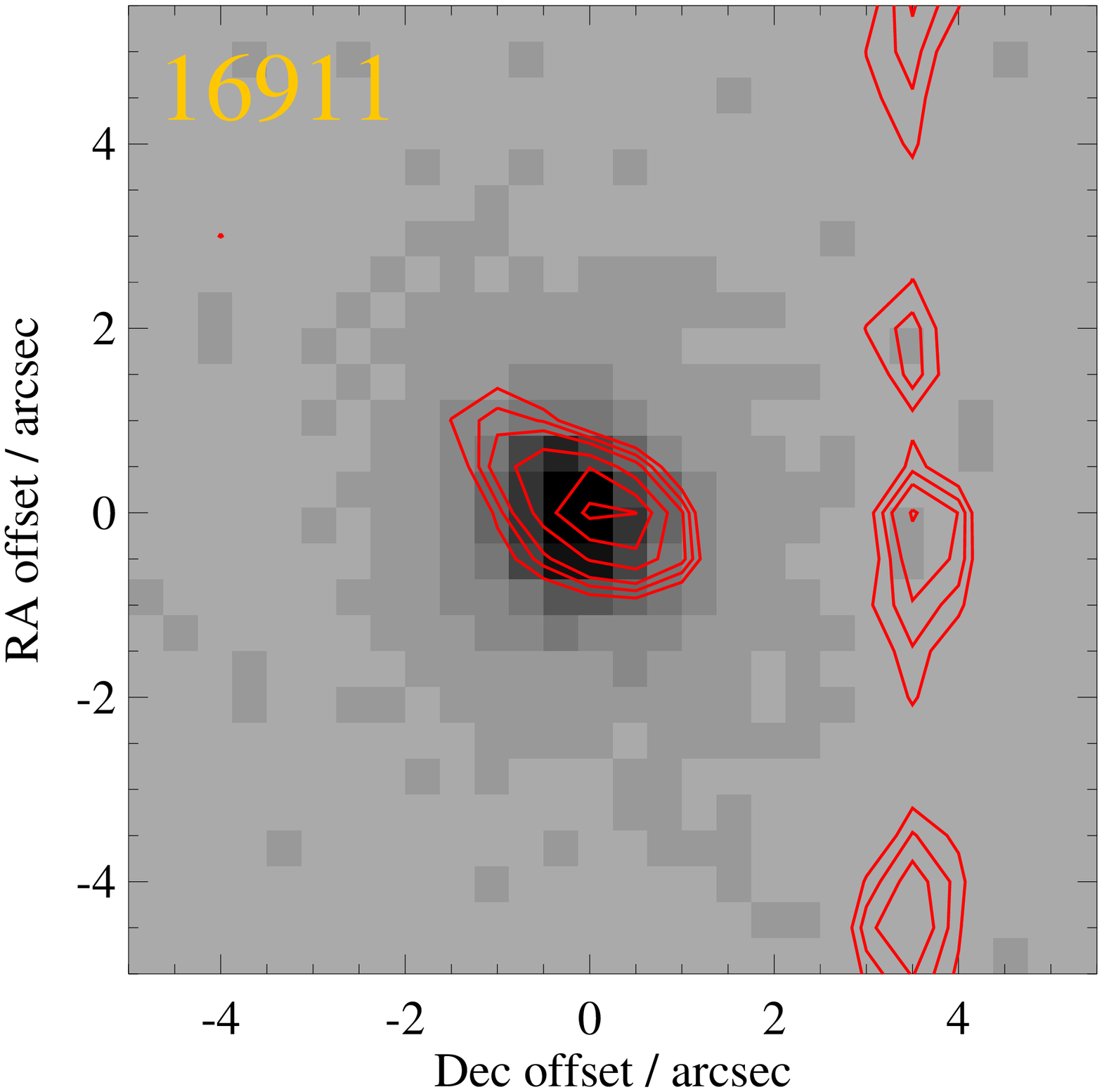} & 
    \includegraphics[width=0.32\textwidth]{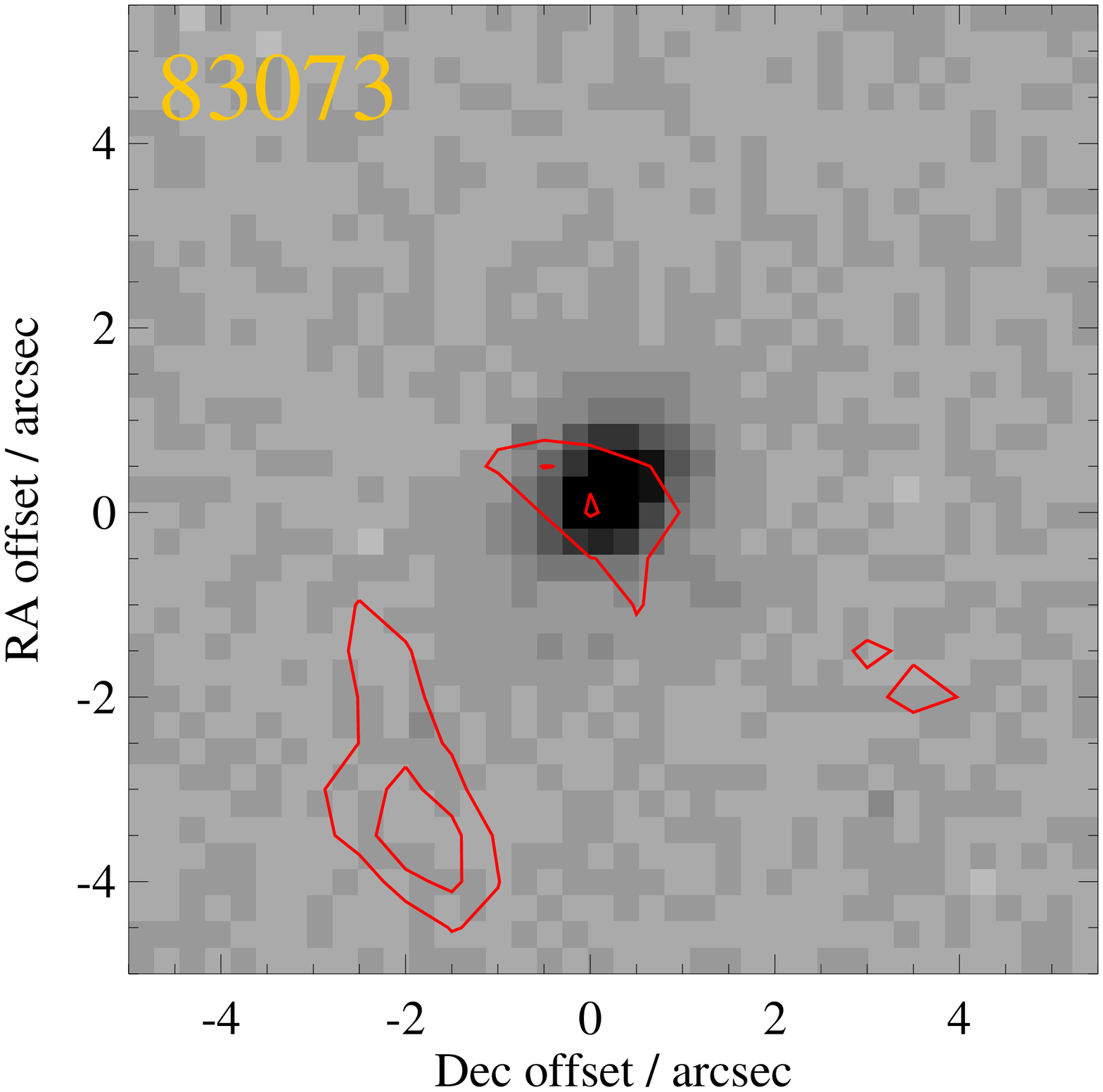} & \\
    \includegraphics[width=0.32\textwidth]{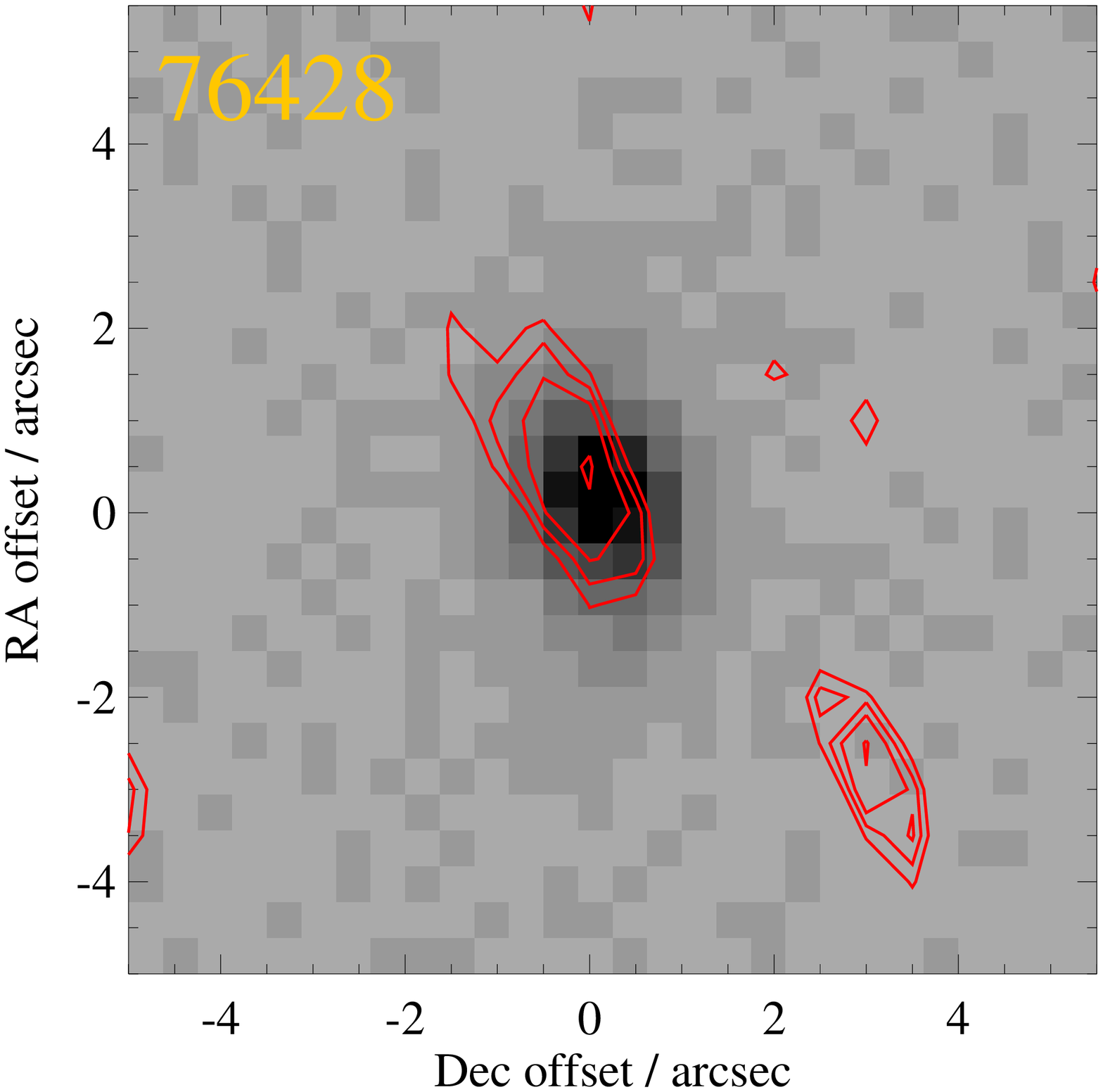} & 
    \includegraphics[width=0.32\textwidth]{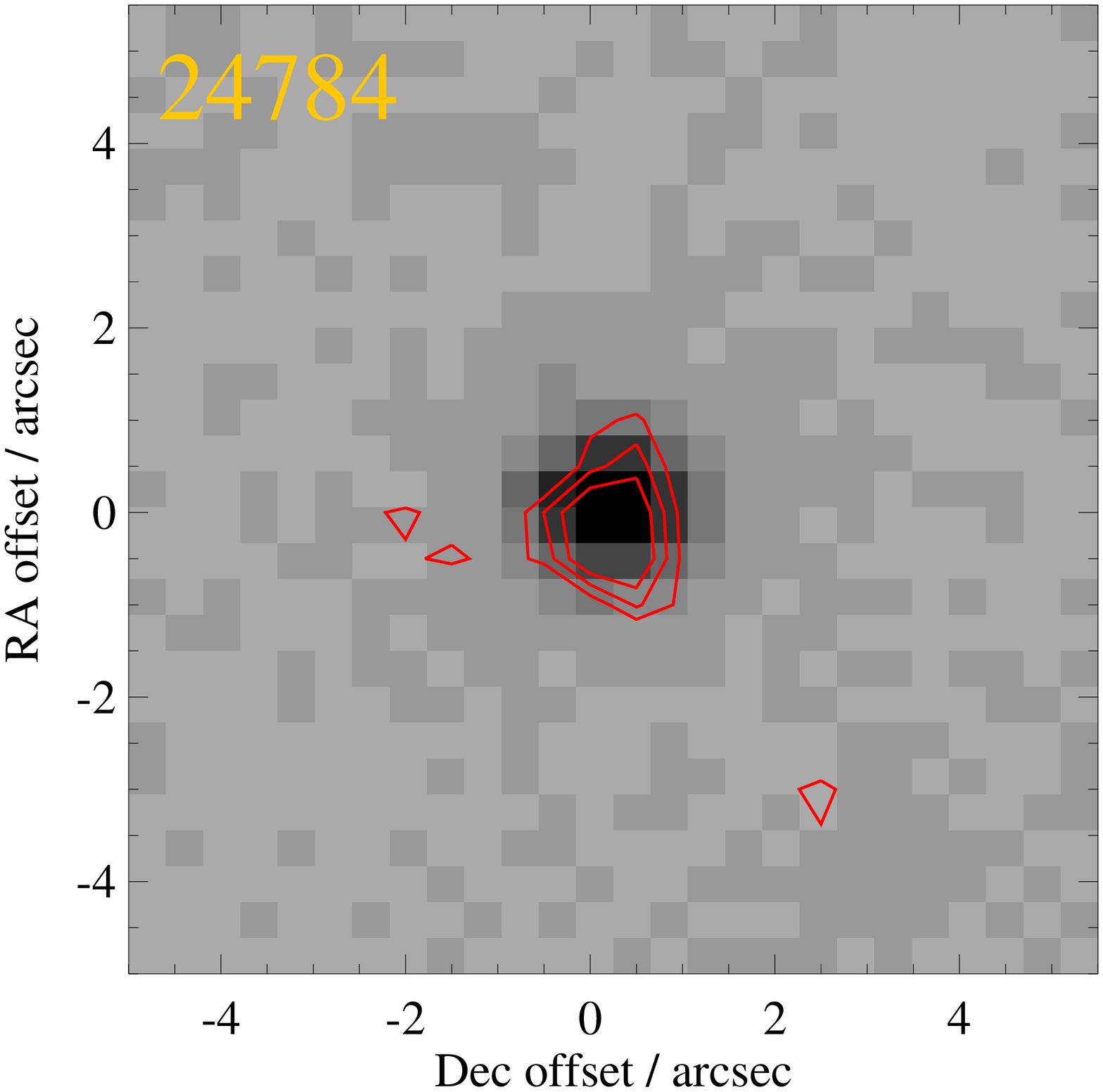} & 
    \includegraphics[width=0.32\textwidth]{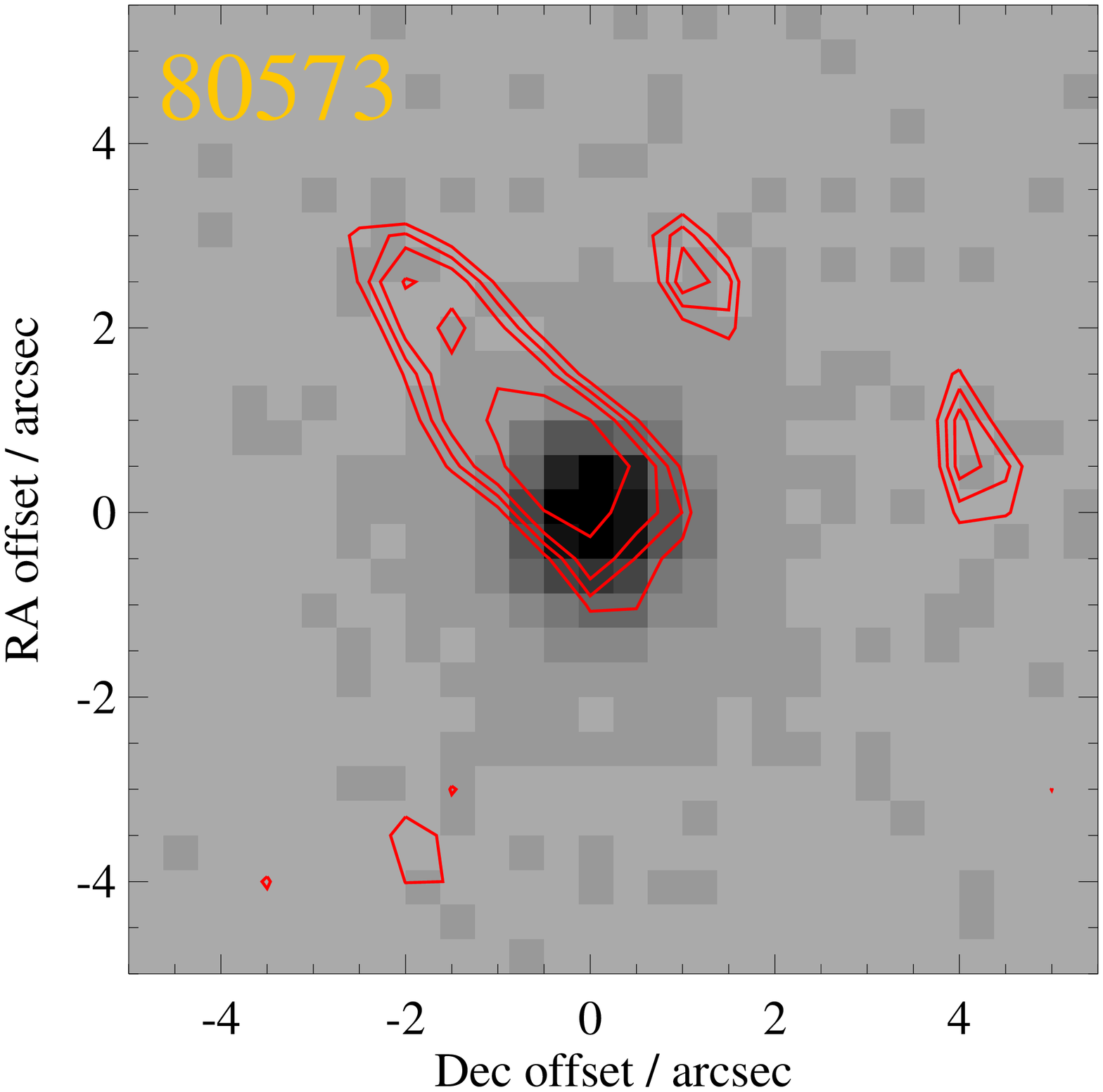} & \\   
    \includegraphics[width=0.32\textwidth]{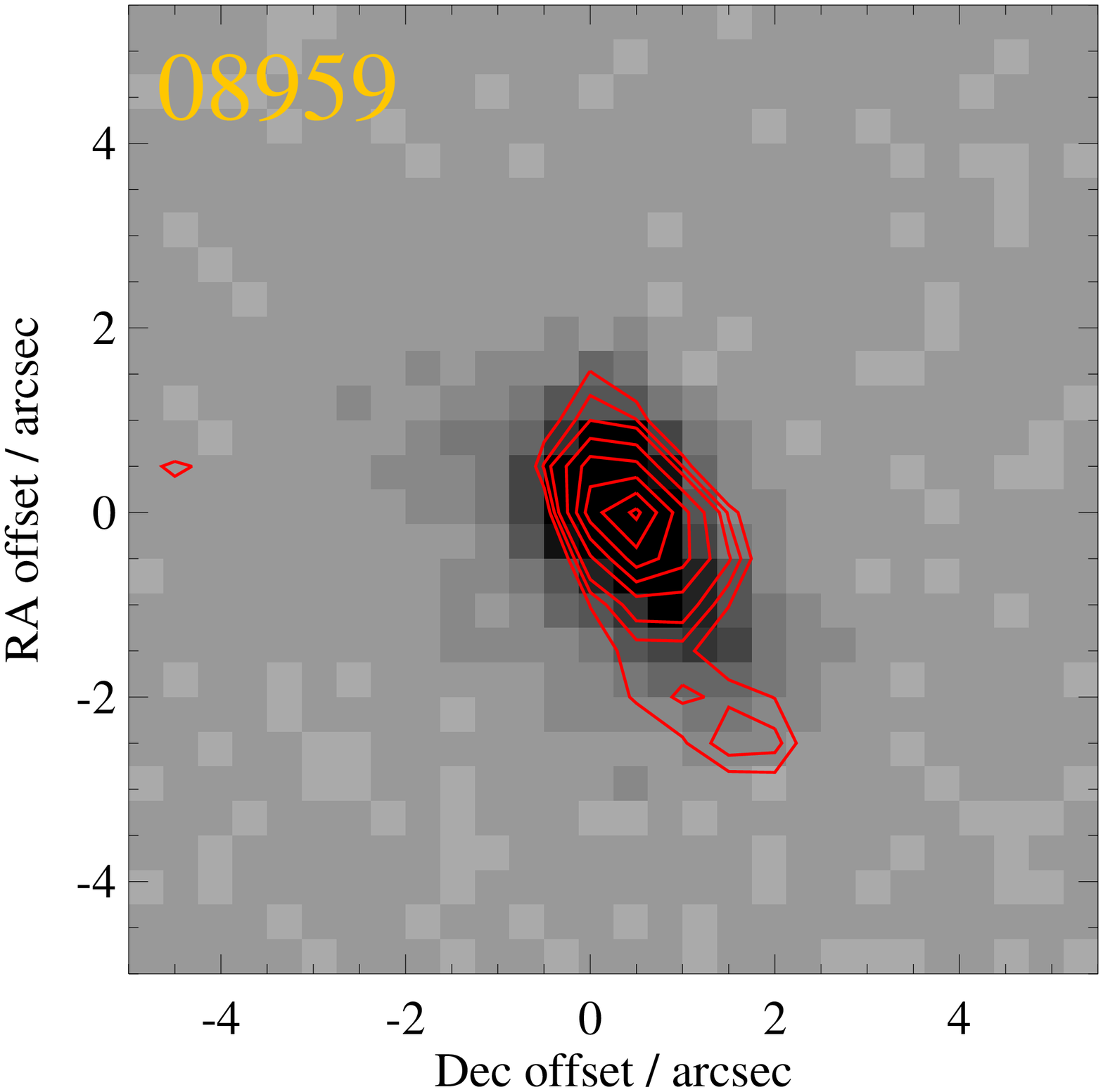} & 
    \includegraphics[width=0.32\textwidth]{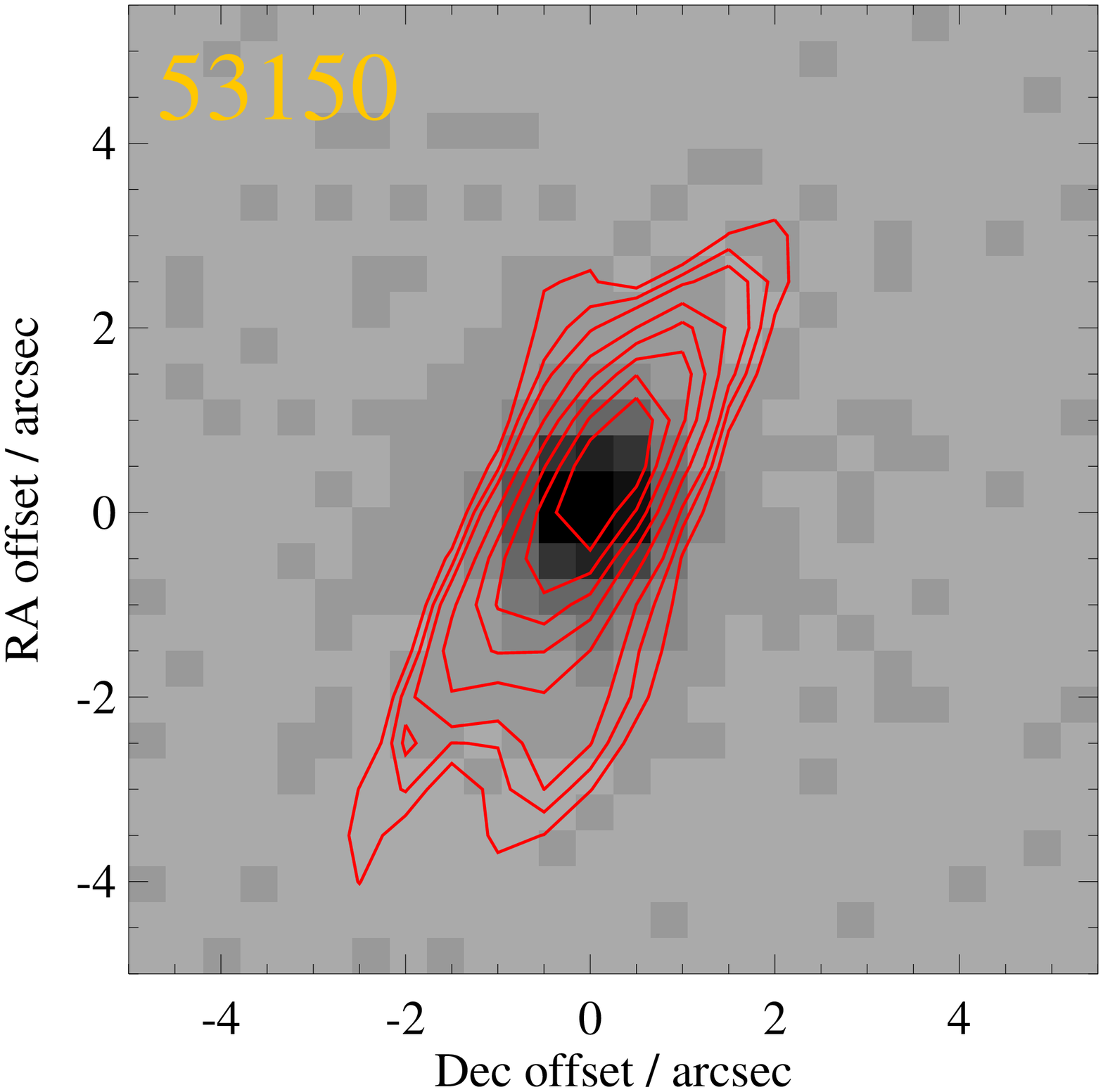} & 
    \includegraphics[width=0.32\textwidth]{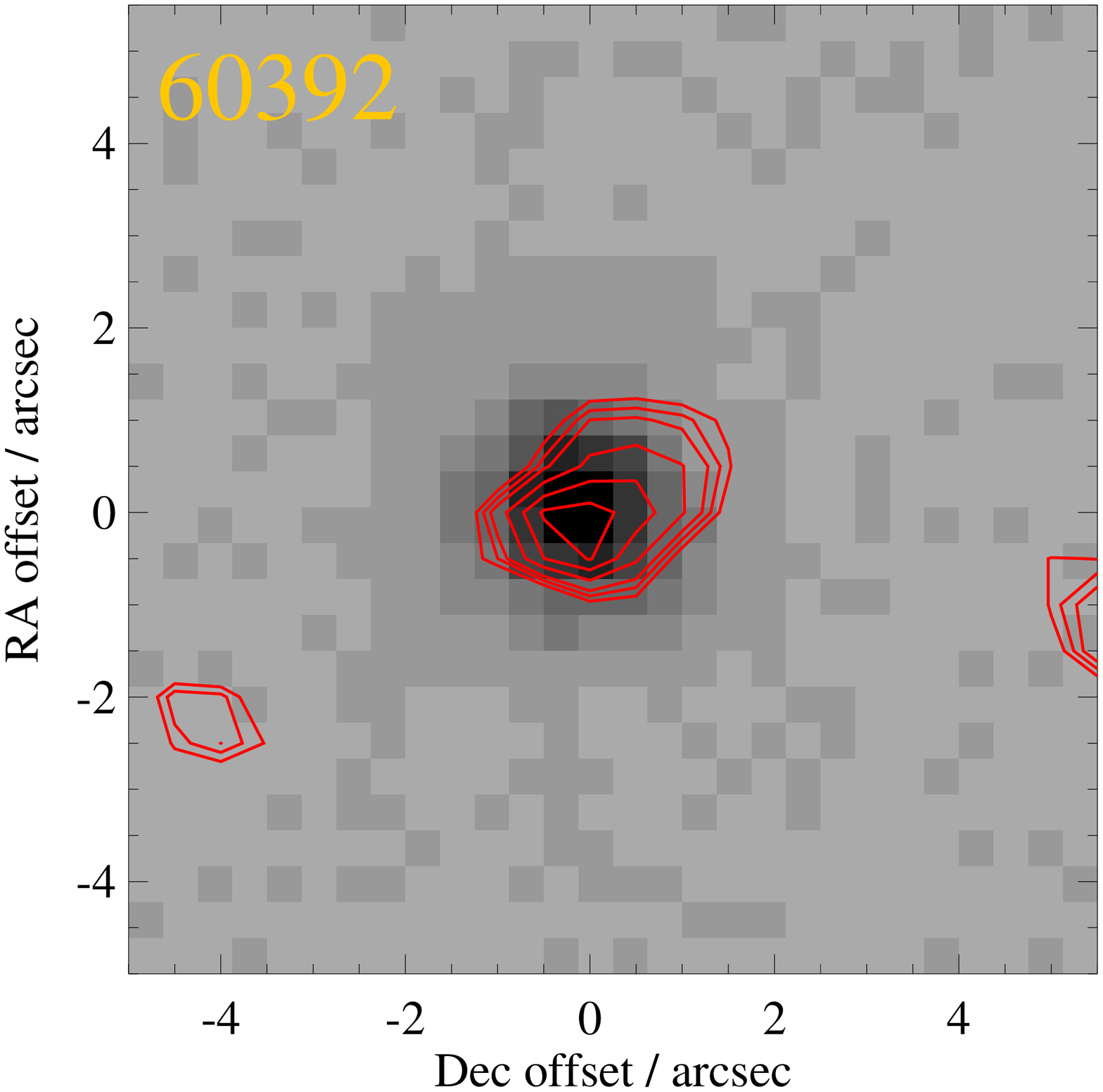} & \\
    \includegraphics[width=0.32\textwidth]{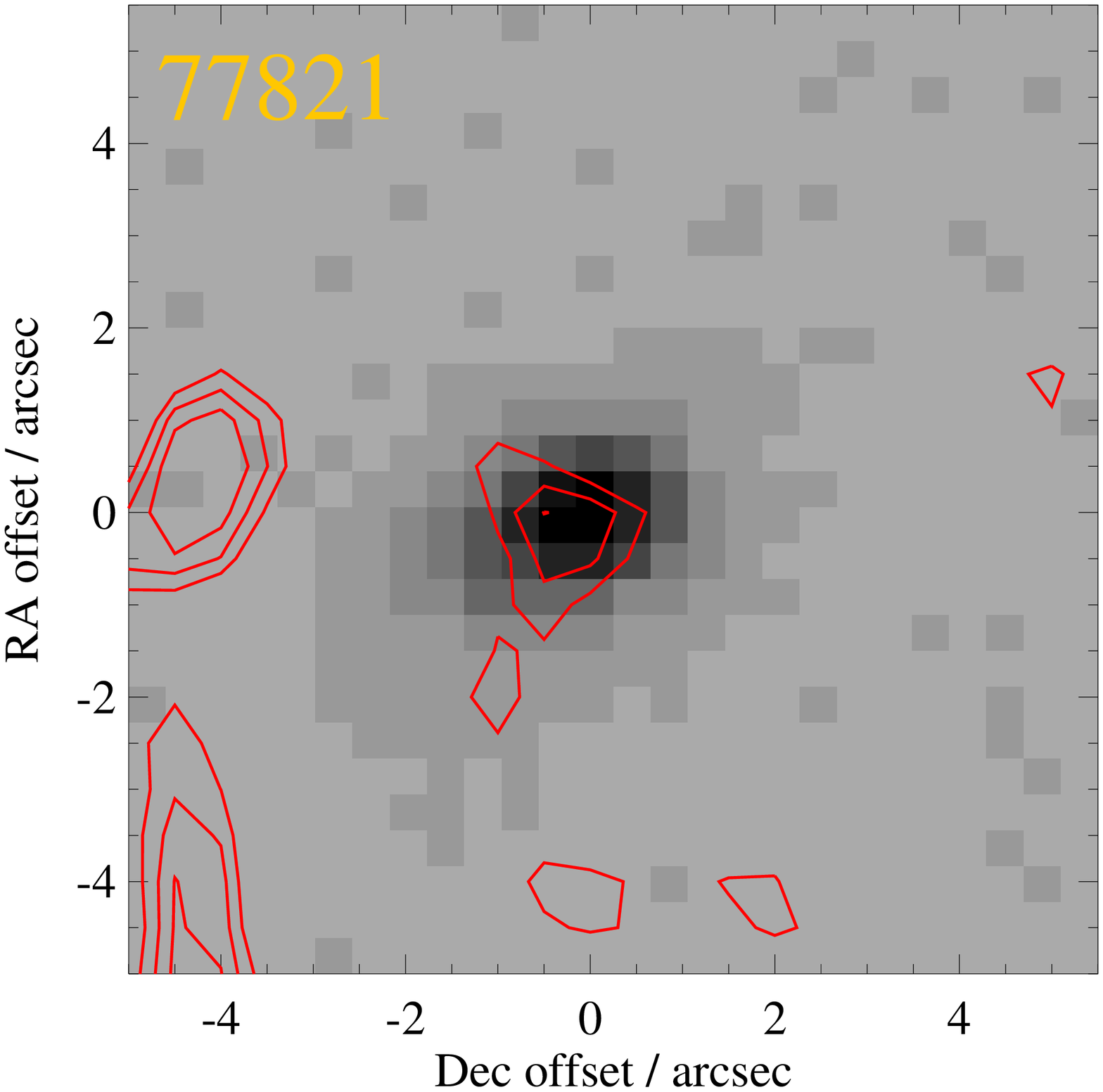} &    
    \includegraphics[width=0.32\textwidth]{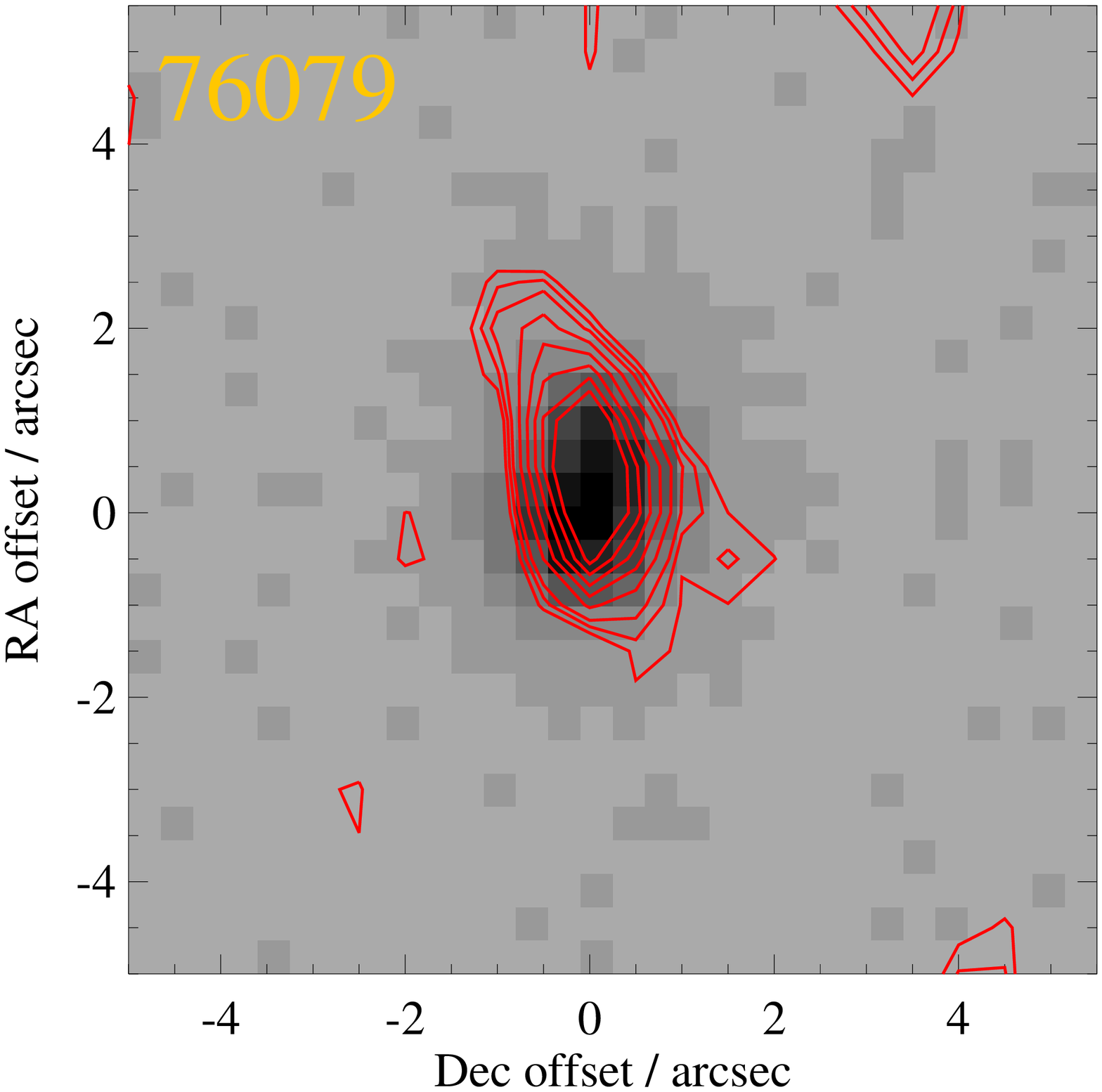} & 
    \includegraphics[width=0.32\textwidth]{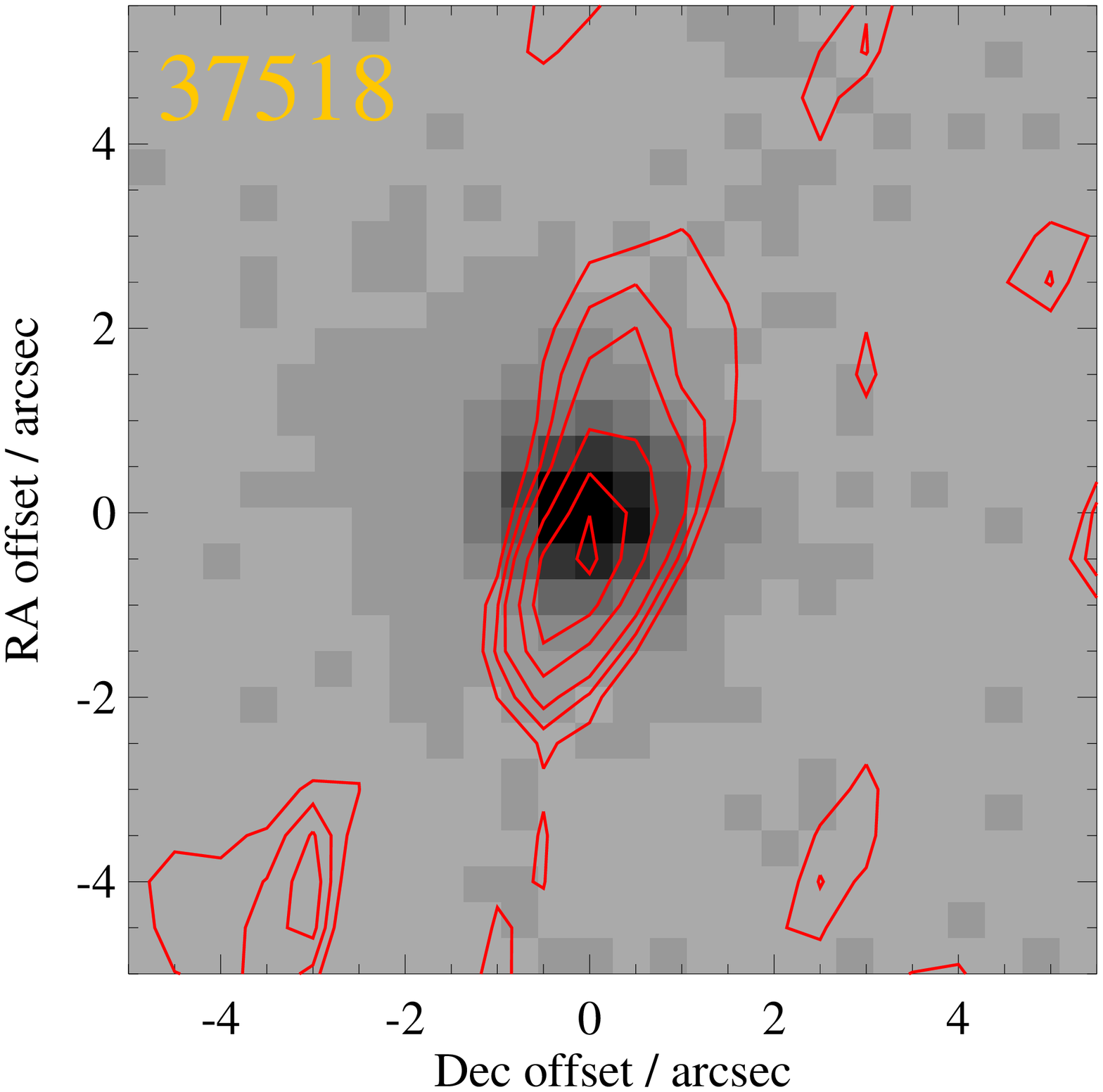} & \\
     \end{tabular}
     \caption{Radio contour plots overplotted on SDSS detections for objects with redshifts between 0.08 and 0.18. The contour levels are indicated at 2, 2.5, 3, 4, and 5$\sigma$.}
     \label{tab:Contours1}
\end{center}
\end{figure*}

\begin{figure*}
\begin{center}
    \begin{tabular}{ c c c c }
    \includegraphics[width=0.32\textwidth]{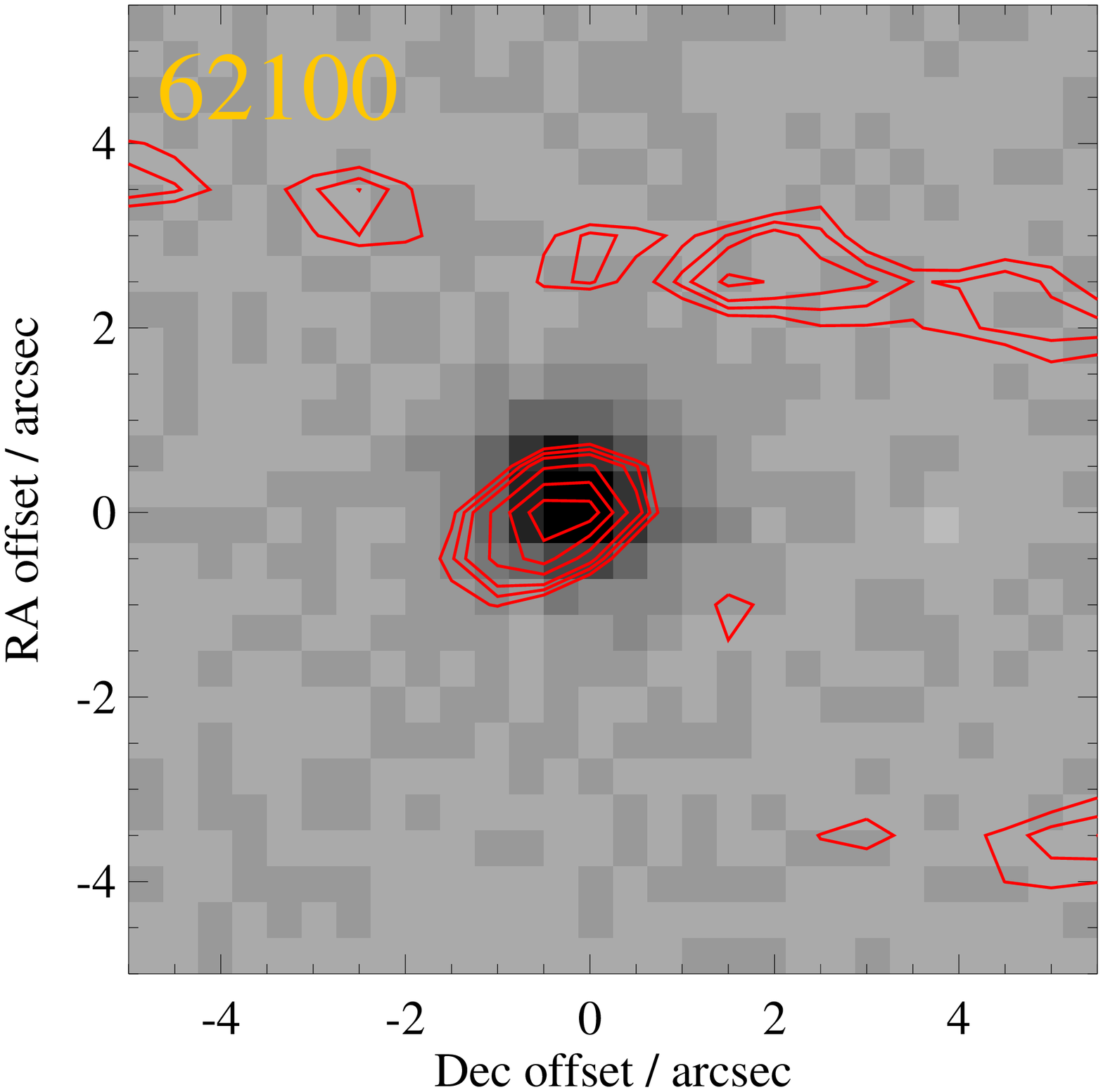} & 
    \includegraphics[width=0.32\textwidth]{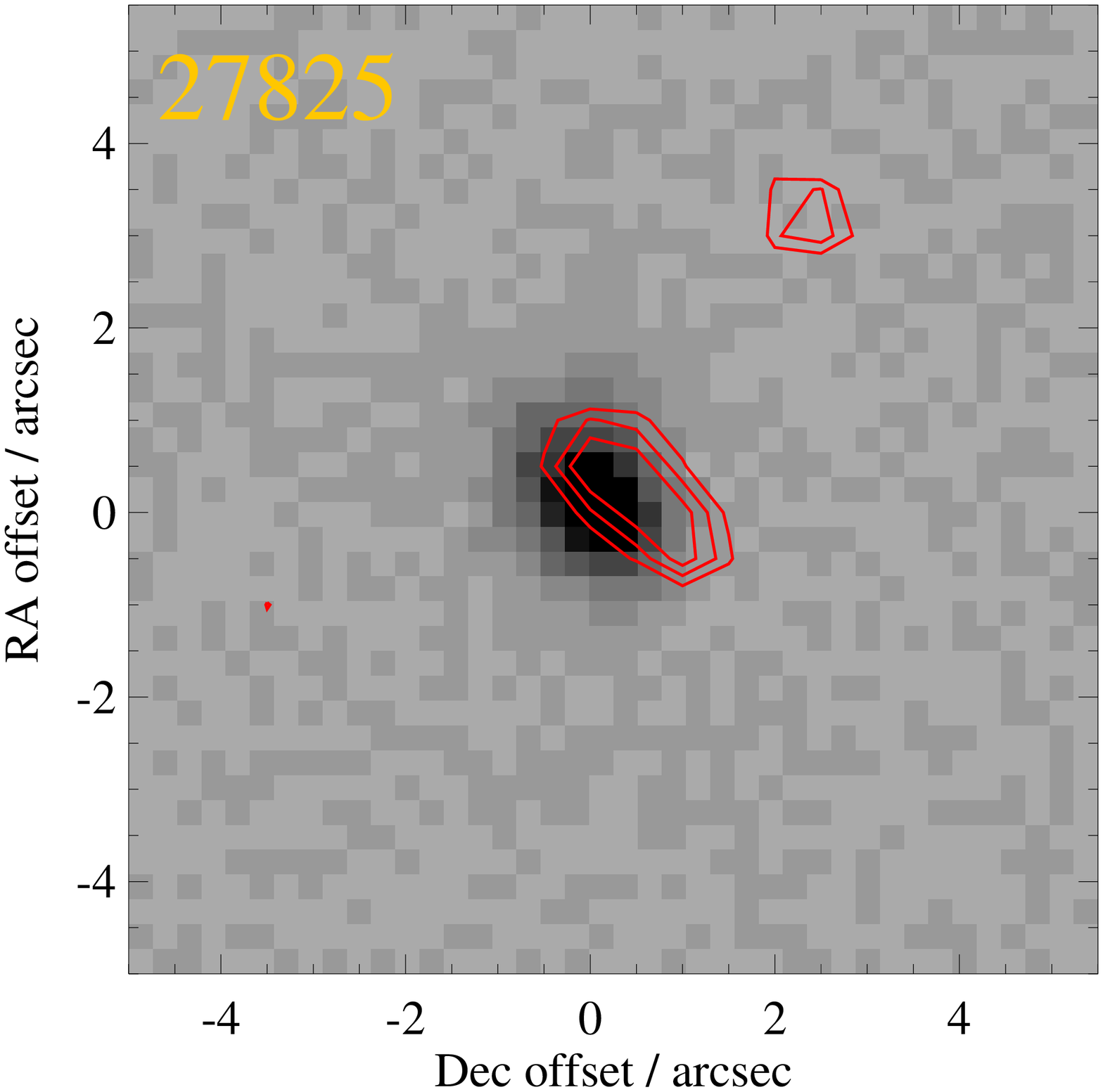} &    
    \includegraphics[width=0.32\textwidth]{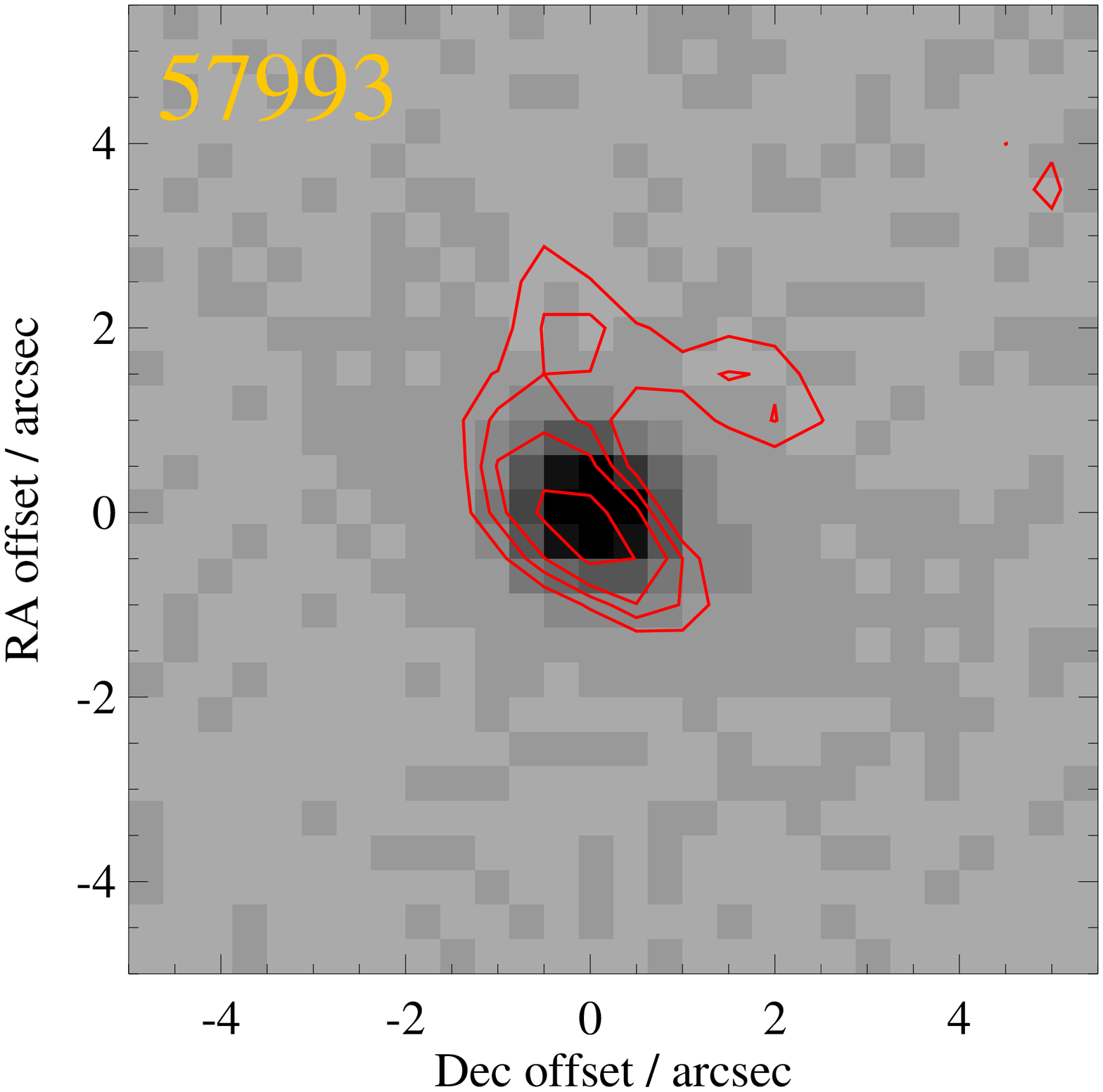} & \\
    \includegraphics[width=0.32\textwidth]{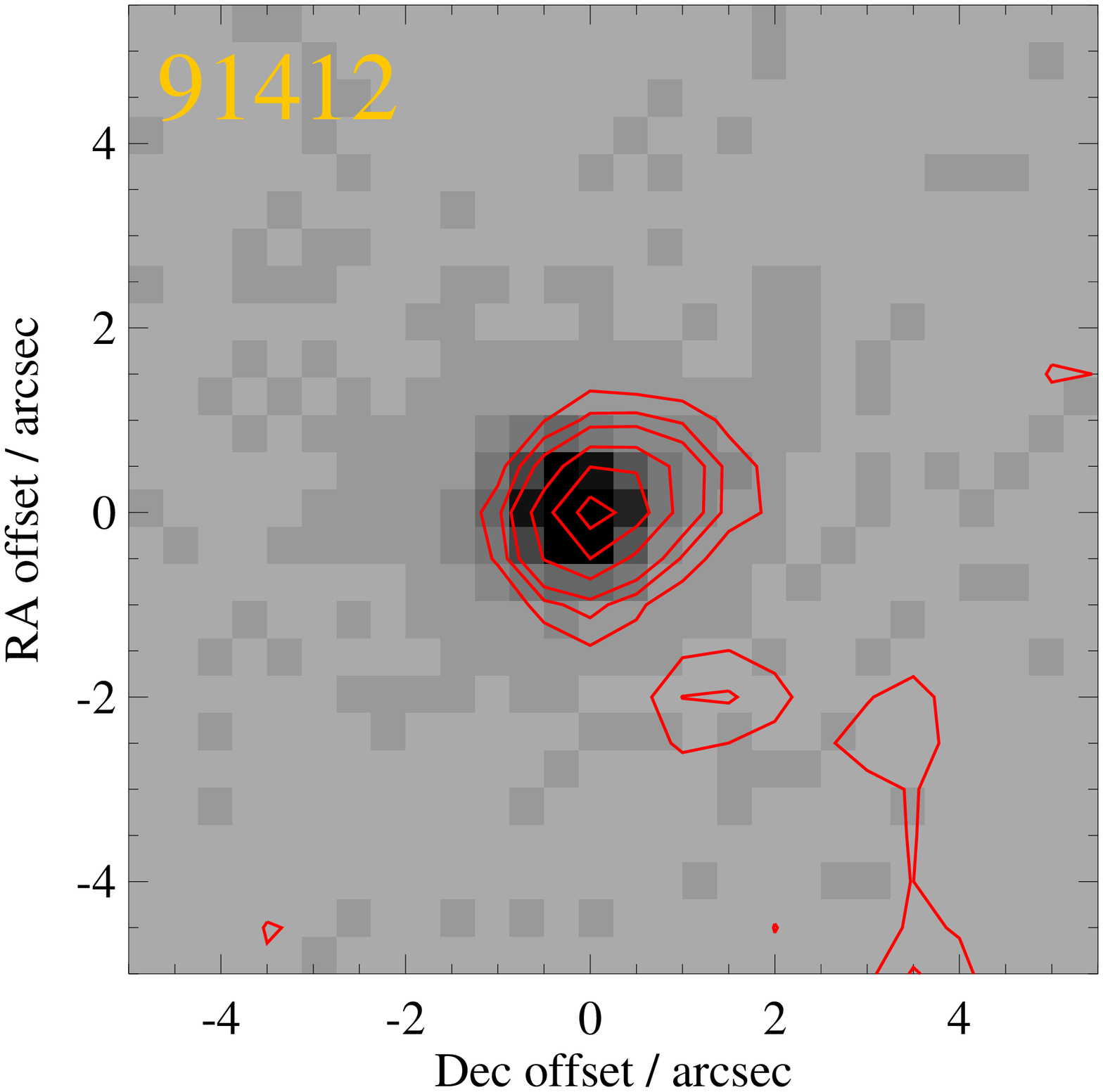} & 
    \includegraphics[width=0.32\textwidth]{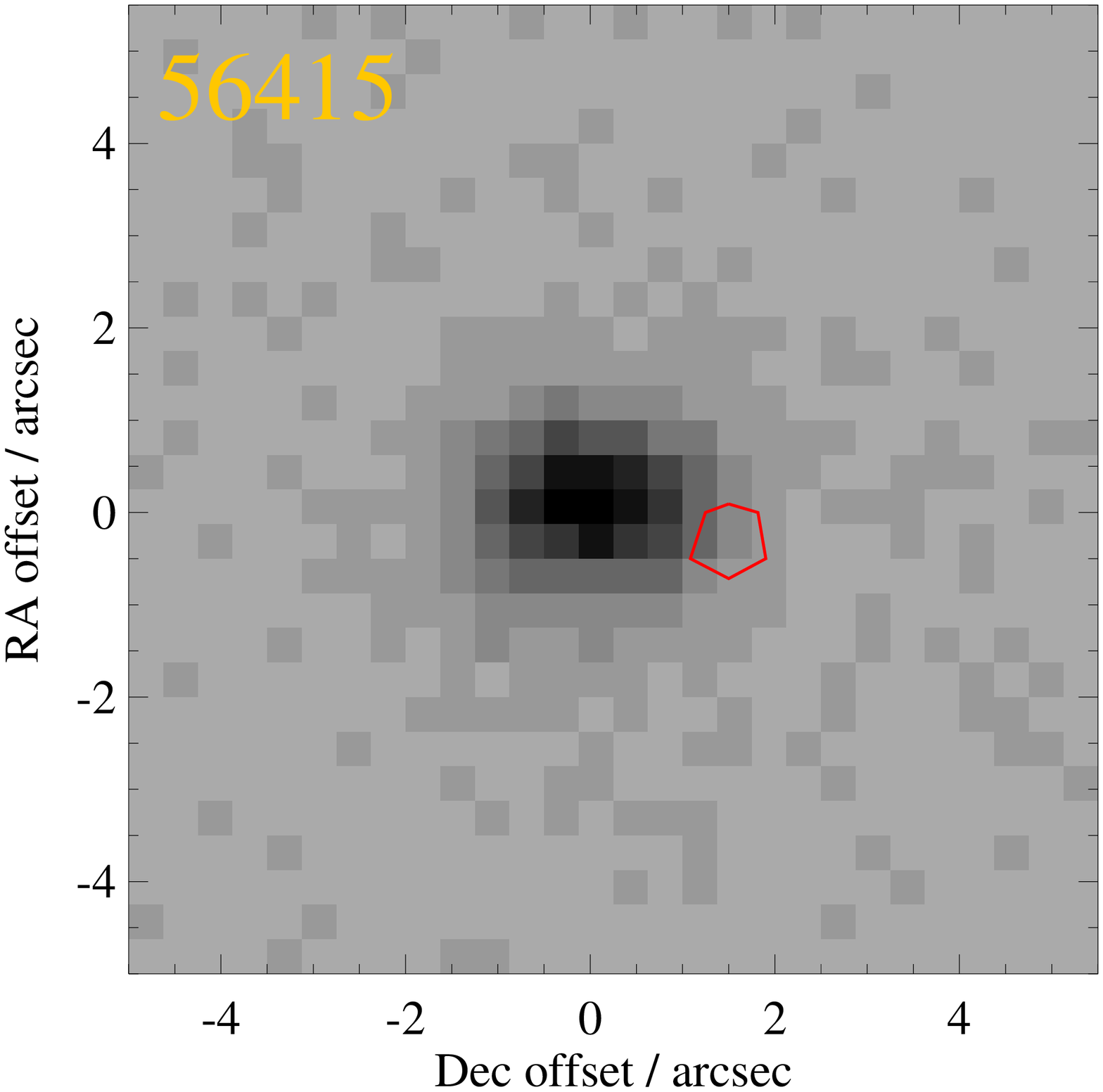} & 
    \includegraphics[width=0.32\textwidth]{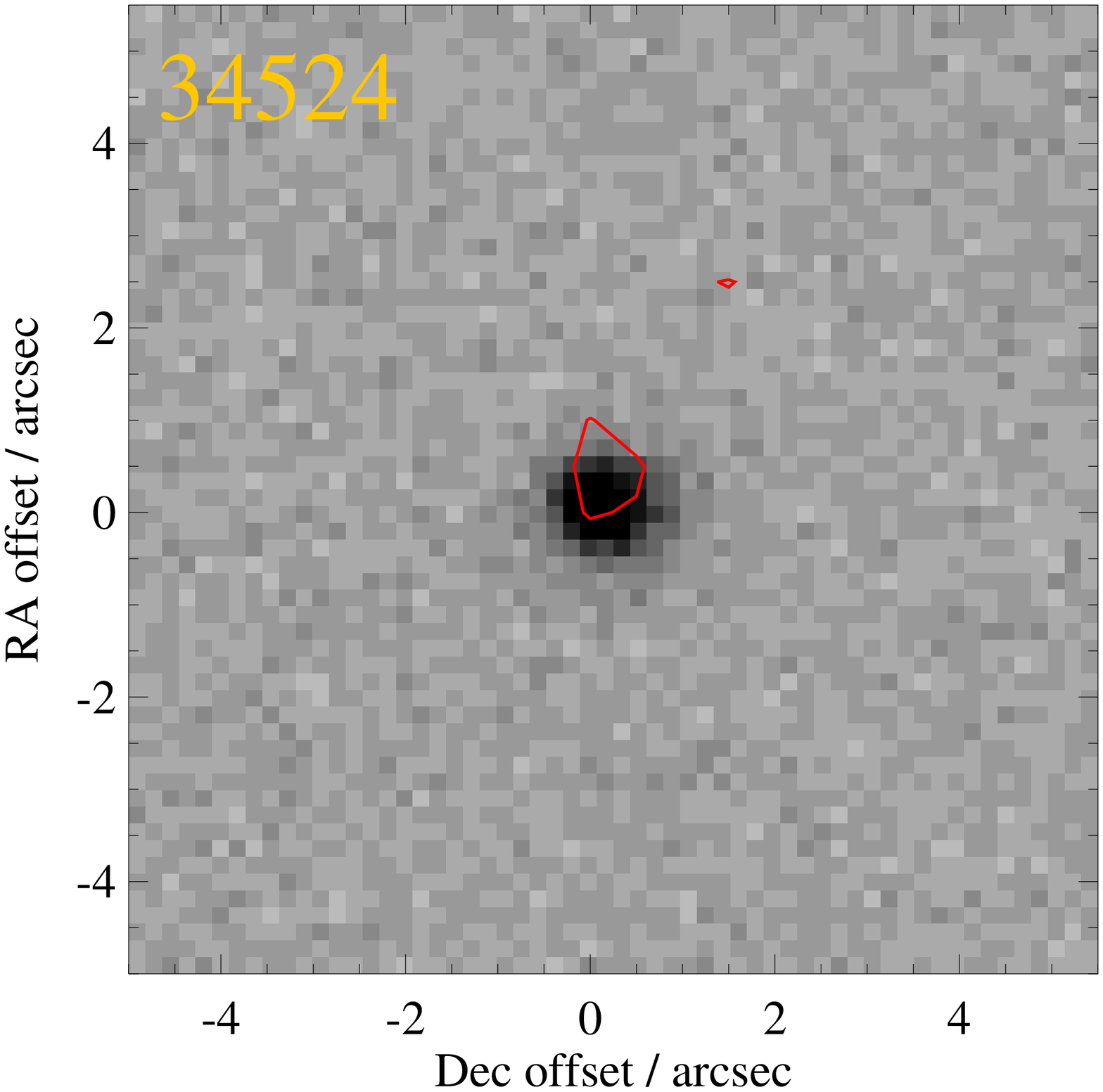} & \\   
    \includegraphics[width=0.32\textwidth]{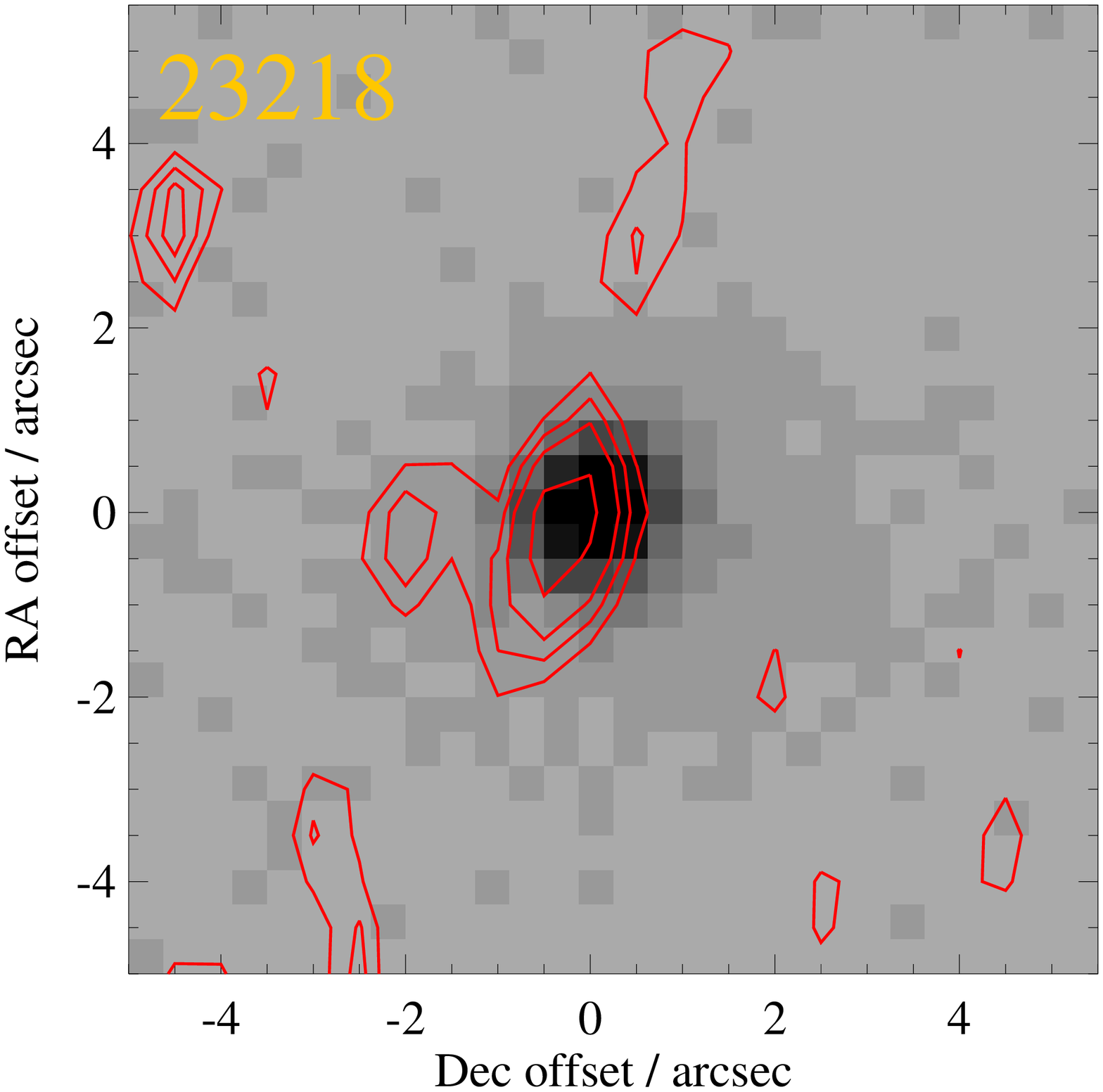} & 
    \includegraphics[width=0.32\textwidth]{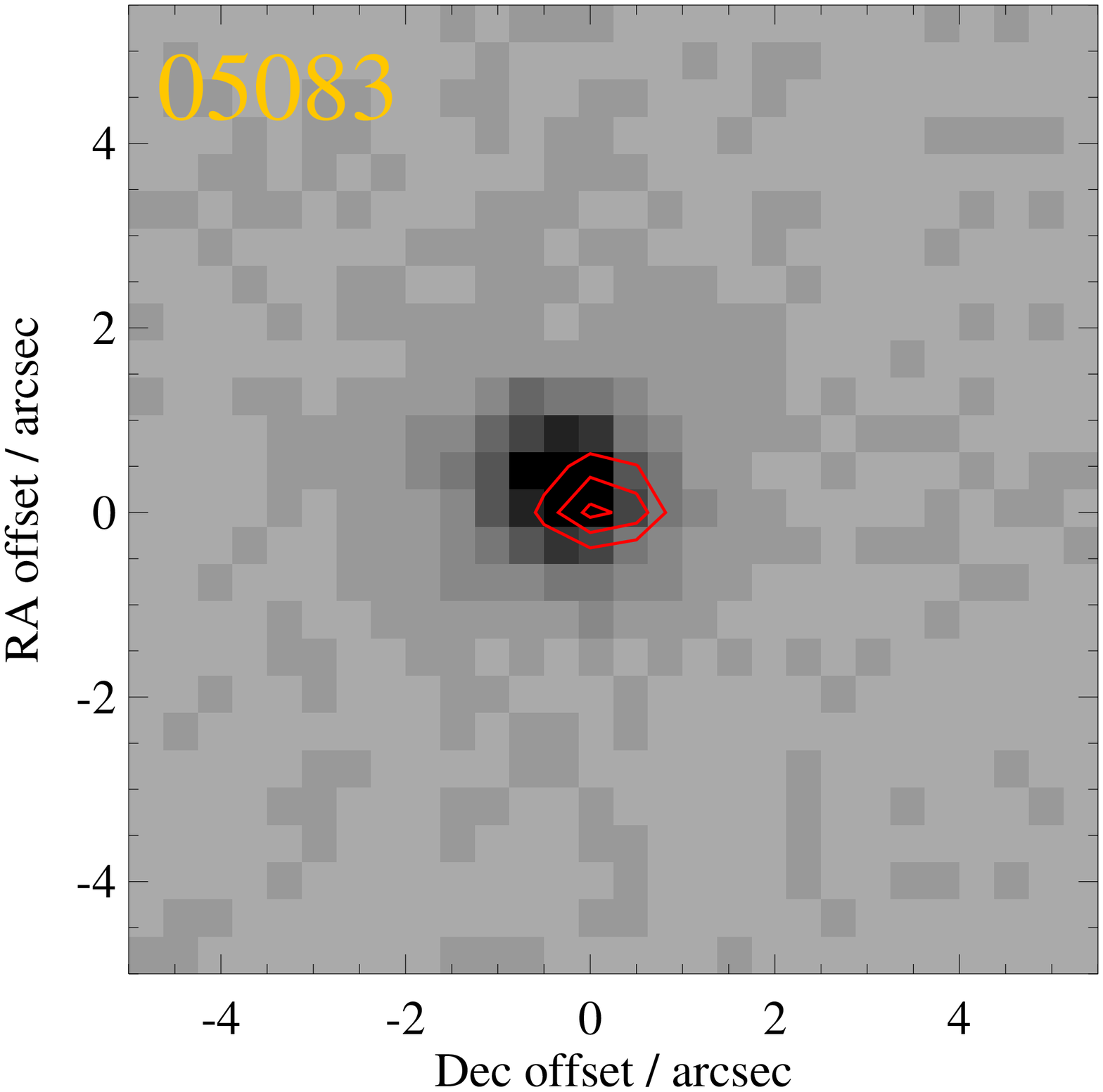} & 
    \includegraphics[width=0.32\textwidth]{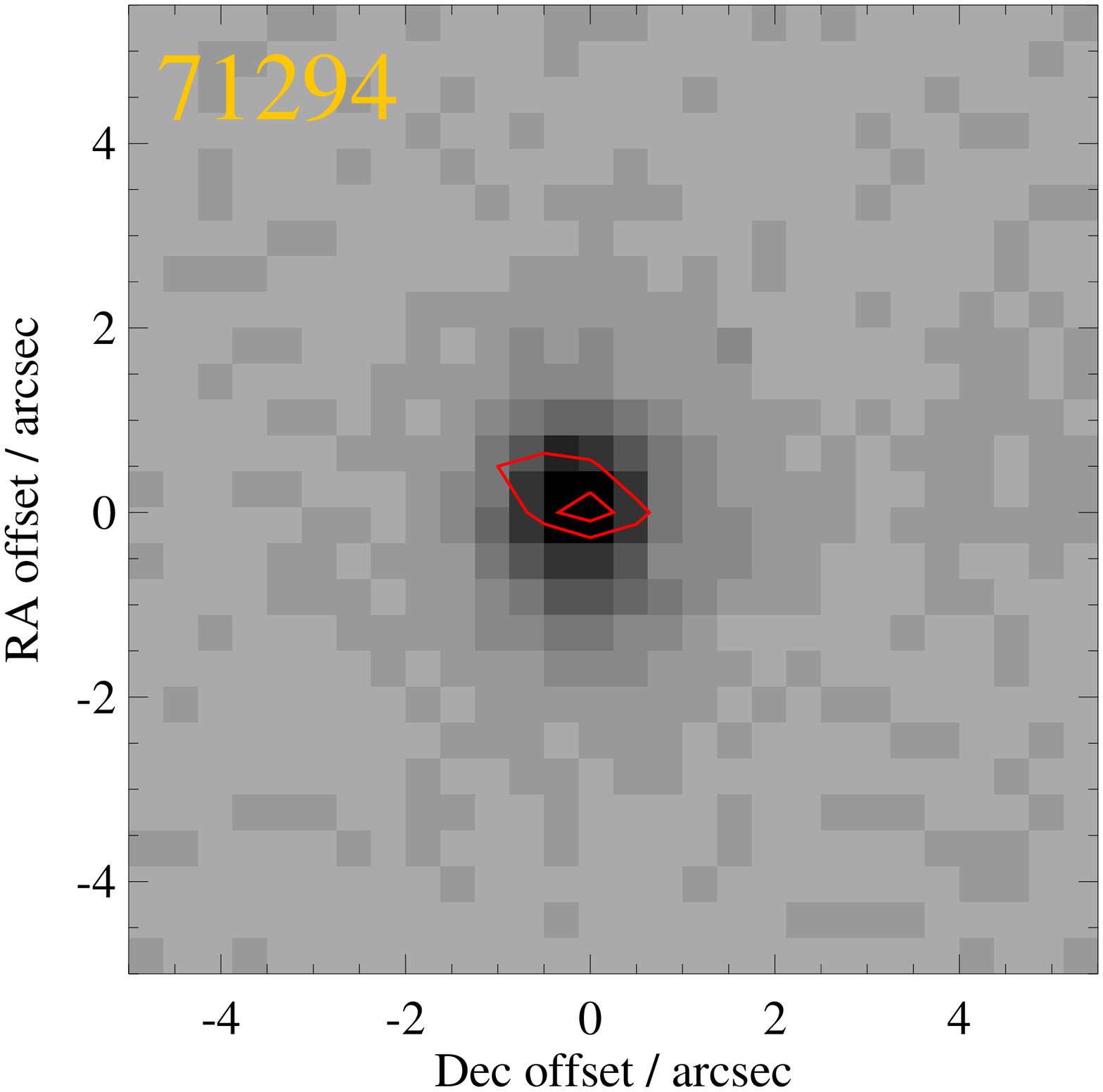} & \\
    \includegraphics[width=0.32\textwidth]{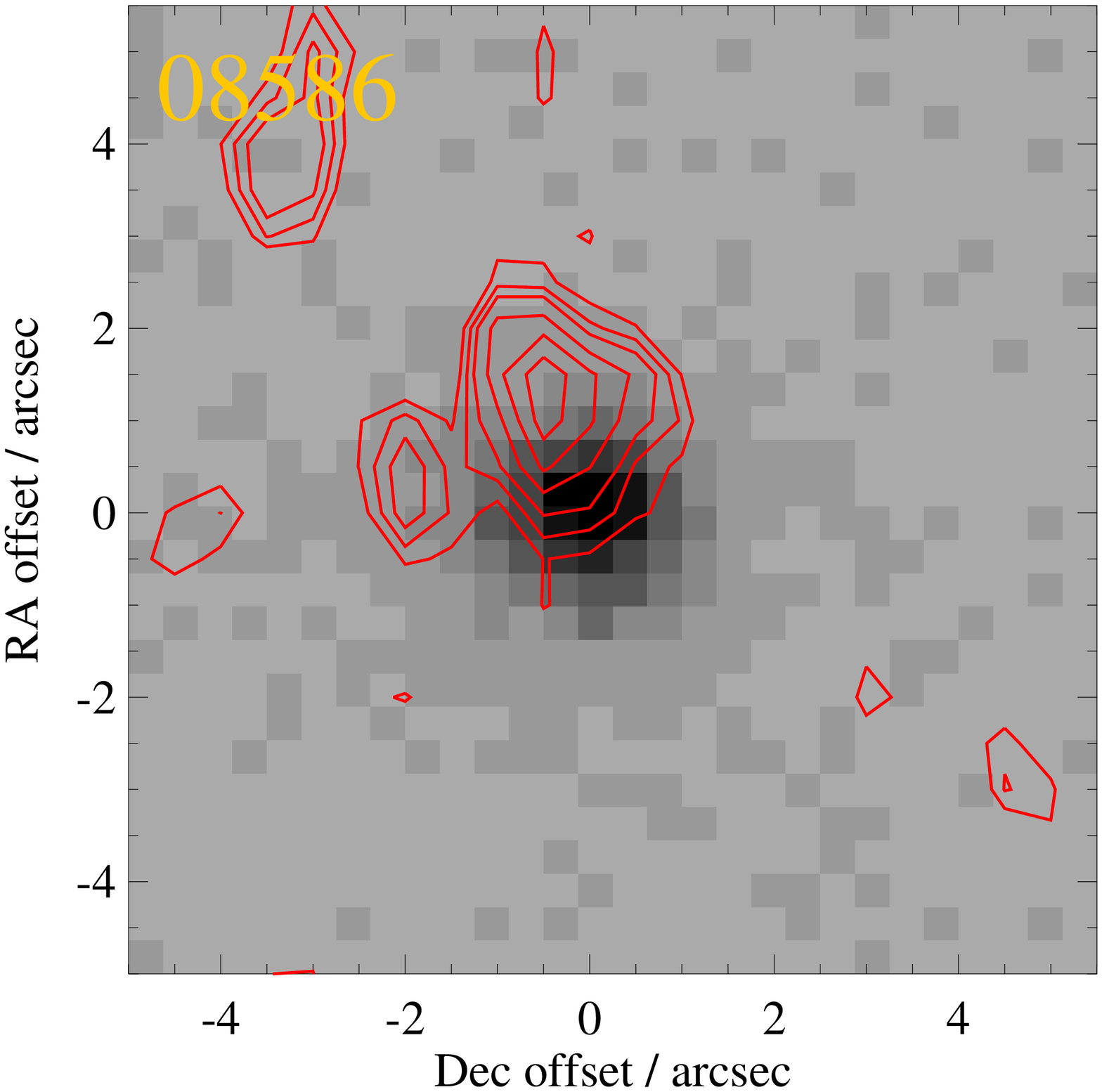} &    
    \includegraphics[width=0.32\textwidth]{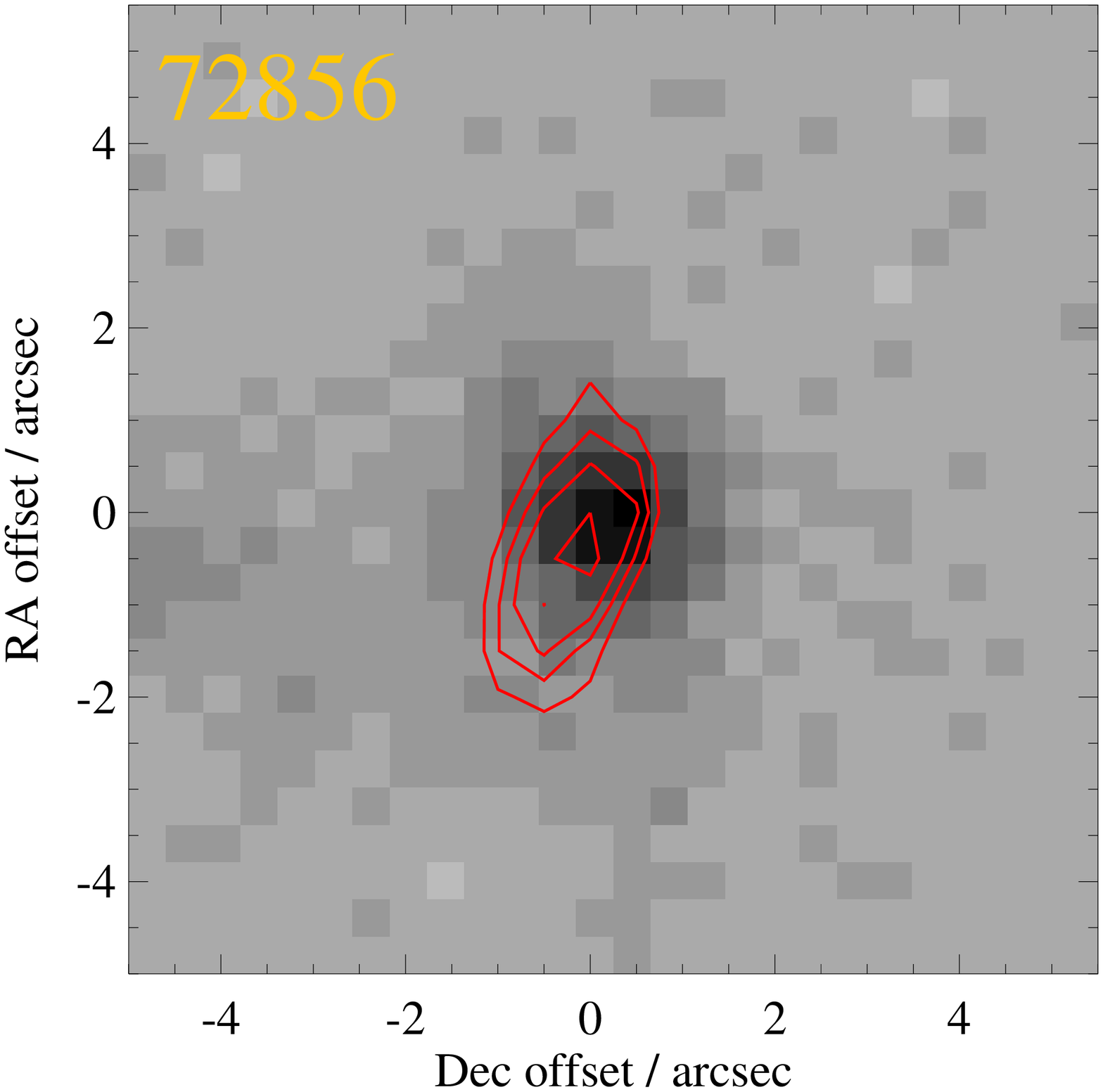} & 
    \includegraphics[width=0.32\textwidth]{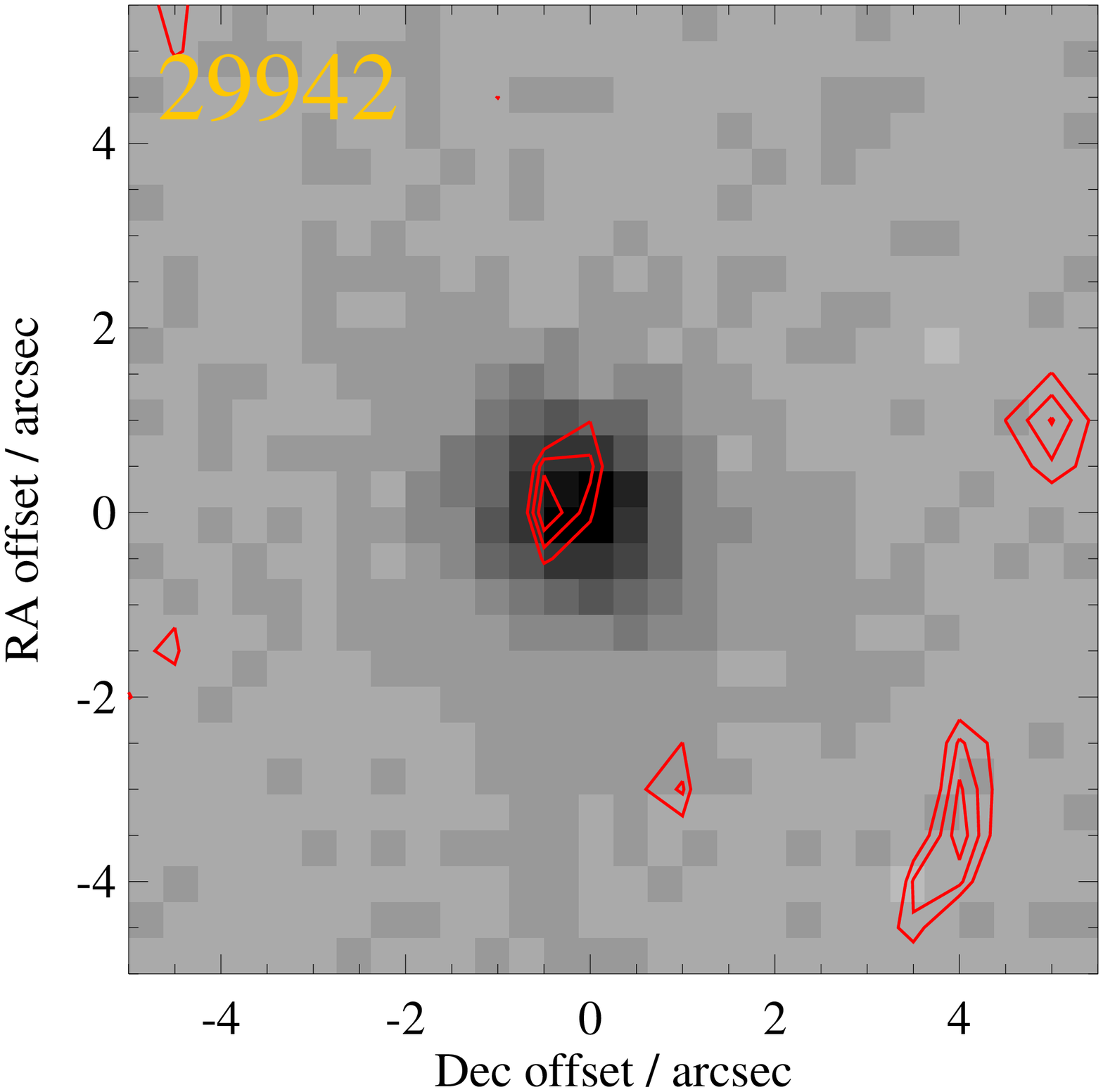} & \\
         \end{tabular}
         \caption{Same as \ref{tab:Contours1} for objects with redshifts between 0.135 and 0.15. The contour levels are indicated at 2, 2.5, 3, 4, and 5$\sigma$.}
         \label{tab:Contours2}
\end{center}
\end{figure*}

The following provides brief descriptions of each of the 32 objects observed in VLA programmes 14A-140, 15A-134, and 16B-104, as well the five undetected APEX targets (4 of which were also observed by the VLA). The descriptions are ordered by increasing redshift of the source. Radio contours overplotted onto SDSS r-band images are shown in figures \ref{tab:extendedContours}, \ref{tab:Contours1} and \ref{tab:Contours2}.

\begin{figure*}
\begin{center}
    \begin{tabular}{ c c c c }    
    \includegraphics[width=0.32\textwidth]{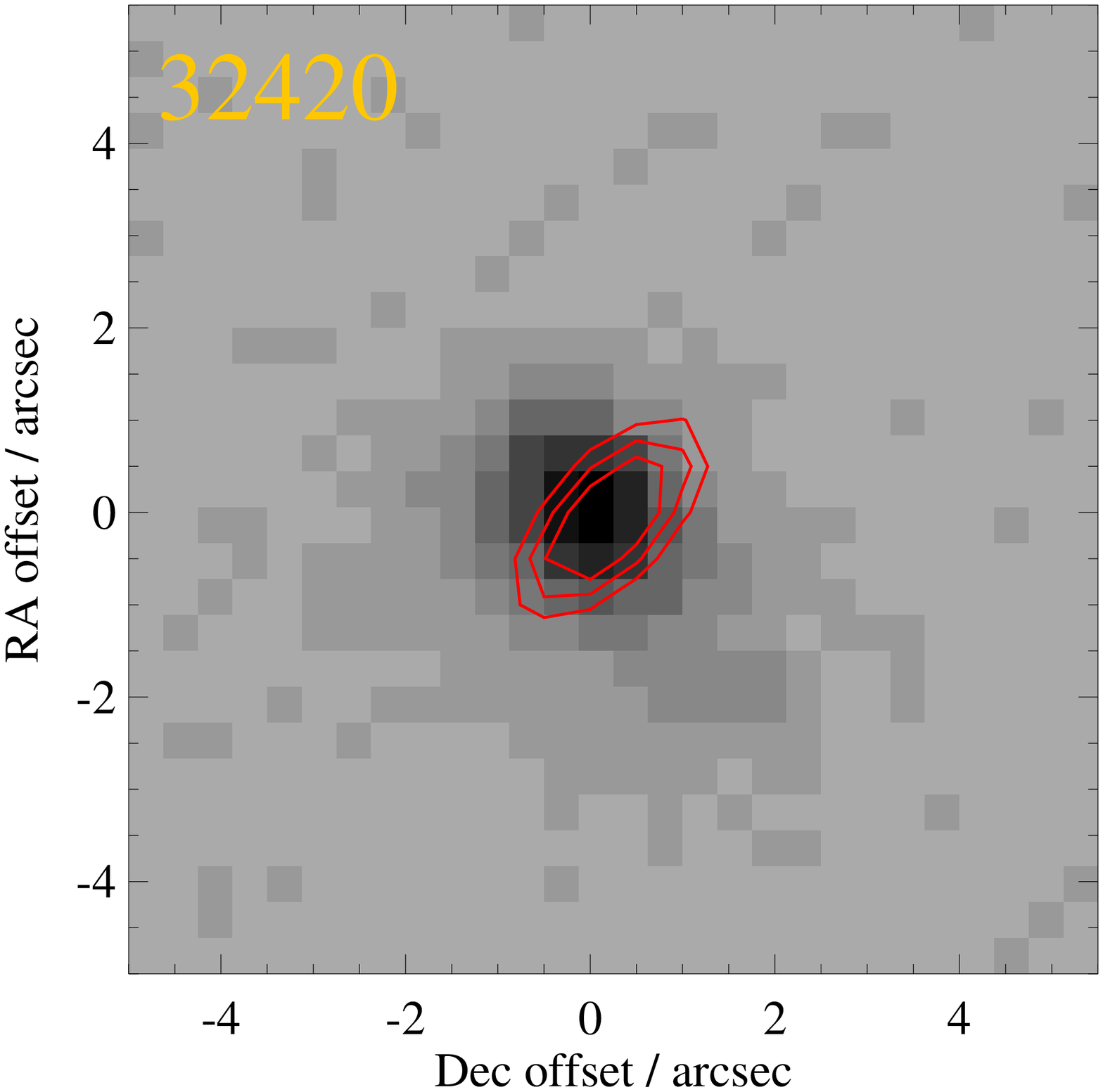} & 
    \includegraphics[width=0.32\textwidth]{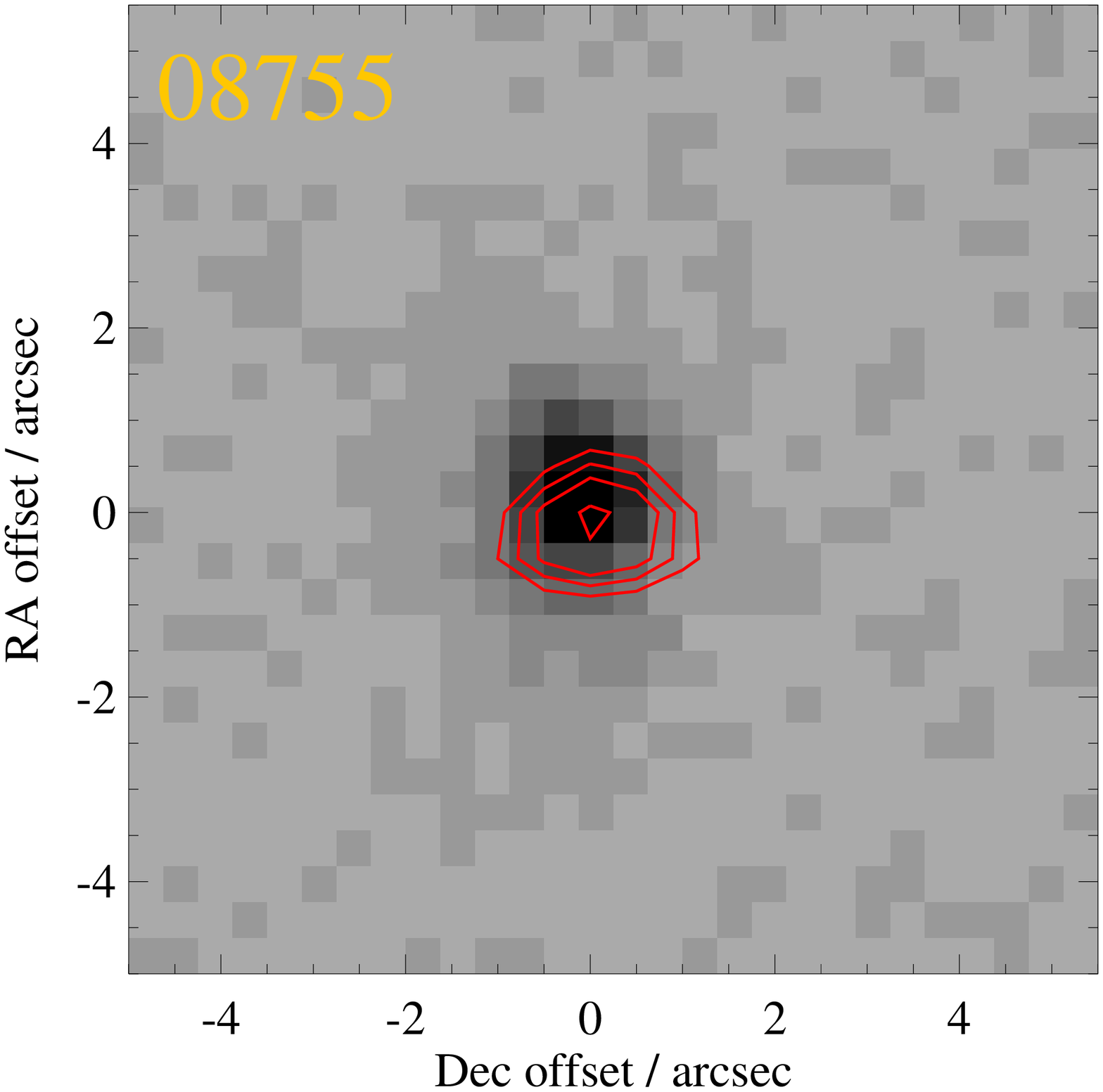} &    
    \includegraphics[width=0.32\textwidth]{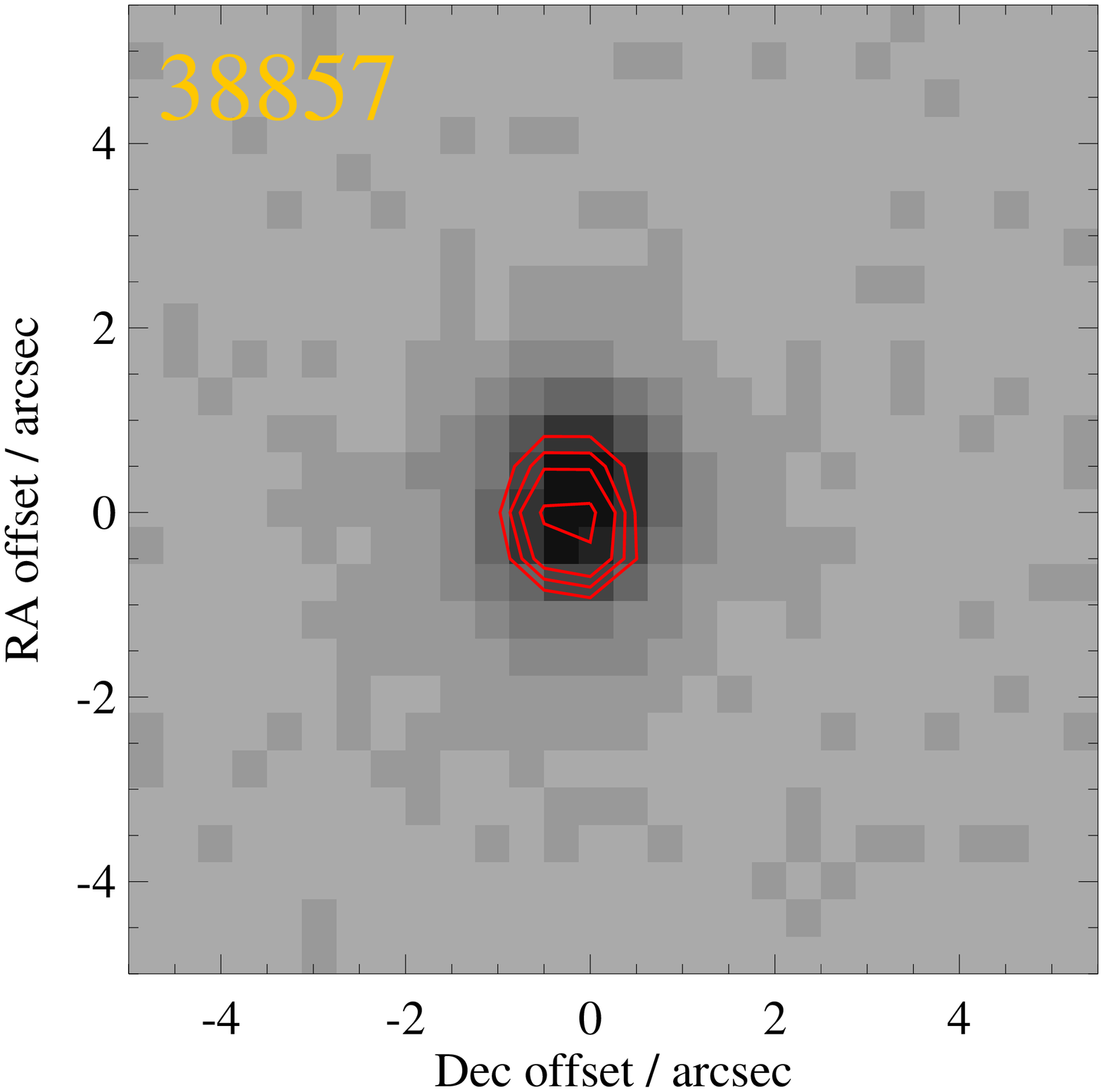} & \\
    \includegraphics[width=0.32\textwidth]{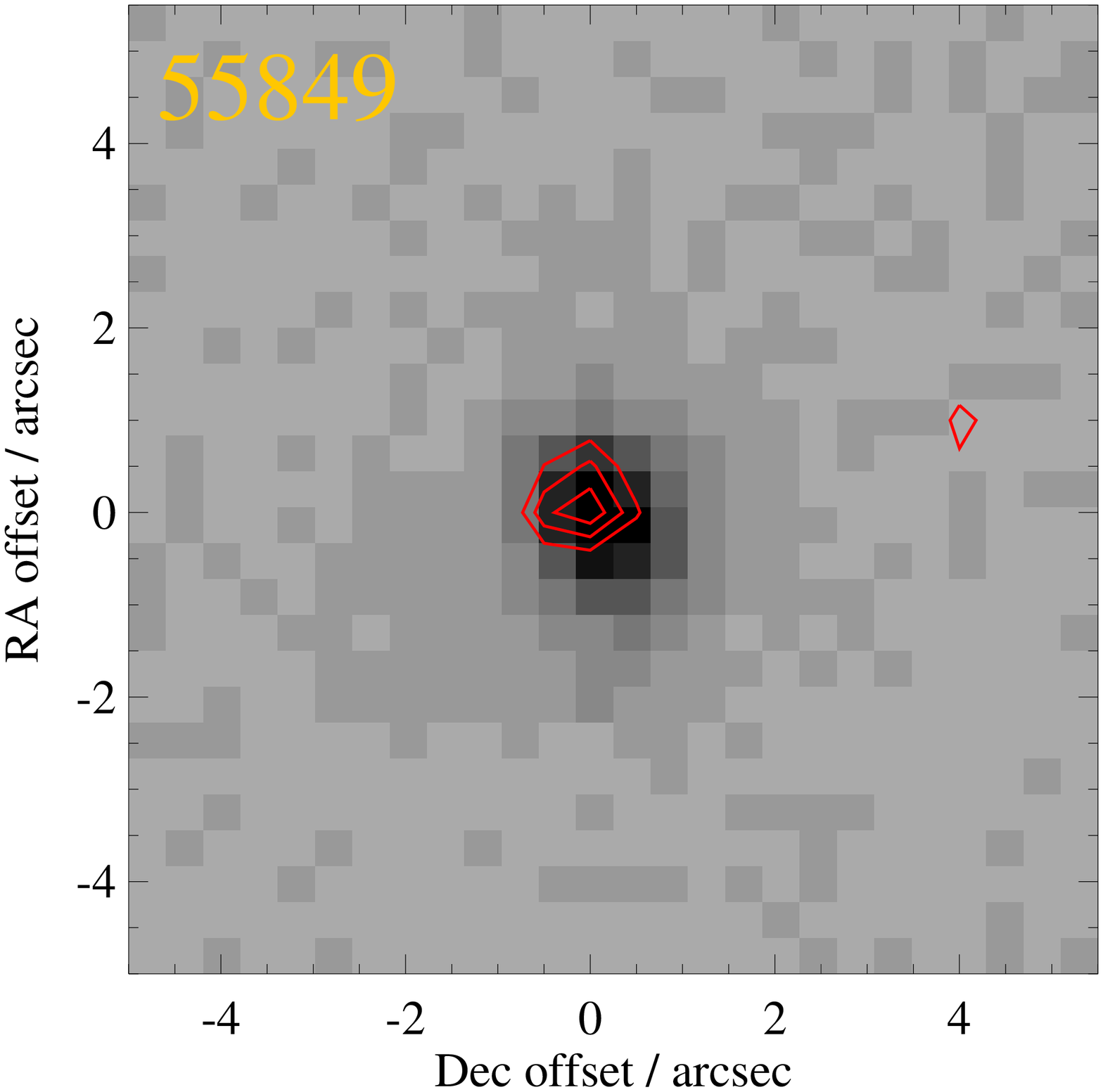} & 
    \includegraphics[width=0.32\textwidth]{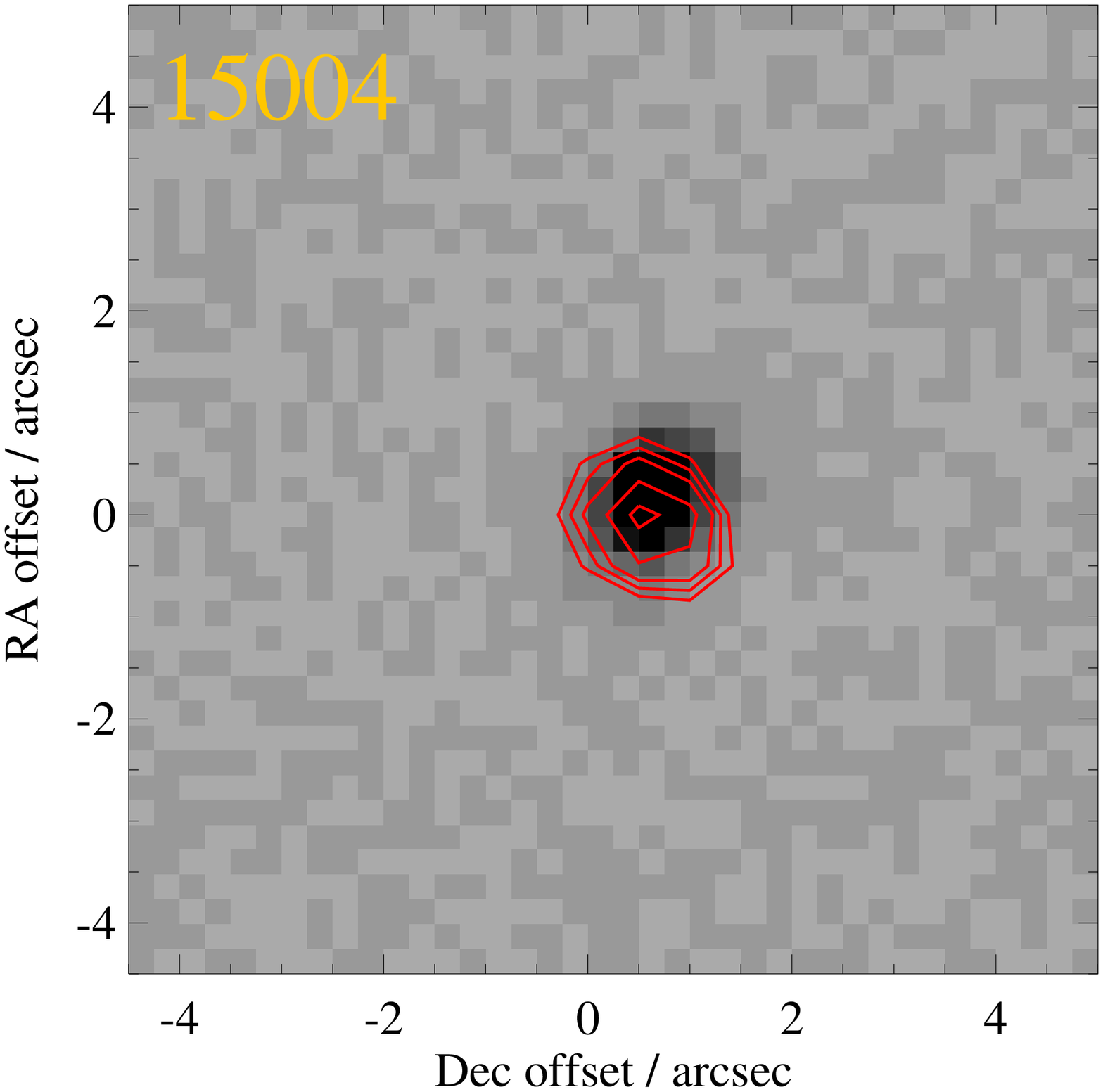} & 
    \includegraphics[width=0.32\textwidth]{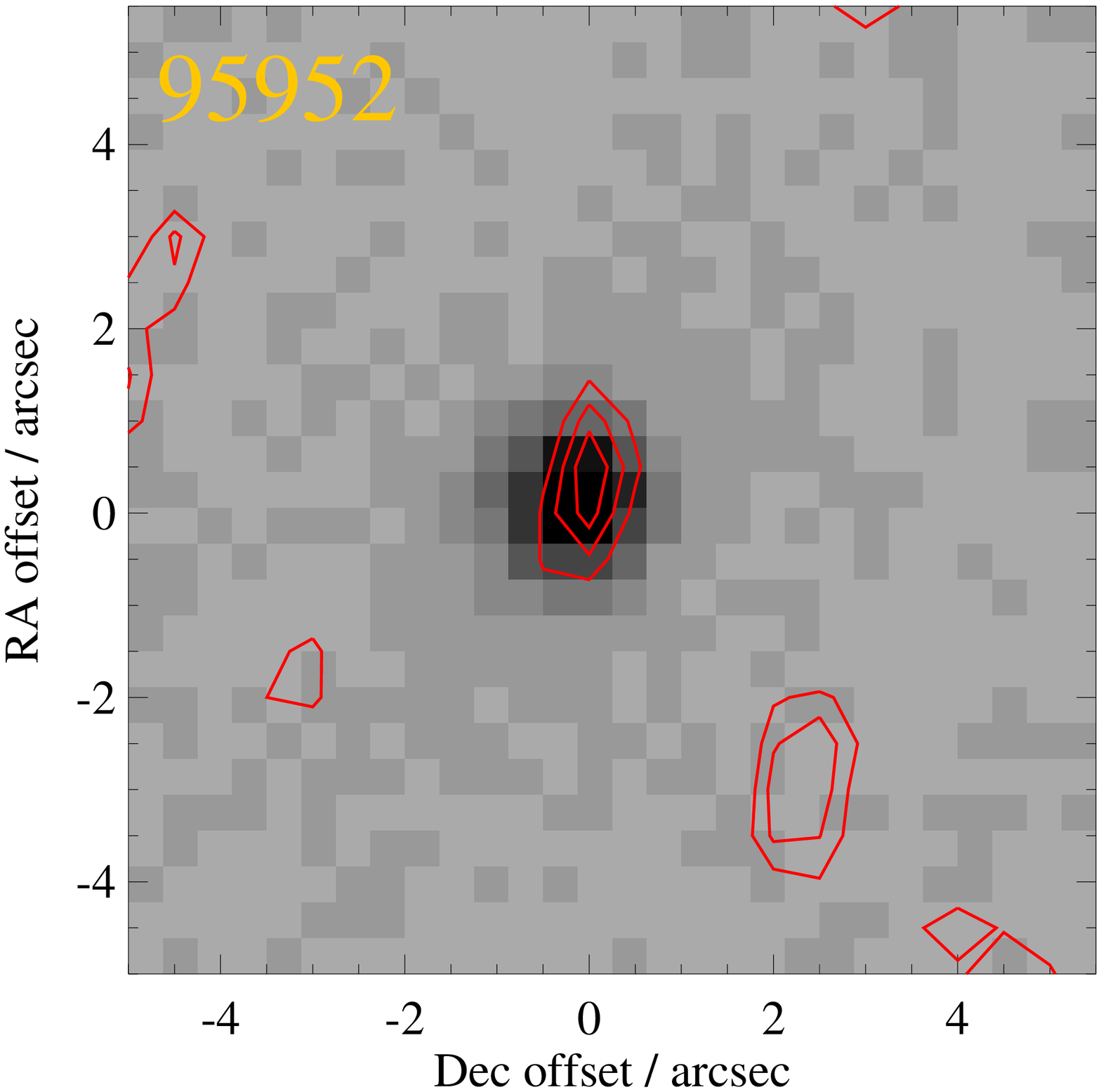} &  \\
    \end{tabular}
    \caption{Same as \ref{tab:Contours1} for objects with redshifts between 0.16 and 0.2. The contour levels are indicated at 2, 2.5, 3, 4, and 5$\sigma$.}
    \label{tab:Contours3}
\end{center}
\end{figure*}

\paragraph*{Object 54061} is the most nearby object with $z=0.074$ and the least massive of our radio sample with log(mass)$\sim$8.0 from SED fitting. Additionally, it is among the youngest objects in our sample with a best-fitting age of $\sim16$ Myr and has a metallicity of 0.23 Z$_{\odot}$. These physical properties make it a good local analogue to high-redshift LBGs. It is the only source in our large LBA sample which is identified in FIRST radio source catalogue with an integrated flux of 570 $\mu$Jy (compared to our measurements of $470 \pm 70$ $\mu$Jy). SDSS g-band imaging shows an elongated structure which is well traced by the radio contours (see figure \ref{tab:extendedContours}). Having been observed by both the VLA and ATCA, its spectral slope is found to be $\alpha = -0.60 \pm 0.25$, consistent with star-formation driven synchrotron emission as well as a strong thermal component. \textsc{CASA}'s `imfit' routine measures an angular size of 4.12 $\times$ 1.67 arcsec (corresponding to $\sim 15$ kpc$^2$). Its radio-derived SFR is 4.3 $\pm$ 0.6 M$_{\odot}$ yr$^{-1}$, resulting in a star formation rate density of 0.4 M$_{\odot}$ yr$^{-1}$ kpc$^{-1}$. With a $\Sigma_{SFR}$ above the critical value of 0.1 M$_{\odot}$ yr$^{-1}$ kpc$^{-1}$ \citep{Heckman2002}, it is thus likely that galaxy-wide superwinds exist in this object. 

\paragraph*{Object 27473} is both one of our most metal-poor sources with Z=0.18 Z$_{\odot}$, as well as one of the most dust-free objects in our radio sample with $E(B-V)=0.05$ from Balmer decrement measurements using SDSS spectroscopy. Together with its SED-derived mass of log(mass/M$_{\odot}$)$\sim$9.2 and age of $\sim$500 Myr this makes this object an acceptable local analogue. It was observed by both the VLA and ATCA, with a measured radio slope of $\alpha = -0.16 \pm 0.57$. Its large measured projected size of 3.97 $\times$ 0.97 arcsec and low radio SFR of 0.8 $\pm$ 0.2 result in the lowest $\Sigma_{SFR}$ of the radio sample of 0.1 M$_{\odot}$ yr$^{-1}$ kpc$^{-1}$. However, it is worth noting that the measured sizes are upper limits, and hence the true $\Sigma_{SFR}$ is likely higher than that derived here. Its radio contours coincident with its SDSS detection are shown in figure \ref{tab:Contours1}. \\

\paragraph*{Object 16911} is an intriguing source and can be seen in figure \ref{tab:Contours1}: despite fulfilling the optical and ultraviolet selection criteria, this source has by far the highest metallicity in our radio sample at 1.62 Z$_{\odot}$ (though it should be noted that the metallicity conversion used is unlikely to produce reliable results in such an extreme object). SED-fitting indicates a mass of log(mass/M$_{\odot}$)$\sim$9.8 and age of $\sim$400 Myr. These physical properties make Object 16911 an unlikely analogue to z $\sim$ 5 LBGs. \\
It was observed by both the VLA and ATCA, producing the steepest of the observed radio slopes with $\alpha = -0.49 \pm 0.29$. Together with its position on the BPT diagram, this indicates the possible presence of an optically-obscured AGN. However, while the VLA observation overall yielded a signal-to-noise ratio of 6.6, when splitting the measurment set into its three component scans, only one scan resulted in a detection with non-detections in the other two. We note that as figure \ref{tab:Contours1} shows, this source has a nearby brighter object whose beam pattern passes close to the source. However, it does not appear to be affecting the source detection or flux measurement. Nonetheless, we conclude that the VLA detection of this source should be considered tentative.\\

\paragraph*{Object 23734} was observed by both the VLA and ATCA with $>6\sigma$ detections in both, giving rise to a radio spectral slope of $\alpha$ = -0.19 $\pm$ 0.21. It had previously been classified as a potential Seyfert 1 galaxies due to its strong emission lines; however, both its spectral slope and high radio star formation rate (SFR=6.0M$_{\odot}$ yr$^{-1}$), indicate a star-formation driven galaxy. Since it has one of the smallest projected area of the sources in this sample (see figure \ref{tab:extendedContours} for its contour plot), the $\Sigma_{SFR}$ determined for this source is among the highest in our sample at 3.5 M$_{\odot}$ yr$^{-1}$ kpc$^{-2}$. Its SED-derived mass of log(mass/M$_{\odot}$)$\sim$8.3 and young dominant stellar population (with an age of $\sim$ 16 Myr), as well as its low dust attenuation ($E(B-V)\sim 0.26$) and metallicity (0.34 Z$_{\odot}$) make Object 23734 a good local analogue to high-redshift LBGs. It is interesting to note that while the optical source is elongated, the radio source is spatially compact.

\paragraph*{Object 83073} was detected at $3.2\sigma$ in our VLA observations. It is a very dust- and metal-poor galaxy with a Balmer-derived $E(B-V)$ value of 0.09, and $\sim0.14$ Z$_{\odot}$. With an SED-derived mass of log(mass) $\sim$ 9.6, and a stellar population age of $\sim 600$ Myr, it is an acceptable, though not ideal, local analogue. Its SDSS detection and radio contours are shown in figure \ref{tab:Contours1}. 

\paragraph*{Object 76428} was observed with the VLA, with an observed SNR of 4.5. It has a radio-derived SFR of 2 M$_{\odot}$ yr$^{-1}$, and one of the lowest $\Sigma_{SFR}$ at $\sim$0.2 M$_{\odot}$ yr$^{-1}$ kpc$^{-2}$. However, as the imfit-derived projected areas are upper limits, this $\Sigma_{SFR}$ is a lower limit (its radio contours are shown in figure \ref{tab:Contours1}). As it lies above the critical value found by \citet{Heckman2002}, it is likely that the galaxy experiences strong superwinds. Using SED-fitting and spectroscopic analysis, its mass and age where found to be log(M/M$_{\odot}$) = 9.4 and 400 Myr, at half Solar metallicity. Since the H$\alpha$-derived SFR in the source is a factor of $\sim4$ higher, a recent starburst which has not yet had time to establish a strong radio continuum appears to be a reasonable interpretation. These physical properties make Object 76428 an acceptable local analogue. 

\paragraph*{Object 24784} is one of two sources in this sample which were observed with VLA, ATCA, and APEX. It was detected at 4.3 $\sigma$ and 5.6 $\sigma$ in the VLA and ATCA data respectively, but undetected in APEX. Radio contours and its SDSS detection are shown in figure \ref{tab:Contours1}. The derived radio spectral slope is consistent with zero (see Table \ref{table:Results_withSpecSlope}), and indicative of star-formation driven radio emission, with a stellar population that has not yet produced sufficient supernovae to establish a strong synchrotron emission spectrum. The SED physical properties derived in \citet{Greis2016} using ultraviolet to near-infrared photometry for this galaxy indicate a very young dominant stellar population at $\sim5$ Myr, the youngest of our radio sources. Additionally it is also one of the most dust-free objects in our radio sample with $E(B-V)=0.06$ from Balmer decrement. Together with an SED-found stellar mass of log(mass/M$_{\odot}$)$\sim$8.5, the physical properties of this source suggest that it can be considered a good local analogue to $z\sim5$ LBGs. 

\paragraph*{Object 80573} is both one of the most massive of our radio sources with log(mass)$\sim$10 from SED-fitting and one of the oldest objects with an SED-derived age of $\sim$1 Gyr. It was detected in our VLA observations with a SNR of 5.6 (see figure \ref{tab:Contours1} for its contour plot), and is one of four objects in our sample whose radio SFR exceeds the H$\alpha$ SFR at 2.7 $\pm$ 0.5 M$_{\odot}$ yr$^{-1}$ and $\sim$ 1.8 M$_{\odot}$ yr$^{-1}$ respectively. This appears consistent with a massive galaxy whose dominant stellar populating is reaching an age at which a supernova-driven radio continuum has been established. It is hence unlikely that this object would make a good local analogue to $z\sim5$ LBGs. 

\paragraph*{Object 08959} produced a 7.3 $\sigma$ detection in our VLA observations. It has one of the smallest projected areas derived from the radio observations, resulting in one of the highest $\Sigma_{SFR}$ of our radio sample at 1.5 M$_{\odot}$ yr$^{-1}$ kpc$^{-2}$. The radio detection is clearly coincident with the SDSS detection, as shown in figure \ref{tab:Contours1}. At an SED-derived age of $\sim$ 100 Myr and an H$\alpha$ SFR of 7.6 M$_{\odot}$ yr$^{-1}$ (a factor of $\sim$ 2 higher than its radio star formation rate), this source appears to show a recent starburst which has not yet established a strong radio continuum. Its stellar mass of log(mass/M$_{\odot}$)$\sim$9.3, very low dust content (with $E(B-V)\sim0.11$), and very low metallicity of 0.18 Z$_{\odot}$ make this source a good local analogue to high-redshift LBGs. 

\paragraph*{Object 53150} was observed by the VLA, producing an 8.8 $\sigma$ detection with an observered radio flux indicating a SFR of $\sim 3.3$ M$_{\odot}$ yr$^{-1}$. With a best-fitting age of $\sim 250$ Myr and a stellar mass of log(mass) $\sim 9.3$, Object 53150 is a good local analogue to distant galaxies. It is very dust-poor with an $E(B-V)$ value of $\sim0.11$ derived from Balmer-decrement measurements, and has a gas-phase metallicity of $\sim 0.35$ Z$_{\odot}$.

\paragraph*{Object 60392} was observed by both the VLA and ATCA. SED-fitting indicates that this is both one of the most massive of our radio sources with log(mass)$\sim$10.3, and the oldest one at log(age)=9.8. This suggests that it is unlikely to be a good local analogue galaxy.\\
Similarly to Object 16911, this source was well-detected with a SNR of 6.8 when combining all scans, but produced non-detections in two out of three scans when split into its three constituent scans. It produced the strongest radio SFR measued in our radio sample, but given the uncertainty on the reliability of the scans, this should be treated with caution. Additionally, it was not detected above 3$\sigma$ in ATCA, making it difficult to put meaningful constraints on its radio spectral slope. We show its radio contours and SDSS detection in figure \ref{tab:Contours1}. 

\paragraph*{Object 77821} was observed with the VLA, producing a 6 $\sigma$ detection. Interestingly, it is, similarly to Object 80573, one of the most massive of our radio sources with log(mass)$\sim$9.9 (from SED-fitting) while also having a higher radio SFR than that calculated from H$\alpha$, with 6.7 $\pm$ 1.1 M$_{\odot}$ yr$^{-1}$ and $\sim$ 3.9 M$_{\odot}$ yr$^{-1}$ respectively. Given its SED-derived age of $\sim 250$ Myr, this suggests that Object 77821 did not recently undergo a starburst, and might in fact have an established supernova-driven continuum. In addition, its high mass makes it a questionable LBA candidate. Its radio contours, coincident with SDSS detection, are shown in figure \ref{tab:Contours1}. 

\paragraph*{Object 76079} produced one of the highest signal-to-noise ratio of our VLA observations, with an $11.3 \sigma$ detection (see figure \ref{tab:Contours1} for its contour plot). Casa's `imfit' procedure was able to deconvolve the source from the beam, measuring an angular size of $(2.11 \pm 0.45)\times(0.48 \pm 0.45)$ arcsec$^2$. SED-fitting indicates that the source has a dominant stellar population aged $\sim 400$ Myr and a stellar mass of log(mass)$\sim 9.7$, making Object 76079 an acceptable local analogue galaxy. It is both dust- and metal-poor with an $E(B-V)$ value of $\sim 0.15$ derived from Balmer-decrement measures, and a metallicity $\sim 0.22$Z$_{\odot}$.

\paragraph*{Object 37518} was detected at $\sim5.8\sigma$ in our VLA observations. The `imfit' procedure was able to deconvolve the source from the beam, indicating an angular size of $(3.41 \pm 1.6)\times(0.99 \pm 0.87)$ (its radio contours and SDSS detection are shown in figure \ref{tab:Contours1}). At an age of $\sim 630$ Myr, this object is one of the oldest of our sample. This might explain why we find good agreement between the derived radio SFR of $2.40\pm0.41$ M$_{\odot}$ yr$^{-1}$ and its H$\alpha$-derived SFR of $\sim2.7$ M$_{\odot}$ yr$^{-1}$. Both its old age and high stellar mass of log(mass)$\sim9.8$ make this galaxy an unlikely local analogue. 

\paragraph*{Object 62100} was observed with the VLA, producing a 4$\sigma$ detection, and with ATCA, where it was marginally detected at 2.4$\sigma$. The resulting radio spectral slope of $0.90 \pm 0.53$ is an outlier in our sample, and should be considered tentative given the low SNR in the ATCA observations. With an SED-derived age of $\sim 20$ Myr, a stellar mass of log(mass)$\sim 8.7$, and a gas-phase metallicity of 0.09 Z$_{\odot}$, this is a good analogue to $z>5$ LBGs. We find that it has a relatively low 1.5 GHz SFR of $\sim$ 0.7 M$_{\odot}$ yr$^{-1}$; however, this does not take into account that the standard SFR calibrators assume a stable stellar population of $>100$ Myr, making it likely that the true SFR within Object 62100 is substantially higher. Its radio detection is coincident with SDSS observations, and a contour plot of the source is shown in figure \ref{tab:Contours2}. We note that a beam residual from a neighbouring source passes close to the object and appears in figure \ref{tab:Contours2}, but does not affect its flux measurement. 

\paragraph*{Object 27825} was observed with the VLA with a 4.4$\sigma$ detection (see figure \ref{tab:Contours2} for its contour plot). At log(mass)$\sim$9.8 and a dominant stellar population age of $\sim 500$ Myr (both from SED-fitting) it is both one of the most massive and oldest galaxies of our radio sources, making it unlikely to be a very good local analogue to the earliest galaxies, but a reasonable one for the $z\sim3$ galaxy population.

\paragraph*{Object 57993} was detected at $4.7\sigma$ in our VLA observations. Casa's `imfit' procedure was able to deconvolve the source from the beam and derived one of the largest angular sizes in our sample, $(4.4 \pm 1.3)\times(3.3 \pm 1.2)$ arcsec$^2$, for it (see figure \ref{tab:Contours2}). With an SED-derived stellar mass of log(mass)$\sim 9.7$ and stellar population age of $\sim 300$ Myr, Object 57993 is a good local analogue galaxy. It has little dust extinction ($E(B-V)\sim0.13$), and a metallicity of $\sim0.40$Z$_{\odot}$.

\paragraph*{Object 91412} was observed by the VLA and detected at 6.0 $\sigma$. It is clearly coincident with its SDSS detection, as shown in figure \ref{tab:Contours2}. With a best-fitting stellar population age of $\sim 200$ Myr, and a stellar mass of log(mass) $\sim$ 9.6, this source is an acceptable local analogue to the earliest galaxies. It is a very metal- and dust-poor galaxy with a gas-phase metallicity of $\sim 0.24$ Z$_{\odot}$ and a Balmer-decrememnt derived dust extinction of $E(B-V)\sim0.05$. With a radio-inferred SFR of $4.96\pm0.82$ M$_{\odot}$ yr$^{-1}$ and H$\alpha$ inferred SFR a factor of $\sim2$ higher, it is plausible that the deficit in radio flux in this source is due to its young stellar age.

\paragraph*{Object 56415} was undetected in our VLA observations, putting an upper limit of $<5.5$ M$_{\odot}$ yr$^{-1}$ on its SFR. We show its SDSS detection in figure \ref{tab:Contours2}. At a dominant stellar population age of $\sim$ 500 Myr and an SED-derived stellar mass of log(mass)$\sim$9.7, this object would be an outlier of the typical $z\sim5$ LBG population. 

\paragraph*{Object 34524} is one of the oldest and most massive sources in our sample with an SED-derived stellar population age of $\sim 800$ Myr and a mass of log(mass)$\sim 9.9$. This object would hence be an extreme outlier of the $z\sim5$ LBG population, but might make a good analogue to the more nearby $z\sim2-3$ galaxy population. The source is not detected above a 3 $\sigma$ limit in our VLA observations, indicating an upper 1.5 GHz radio SFR of $<2.5$ M$_{\odot}$ yr$^{-1}$. Its SDSS detection is shown in figure \ref{tab:Contours2}. 

\paragraph*{Object 23218} was detected at 4.9 $\sigma$ in VLA observations. At $(3.6 \pm 1.1)\times(3.2 \pm 1.3)$ arcsec$^2$, its deconvolved angular size (according to casa's `imfit' procedure; see figure \ref{tab:Contours2} for a contour plot) is one of the largest in the sample. Together with a radio-derived SFR of $\sim 3.2$ M$_{\odot}$ yr$^{-1}$, this indicates a low $\Sigma_{SFR}$ of $\sim 0.06$ M$_{\odot}$ yr$^{-1}$ kpc$^{-2}$ for this source. Its SED-derived mass of log(mass)$\sim 9.9$, and stellar population age of $>600$ Myr make Object 23218 an extreme object compared to the typical $z\sim5$ LBG population.

\paragraph*{Object 05083} is the second of our sources which was observed by VLA, ATCA, and APEX. It is one of the most massive and oldest of our radio sources with log(mass)$\sim$9.9 and an age of $\sim 500$ Myr. It has low dust reddening with an $E(B-V)$ value derived from the Balmer decrement of 0.12, and a low metallicity of 0.34 Z$_{\odot}$, making it a potentially acceptable local analogue. It was observed, but not detected, using APEX, with a total observing time of 1.2 hours and an rms of 3.9 $\mu$Jy. Interestingly, however, it was detected in both VLA and ATCA observations with a signal-to-noise ratio of 4.1, 3.4 and 2.5 at 1.5\,GHz, 5.5\,GHz and 9\,GHz respectively. From these measurements, the spectral slope between 1.5 and 5.5 GHz could be constrained to be $-0.52 \pm 0.42$. Together with its inferred 1.5 GHz star formation rate of $\sim6.1$ M$_{\odot}$ yr$^{-1}$, this supports the interpretation of this source as a recent starburst. Its radio detection is clearly coincident with its SDSS observation, as shown in figure \ref{tab:Contours2}. 

\paragraph*{Object 71294} was observed by both the VLA and ATCA, with detections in both bands, allowing us to constrain its radio spectral slope to $-0.6 \pm 0.42$. This is consistent with a radio spectrum arising from thermal emission, as well as sychrotron radiation due to a recent starburst. This interpretation is given further support by the object's SED-derived age of $\sim50$ Myr. Being one of our most metal-poor objects in our radio sample with Z = 0.16 Z$_{\odot}$ makes this a good analogue galaxy to the distant LBG populaiton. Its radio contours and SDSS detection are shown in figure \ref{tab:Contours2}. 

\paragraph*{Object 08586} was observed with the VLA, producing a 5.6 $\sigma$ detection and indicating a radio SFR of $\sim 2.7$ M$_{\odot}$ yr$^{-1}$. Casa's `imfit' procedure was able to deconvolve the object from the beam (see figure \ref{tab:Contours2}), determining an angular size of $(2.17 \pm 0.72) \times (1.74 \pm 0.9)$ arcsec$^2$. However, with an SED-derived stellar mass of log(mass) $\sim 9.9$, and a best-fitting stellar population age of $\sim$ 500 Myr, this source is an likely to constitute an outlier of the typical $z\sim5$ LBG sample.

\paragraph*{Object 72856} is one of the most dust-poor objects in our radio sample with $E(B-V)= 0.09$ from Balmer decrement measurements. Its SED-derived age of 100 Myr, stellar mass of log(mass)$\sim$9.4, and metallicity of 0.24 Z$_{\odot}$ make it an acceptable LBA candidate. Both APEX and VLA observations of this source resulted in non-detections, producing an upper 1.5 GHz SFR limit of $<2.4$M$_{\odot}$ yr$^{-1}$. Figure \ref{tab:Contours2} shows its radio contours overplotted on its SDSS detection. 

\paragraph*{Object 29942} is a relatively massive source in our sample with an SED-derived stellar mass of log(mass)$\sim$9.7. It is also among the older sources in this sample (with an SED-derived age of 500 Myr), making it an unlikely local analogue. It was undetected in VLA observations, producing an upper 1.5 GHz SFR limit of $<2$M$_{\odot}$ yr$^{-1}$ (see figure \ref{tab:Contours2} for a contour plot of this source). 

\paragraph*{Object 32420} is both one of the most massive of our radio sources with log(mass)$\sim$10.2 found using SED-fitting, and the most highly star forming galaxy in our radio sample with H$\alpha$ SFR = 41.1M$_{\odot}$ yr$^{-1}$. Its stellar population age of $\sim 500$ Myr, together with its other physical properties, makes it an unlikely LBA candidate. The galaxy was observed with both APEX and the VLA, resulting in a non-detection at 345 GHz, but a 5$\sigma$ detection at 1.5GHz (see figure \ref{tab:Contours3} for its radio contours and SDSS detection). Its 1.5 GHz flux indicates a SFR of $11.0 \pm 2.2$ M$_{\odot}$ yr$^{-1}$. 

\paragraph*{Object 08755} is our most distant source observed with both the VLA and ATCA, producing strong ($>7\sigma$) detections in both (see figure \ref{tab:Contours3} for its radio contours). From this, a radio spectral slope of $-0.36\pm0.20$ could be determined. With a stellar mass of log(mass)$\sim$9.7 and a dominant stellar population age of $\sim 125$ Myr, this source is an acceptable local analogue to the distant galaxy population. Its strong VLA-detected flux resulted in Object 08755 having the highest SFR of the sources in our radio sample. 

\paragraph*{Object 38857} was detected in our VLA observations, with one of the highest inferred radio SFRs (of $12.0\pm1.8$ M$_{\odot}$ yr$^{-1}$). Its stellar mass of log(mass)$\sim$9.7, SED-derived age of $\sim$ 100 Myr, low dust obscuration and $\sim 0.28$ Z$_{\odot}$  make this source an acceptable analogue to the distant galaxy population. It is clearly coincident with its SDSS detection, as shown in figure \ref{tab:Contours3}. 

\paragraph*{Object 55849} was not detected above the noise level in our VLA observations, producing an upper limit on the SFR within it of $<5.8$ M$_{\odot}$ yr$^{-1}$. With a stellar mass of log(mass)$\sim$9.7 and a stellar population age of $\sim250$ Myr, this is an unlikely analogue to $z\sim5$ LBGs. See figure \ref{tab:Contours3} for its SDSS detection. 

\paragraph*{Object 15004} produced a very strong detection of 11 $\sigma$ in our VLA observations. In both H$\alpha$ and radio, it is among the most star forming of our detected radio sources. We note that this source had been suggested to be a Seyfert 1 on SIMBAD due to its strong emission lines. Its stellar mass of log(mass)$\sim$9.6 and population age of $\sim 200$ Myr make this a plausible local analogue galaxy with low dust ($E(B-V)=0.15$) and metallicity (0.26 Z$_{\odot}$). Figure \ref{tab:Contours3} shows the radio contours overplotted on the SDSS image. Both show a compact source with an upper limit on the projected area of 1.4 $\times$ 1.09 arcsec$^2$. 

\paragraph*{Object 92239} is one of our most metal-poor and dust-poor of our radio objects with Z = 0.17$_{\odot}$ and $E(B-V) = 0.04$ respectively, while also being one of the most massive of our radio sources with log(mass)$\sim$9.8 as found from SED-fitting. Its dominant stellar population age of $\sim300$ Myr, in conjunction with its high mass, makes this an unlikely local analogue for high-redshift LBGs. The source was observed, but not detected, using APEX.

\paragraph*{Object 95952} is the most distant of our radio objects at z=0.198. It was detected, though not resolved, at $\sim 3 \sigma$ in our VLA observations (see figure \ref{tab:Contours3} for its SDSS detection and radio contours). Its SED-derived age of 150 Myr, stellar mass of log(mass)$\sim$9.6, and $\sim 0.4$ Z$_{\odot}$ make it an acceptable local analogue. It has an above-average H$\alpha$ SFR of 10.4 M$_{\odot}$ yr$^{-1}$, and a radio SFR of $\sim 3.3$ M$_{\odot}$ yr$^{-1}$. The dust extinction calculated from SDSS spectroscopy indicates a low extinction with $E(B-V)= 0.10$ from Balmer decrement measurements.

\label{lastpage}

\bsp
\end{document}